\documentclass{article}
\usepackage{graphicx} 
\usepackage{amsmath}
\usepackage{amsfonts}
\usepackage{amssymb}
\usepackage{xcolor}
\usepackage{overpic}
\usepackage{soul}
\usepackage{csquotes}
\usepackage{authblk}
\usepackage[breaklinks=true]{hyperref}
\usepackage[style=numeric, sorting=none, backend=biber]{biblatex}
\usepackage{multirow}
\usepackage{subcaption}
\newcommand{\labelphantom}[1]{%
{\phantomsubcaption%
\label{#1}}%
}%
\title{Compact Experimental Negative TriAngUlarity Reactor (CENTAUR): A design study for a compact, affordable breakeven tokamak}
\author[1,2]{The CENTAUR Collaboration} 
\author[1*]{S. W. Freiberger} 
\author[1]{E. Bursch}
\author[1]{J. Chiriboga}
\author[2]{H.J. Farre-Kaga}
\author[1]{E. Felske}
\author[1]{S. Guizzo}
\author[1]{J. Labbate}
\author[1]{S. Seethalla}
\author[1]{F. Sheehan}
\author[1]{J. L. Xia}
\author[1]{A. Braun}
\author[1]{D.A. Burgess}
\author[2]{N. Chen}
\author[1]{J. Halpern}
\author[1]{M. Haque}
\author[1]{A. Hyder}
\author[1]{A. Lachmann}
\author[1]{R. Lopez }
\author[2]{K. Orr}
\author[1]{K. Richardson}
\author[1]{M. Russo}
\author[1]{A. Veksler}
\author[1]{C. J. Hansen}
\author[4]{A. Holm}
\author[1]{N. Leuthold}
\author[3]{O. Meneghini}
\author[1]{A. O. Nelson}
\author[1]{M. Pharr}
\author[3]{T. Slendebroek}
\author[1]{I. G. Stewart}
\author[4]{F. Scotti}
\author[1]{M. Tobin}
\author[1]{H. Wilson}
\author[1]{C.F.B. Zimmermann}
\author[1]{C. Paz-Soldan}

\affil[1]{Columbia University, New York, NY}
\affil[2]{Princeton University, Princeton, NJ}
\affil[3]{General Atomics, San Diego, CA}
\affil[4]{Lawrence Livermore National Laboratory, Livermore, CA}
\date{April 2026}

\begin{document}
\maketitle

\begin{abstract}
\noindent 
This work presents the compact experimental negative triangularity reactor (CENTAUR), a low overnight cost, high-field tokamak, breakeven reactor design, achieving a predicted total fusion power of 40~MW and scientific energy gain of 1.3. Ballooning stability calculations confirm that the device's pedestal is within the first stability regime, which is consistent with the expected ELM-free operation associated with negative triangularity (NT) plasmas. The geometry of the NT divertor allows for high fraction of radiated power (13.5$\%$) between the separatrix and plasma facing components. Heat transport modeling based on simulations of the edge region show heat loads into plasma facing components well below material limits. The magnet system employs rare-earth barium copper oxide (REBCO) high-temperature superconductors in 18 toroidal field coils, an hourglass-shaped central solenoid, and six poloidal field coils to support high-field ($B_0=10.9$~T) plasma confinement, shaping, and current drive. Neutronics analysis shows that a 12~cm $B_4C$ shield keeps superconducting magnet heating below the 33~K quench limit during 10~s, 40~MW DT pulses. With this shielding, the modeled fluence indicates HTS components can survive more than ten times the 3000-pulse design lifetime. Iteration of economic analysis in tandem with the technical design process allows CENTAUR to achieve its overnight cost goal of $\$$2B determined using a custom costing model that predicts a total overnight cost of $1.6$B$\pm0.2$B.

\end{abstract}

 \section{Introduction and Overview}
\label{sec:intro}
Most experimental tokamaks operate with a positive triangularity equilibrium geometry and achieve maximum performance in High-confinement mode (H-mode). However, H-mode plasmas require high power through the edge region and are prone to periodic, uncontrolled exhaust of heat and impurities in the form of Edge Localized Modes (ELMs) \cite{Scotti_2024, nelson_2023, Austin_2021}. These events can cause damage to plasma facing components and reduce the operational lifetime of any such device. Negative triangularity (NT) plasmas operate in an ELM-free regime while providing the requisite plasma performance for reactor designs. Further, high-field operation in positive triangularity exposes plasma facing components to extremely high load, creating complex engineering problems for power handling \cite{kuang_2020}. NT offers the dual advantages of increased radiation, which may be beneficial for detached operation, and increased flexibility in poloidal field (PF) coil placement \cite{medvedev_2015}. The increased capacity to radiate power between the separatrix and the divertor plate protects plasma facing materials from high heat loads. Further, the negative triangularity geometry allows for a broader strike point in the divertor. Device longevity is essential to the economic viability of a future power plant, so the exploration of NT tokamak designs is a valuable area of fusion technology development \cite{Slendebroek_2026,GUIZZO2025115257}.

This design study explores the potential for an NT tokamak to attain physics gain (Q$>$1) at a sub-$\$$2B operational cost. This device would serve as a stepping stone towards a larger fusion power plant, such as the proposed Modular Adjustable Negative Triangularity ARC-class (MANTA) device, bringing fusion closer to grid deployment \cite{the_manta_collaboration_manta_2024}. The study expands on a tradition of fusion design classes, which produced the first ARC design \cite{sorbom2015} and the design study for the MANTA device. Unlike in positive triangularity devices that achieve maximal operating performance in high-confinement regimes (H-mode), negative triangularity (NT) plasmas can achieve high performance while remaining in a naturally ELM-free state \cite{nelson_2024}. 

\begin{figure}[htbp]
    \centering
    \includegraphics[width=\linewidth]{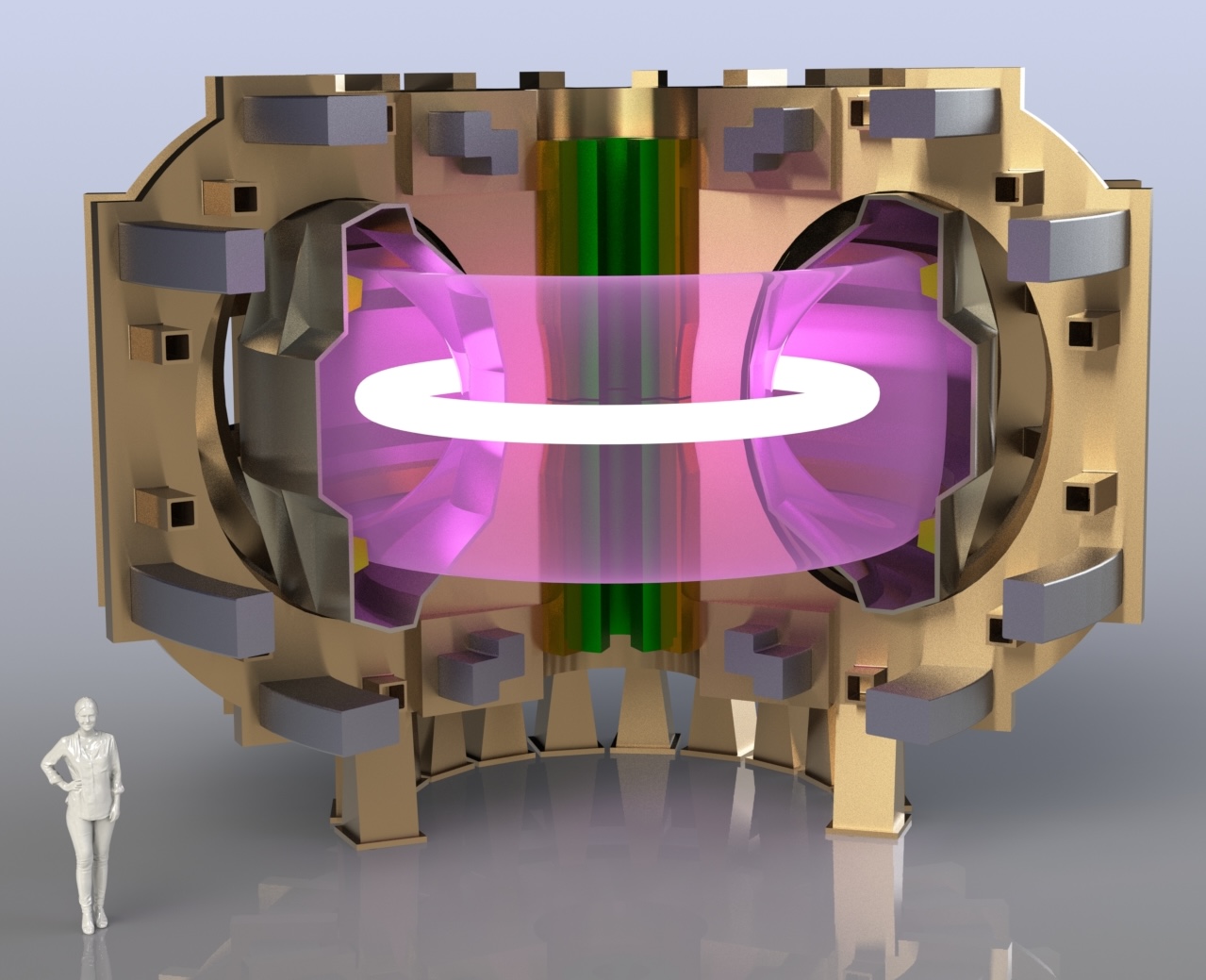}
    \caption{Full CAD render of the CENTAUR design. The volume in pink is representative of the plasma. The metal rendered in gold is the TF casing structure. The gray metal volumes are PF coils, and the green section at the center is the central solenoid.}
    \label{fig:CADrender}
\end{figure}

This paper presents a self-consistent design study of a compact, high-field, breakeven negative triangularity tokamak. The analysis is presented from the plasma core outwards. Section \ref{sec:core} describes the initial scoping of the core scenario and analysis with a more complete integrated core modeling workflow. Section \ref{sec:power} presents core-edge integration and analysis of heat transport between the separatrix and the divertor plates as well as heat transfer modeling for a simplified divertor plate and support structure. Section \ref{sec:magnet} presents a novel configuration of toroidal field coils to minimize system stresses in an NT geometry and an ``hourglass" center stack design to provide the requisite flux swing for startup. This section further presents optimized poloidal field coil placement as well as modeling of stresses from vertical displacement events. Analysis of the impact of neutron damage on system lifetime and neutron heating on magnet performance are discussed in section \ref{sec:neutron}. Finally, section \ref{sec:econ} provides costing analysis based on the that in \cite{the_manta_collaboration_manta_2024}, concluding that this device is well within the design target overnight cost of \$2B.

\begin{table}[htbp]
    \centering
    \caption{CENTAUR Key Design Parameters}
    \label{tab:centaur-params}
    \begin{tabular}{|c|c|c|} \hline
  Parameter & Symbol & Value \\ \hline
 \text{Fusion Power} & $P_{\text{fus}}$ & 40 MW \\ \hline
 \text{ICH coupled power} & $P_\text{ICH}$ & 30 MW \\ \hline
 \text{Scrape-off layer power} & $P_\text{SOL}$ & 9.3 MW \\ \hline
 \text{Radiated power (core \textbar{} SOL)} & $P_\text{rad}$ & 23 MW \textbar{} 5.4 MW \\ \hline
 \text{Radiated power fraction (core \textbar{} SOL)} & $f_\text{rad}$ & 0.62 \textbar{} 0.135 \\ \hline
 \text{Plasma gain} & Q & 1.3 \\ \hline
 \text{Major radius} & $R_0$ & 2 m \\ \hline
 \text{Minor radius} & a & 0.72 m \\ \hline
 \text{Elongation} & $\kappa_{edge}$ & 1.65  \\ \hline 
 \text{Triangularity} &  $\delta$ & -0.55  \\ \hline
 \text{Plasma volume} & $V_{p}$ & 29.7 m$^3$ \\ \hline
 \text{Plasma surface area} & $A_p$ &  2.34 m$^2$ \\ \hline
 \text{On-Axis Toroidal magnetic field} & $B_0$ & 10.9 T\\ \hline
 \text{Plasma current} & $I_p$ & 9.6 MA \\ \hline
 \text{Bootstrap fraction} & $f_{\text{BS}}$ & 0.103 \\ \hline
 \text{Avg. ion temperature} & $\langle T_i \rangle$ &  5.15 keV \\ \hline
 \text{Avg. electron temperature} & $\langle T_e \rangle$ & 4.15 keV \\ \hline
 \text{Avg. density} & $\langle n \rangle$ & $3.21 \cdot 10^{20} \; \text{m}^{-3}$  \\ \hline
 \text{On-axis ion temperature} & $T_{i,0}$ & 14.8 keV \\ \hline
 \text{On-axis e\textsuperscript{-} temperature} & $T_{e,0}$ & 9.26 keV \\ \hline
 \text{On-axis e\textsuperscript{-} density} & $n_0$ &  $4.33 \cdot 10^{20} \; \text{m}^{-3}$  \\ \hline 
 \text{Average Greenwald fraction} & $f_\text{GW}$ & 0.61 \\ \hline
 $f_\text{GW}$ \text{at separatrix} & $f_\text{GW, sep}$ & 0.13 \\ \hline
 \text{Pulse length} & $\tau_\text{pulse}$ & 10 s \\ \hline
 \text{Normalized beta} & $\beta_N$ & 1.5 \\ \hline
 \text{Average effective charge ($\rho=[0,0.96]$)} & $Z_\text{eff}$ & 1.43 \\ \hline 
 \text{Safety factor at }$\Psi_N = 0.95$ & $q_{95}$ & 2.58 \\ \hline
  \text{Safety factor at }$\Psi_N = 0$ & $q_{0}$ & 1.04 \\ \hline

 \text{Energy confinement time} & $\tau_E$ & 0.41 s\\ \hline
 \text{H\textsubscript{98,y2} confinement factor} & $H_{98, y2}$ & 0.54 \\ \hline
 \text{H\textsubscript{NT, 24} confinement factor \cite{lunia2025energyconfinementtimescaling}} & $H_{NT}$ & 0.87 \\ \hline
 \text{Loop voltage} & $V_\text{loop}$ & 0.77 V \\ \hline
    \end{tabular}
\end{table}

\begin{figure}[htbp]
    \centering
    \label{fig:TM_equil}
    \includegraphics[width=0.8\linewidth]{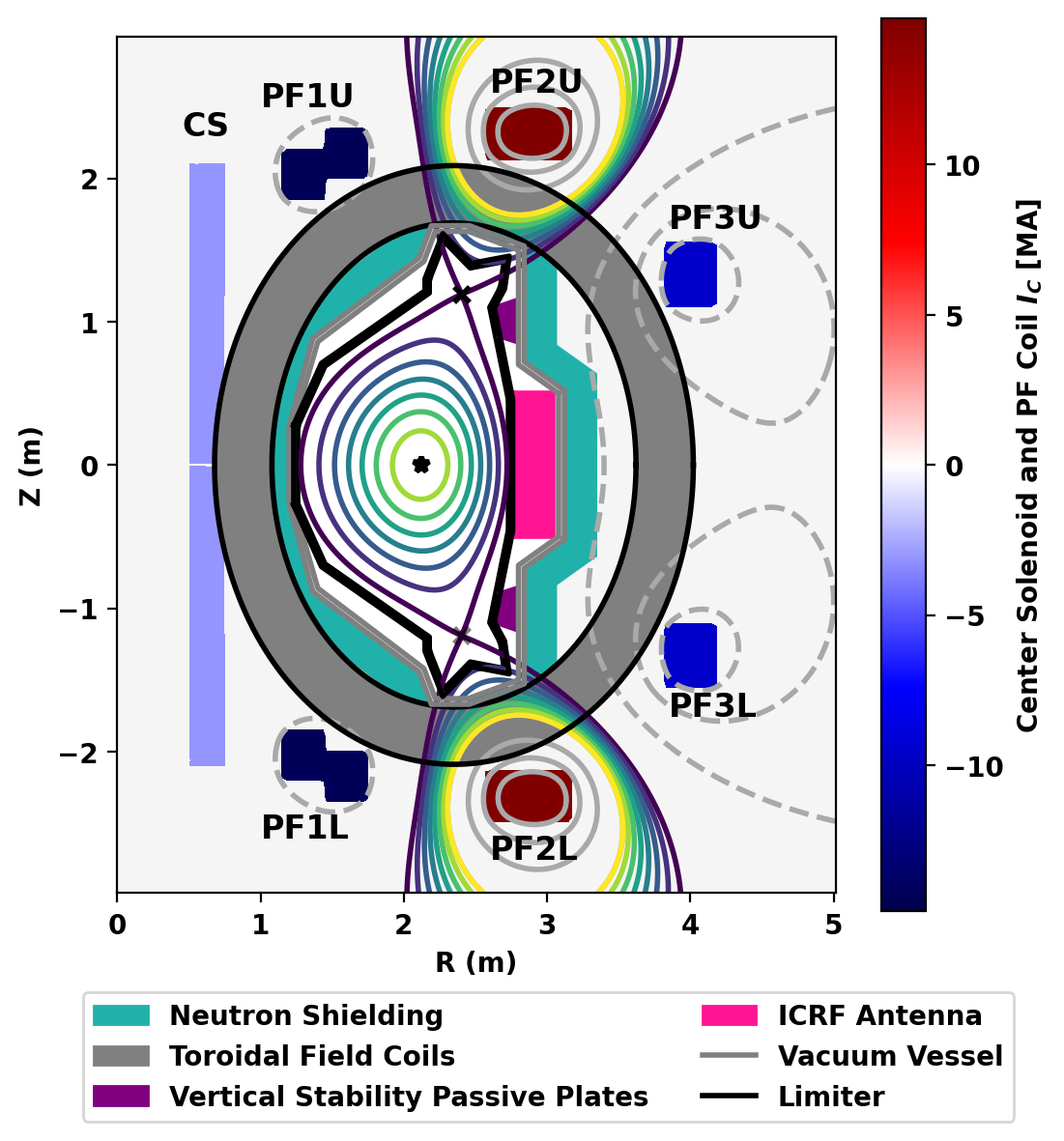}
    \caption{Poloidal cross section showing the primary components and the equilibrium solution produced with TokaMaker \cite{Hansen_2024_tokamaker}}
    \label{fig:tm_cross_section}
\end{figure}

\section{Core Plasma Scenario}
\label{sec:core} 
Negative triangularity has significant operational advantages that make it an attractive candidate for the operating scenario in a fusion pilot plant. Notably, NT has demonstrated energy confinement times ($\tau_E$) and normalized beta ($\beta_N$) comparable to H-mode operation while remaining robustly ELM-free \cite{10.1063/1.5091802, PhysRevLett.122.115001}. These properties motivate the development of breakeven class NT devices as a step toward a fusion power plant. However, building a $\mathrm{Q > 1}$ experimental device requires the identification and modeling of a viable core operating point. This work identifies one possible core operating scenario, providing a concrete starting point for further evaluation of NT as a path toward fusion energy. 

\subsection{0-D Scoping of the Core Operating Regime}
The open source code CFSpopcon \cite{Body2024-bn} solves the global power-balance equation \cite{Houlberg_1982}, generating a 0-D POPCON (Plasma OPerational CONtours) analysis for broad scoping of the initial core operating scenario. Inputs for the POPCON include CENTAUR's argon impurity concentration, the H98y2 confinement factor, major radius, plasma current, and toroidal magnetic field. 

\begin{figure}[htbp]
    \centering
    \includegraphics[width=0.8\linewidth]{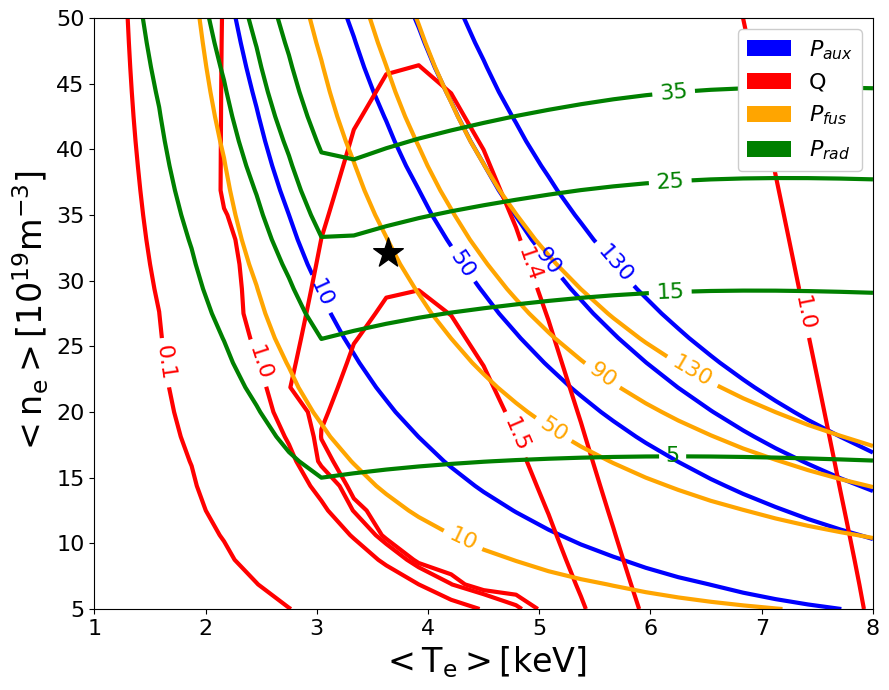}
    \caption{POPCON visualization of parameter space surrounding CENTAUR's operating point. The intended operating point is marked with a black star.}
    \label{fig:popcon}
\end{figure}

Using the final operating point's H98y2 confinement factor, the resulting POPCON for CENTAUR in Figure \ref{fig:popcon} maps fusion power, plasma gain, auxiliary power, and radiative power across electron temperature-density space. The final operating point is depicted as a black star. Notably, the parameter space permits higher fusion power at increased temperature or higher plasma gain at lower densities, giving flexibility for future optimization as NT performance scalings are expanded and validated.

\subsection{Plasma Core Workflow}

\begin{figure}[htbp]
    \centering
    \includegraphics[width=0.8\linewidth]{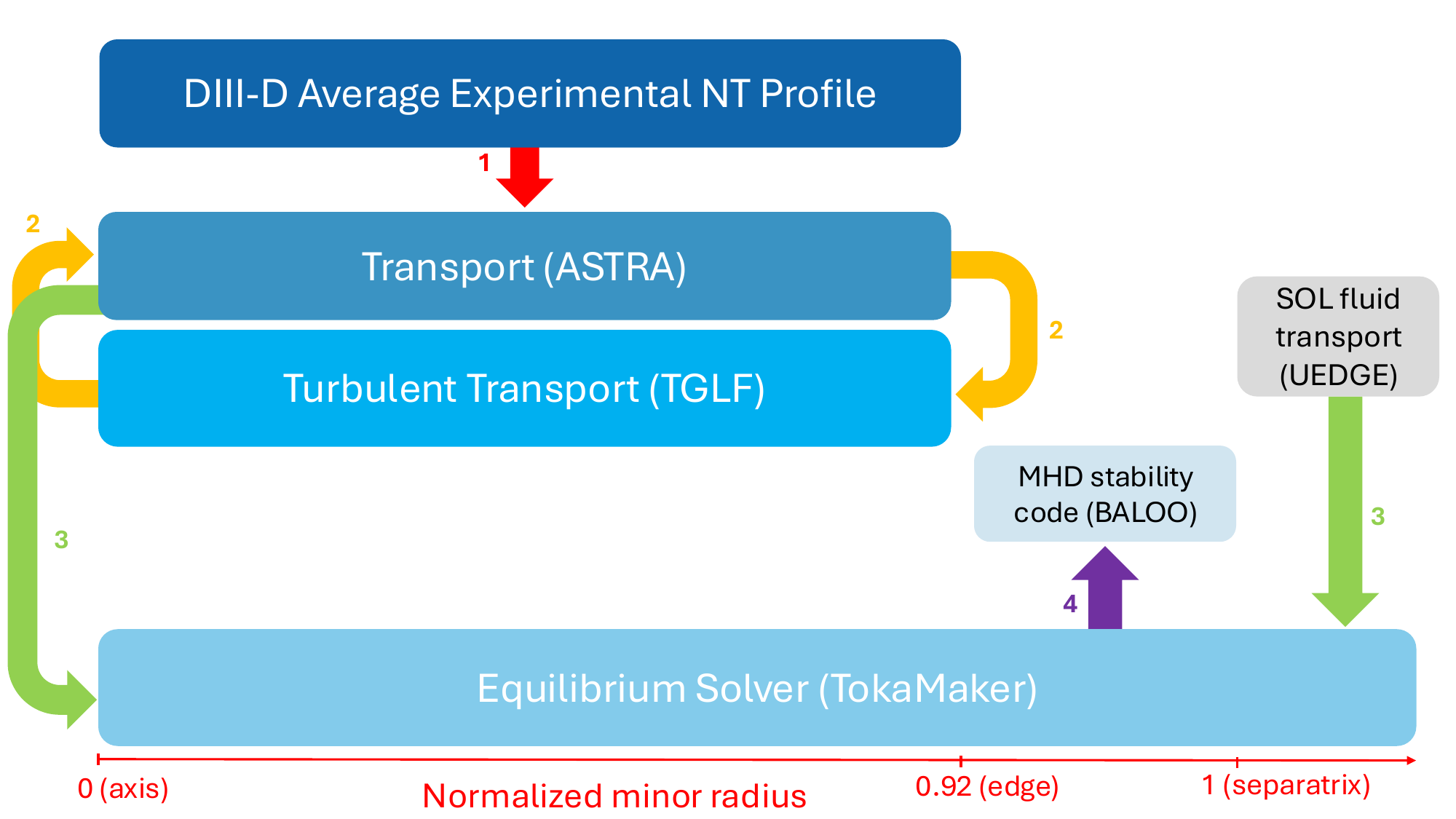}
    \caption{Integrated modeling flowchart showing the codes used over various $\rho$ regions and the iterative steps between them.}
    \label{fig:integrated_model}
\end{figure}

Building on the initial operating point scoped by the 0-D POPCONs, the core integrated workflow employs higher-fidelity codes to verify and refine this solution into a realistic, self-consistent core scenario, as depicted in Figure \ref{fig:integrated_model}. This workflow uses the following code bases: ASTRA~\cite{pereverzev2002astra}, a 1.5D modular plasma transport code; TGLF, a quasi-linear turbulence transport code \cite{10.1063/1.2044587}; TokaMaker, an open source Grad-Shafranov equilibrium solver \cite{HANSEN2024109111}; UEDGE, a 2D fluid edge transport code \cite{rognlienFullyImplicitTime1992}; and BALOO, a ballooning stability code \cite{10.1063/1.872193}.

This medium fidelity core modeling workflow begins with a characteristic DIII-D NT electron density profile from experimental data \cite{Thome_2024}, scaled to the the initial operating point determined through POPCON analysis. This initial profile and a Gaussian approximation of an ion cyclotron resonant heating (ICRH) profile then act as inputs to ASTRA. Using the TGLF submodule to model radial transport, the profiles time-evolve to steady state. The convergence criterion is that the change in on-axis electron and ion temperature between successive iterations to be less than 0.01~keV. Profiles from ASTRA are truncated at $\rho = 0.92$, where the density and temperature are imposed as interface boundary conditions for UEDGE. UEDGE then solves the scrape-off layer and divertor plasma from the separatrix outward to the wall.

\begin{figure}[htbp]
  \centering
  \begin{overpic}[width=0.6\linewidth]{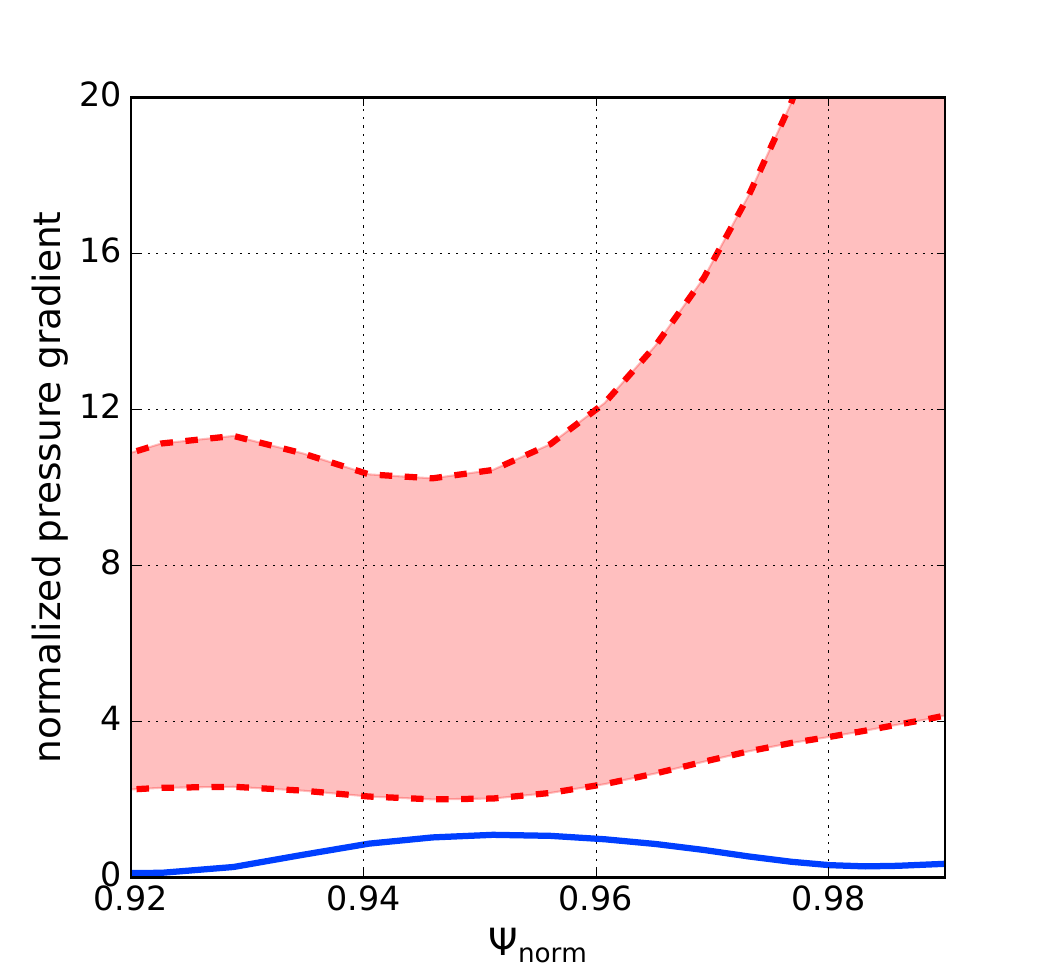}
  \footnotesize{
    \put(20,65){ \textbf{2nd stability}}
    \put(20,30){\textcolor{red}{\textbf{ballooning unstable}}}
    \put(60,15){ \textbf{1st stability}}
    }
    \end{overpic}
    \caption{The normalized pressure gradient (blue curve) occupies the 1st stability region below the ballooning unstable curve (red) calculated by the BALOO code.}
    \label{fig:baloo}
\end{figure}

Due to uncertainty and lack of fully predictive NT edge models \cite{nelsonCharacterizationELMfreeNegative2024b}, a cubic spline connects the ASTRA edge at $\rho = 0.92$ to the separatrix values from UEDGE. Here, the profile shape is modeled after the profile form realized in a large database of high-power NT discharges on DIII-D \cite{nelsonCharacterizationELMfreeNegative2024b}. BALOO evaluates the stability of this profile to verify that the profile is infinite-n ballooning stable. As shown in Figure \ref{fig:baloo}, the normalized pressure gradient for CENTAUR is well below the 1st stability limit for infinite n-ballooning modes, as expected for NT ELM-free regimes. 

Future work could expand on this study in several important directions, including more realistic heating profiles, higher-fidelity turbulence modeling, and input from broader NT experimental validation. Self-consistent ICRH simulations (instead of the prescribed Gaussian heating profiles) would allow for higher-fidelity treatment of heating deposition. In addition, coupling the present framework with gyrokinetic simulations could provide more accurate treatment of turbulence and transport, particularly in the region near the last closed flux surface (LCFS), allowing for a more accurate characterization of profiles in the plasma edge. Beyond heating and turbulence, introducing density evolution into the simulations through the inclusion of a fueling source, would enable a more complete description of coupled particle and energy transport. A current scan could also provide insight on how plasma performance and stability vary with both the magnitude and profile of the plasma current. Finally, further NT experiments across a wider range of device geometries and magnetic field configurations would be valuable for validating the robustness of NT core predictions and assessing their generality across different tokamak regimes.

\begin{figure}[htbp]
    \centering
    \includegraphics[width=1\linewidth]{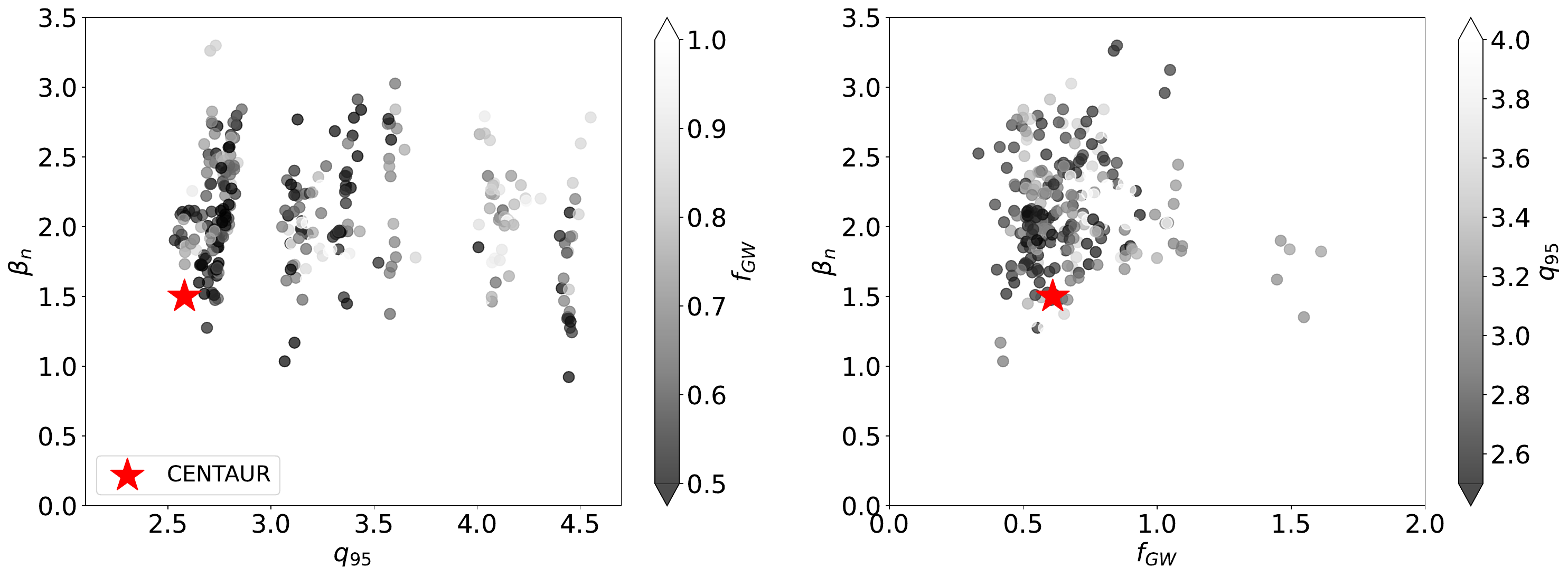}
    \caption{Operating point indicated by stars at $\beta_N=1.5$, $f_{GW}=0.61$, and $q_{95}=2.58$. Operational space data from supplementary data supplied with the work by Paz-Soldan et. al. \cite{paz-soldan_2024}. Stars indicate where the CENTAUR operating plot falls within operating space of strongly shaped, negative triangularity plasmas on DIII-D.}
    \label{fig:beta_gw_operating_space}
\end{figure}

\subsection{Transport simulation with TGLF in ASTRA}
The ASTRA modeling \cite{ASTRA_Reference_Paper,Fable_2013} framework, coupled to the TGLF turbulence code \cite{Staebler_2007}, simulates transport in CENTAUR. ASTRA solves the time-dependent 1D transport equations with turbulent fluxes provided by TGLF. All TGLF runs use the SAT2 saturation rule, which has been shown to be sufficient for NT plasmas \cite{mcclenaghan2024examining, Aucone_2024}. The simulation includes electrons, a 50/50 mix of deuterium/tritium (D-T), and argon with an average impurity density of $n_{\text{Ar}} = 8.1 \cdot 10^{17} \mathrm{m}^{-3}$. Preliminary scans of the impurity density find that CENTAUR's performance remains relatively robust, with increasing the average impurity density by 11\% only resulting in a decrease of fusion gain by 8\%. Further modeling of impurity transport is out-of-scope for the present analysis. This argon impurity is implemented in ASTRA through a prescribed average core $Z_\text{eff}$ of 1.4, which monotonically increases from 1.0 at $\rho = 1.0$ to 1.6 at $\rho = 0.92$. Bremsstrahlung and line radiation are also self-consistently included. The temperature profile is evolved until to a steady state, with the density profiles held fixed to an edge electron density of $ n_{e, \mathrm{ped}}  = 2.55 \cdot 10^{20} \; \mathrm{m}^{-3}$ and edge electron temperature of $T_{e, \mathrm{ped}} = 1.8 \;\mathrm{keV}$ at $\rho = 0.92$. These boundary conditions are selected to reflect a pedestal sufficient to achieve $Q>1$, as the NT edge pressure pedestal is closely correlated with overall plasma performance \cite{wilsonCharacterizingNegativeTriangularity2025a}. The ASTRA-TGLF simulation is limited to $\rho = 0.92$ as TGLF does not capture pedestal physics and cannot properly model the edge region in NT; similar approaches have been adopted in previous ASTRA-TGLF studies of NT plasmas with simulation boundaries extended to $\rho \approx 0.94-0.95$ \cite{Aucone_2024, Mariani_2024}. 

\begin{figure}[htbp]
  \centering
  \begin{overpic}[width=0.5 \linewidth]{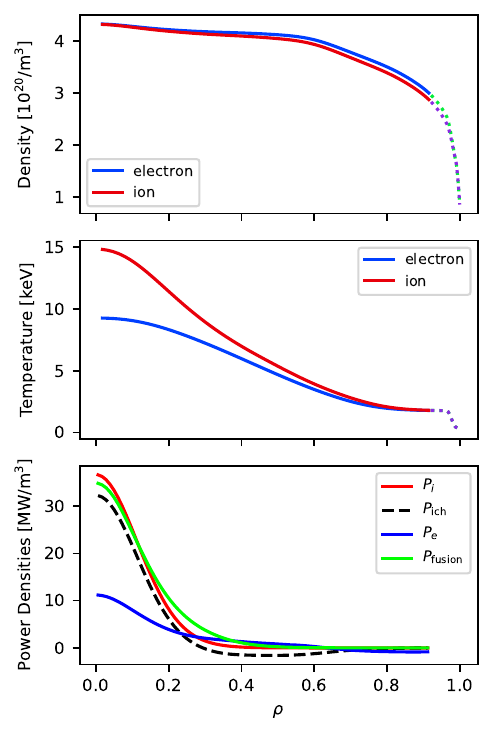}
  \end{overpic}
\caption{Selected ASTRA profiles as a function of $\rho$. 
The dotted segments for $\rho > 0.92$ indicate values beyond the ASTRA computational boundary: (a) electron and ion density profiles,
(b) electron and ion temperature profiles,
(c) power density profiles, including auxiliary heating, ion and electron heat fluxes, and fusion power}
  \label{fig:profiles}
\end{figure}

Figure \ref{fig:profiles} displays the resulting density, temperature, and power density profiles from the converged ASTRA-TGLF simulation. Figure \ref{fig:profiles}(a) shows the ion (blue curve) and electron density profiles (red curve). The electron density profile and $Z_{eff}$ are held fixed, while the ion and impurity densities are calculated to enforce quasi-neutrality; the slight deviation between the electron and ion densities is due to argon impurity density. The ion density represents the combined deuterium and tritium density. Figure \ref{fig:profiles}(b) shows the temperature profile where on-axis $T_{e,0} = 9.26 \; \mathrm{keV}$ and $T_{i,0} = 14.8 \; \mathrm{keV}$ are predicted. The ion temperature profile has stronger peaking in the core relative to the electron temperature due to the prescribed axis-centered ICRH heating profile, shown in \ref{fig:profiles}(c) (black dashed line). Figure \ref{fig:profiles}(c) depicts the power density profiles, including the ion and electron heat fluxes, auxiliary heating, and fusion power. The auxiliary power (black dashed line) is on the same order of the fusion output power (green line), which is expected for a $Q \approx 1$. The ICRH heating deposition mainly contributes to the ions, which explains why the total ion flux (red line) is larger than the electrons (blue line). The peak fusion power density at the core is $P_{\mathrm{fus},0} = 35 \; \mathrm{MW/m^3}$. 

A more consistent and radially resolved analysis of the core impurity transport is left for future work and would require using neoclassical and impurity transport codes as discussed by Fajardo et al. \cite{fajardoAnalyticalModelCombined2023}. Nonetheless, the final operating point, as seen in Figure \ref{fig:beta_gw_operating_space}, falls within the NT operating space defined by the DIII-D database of stationary, strongly-shaped, diverted NT plasmas; confirming that the simulated operating point is consistent with experimentally achieved NT conditions. 

\subsection{Vertical Stability}

\begin{figure}[htbp]
    \centering
    \begin{overpic}[width=0.75\textwidth]{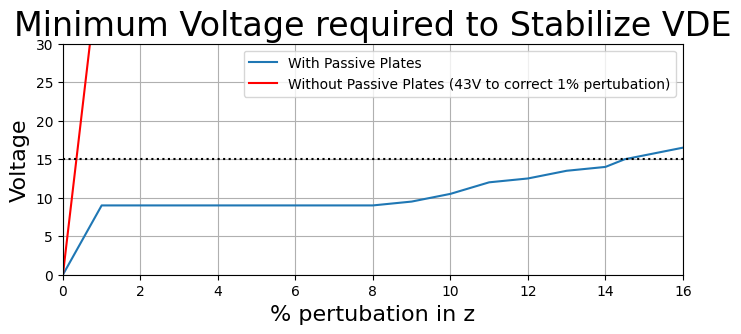}
    \put(-7, 24){\footnotesize{\textbf{(a)}}}
    \end{overpic}
    \begin{overpic}[width=0.9\textwidth]{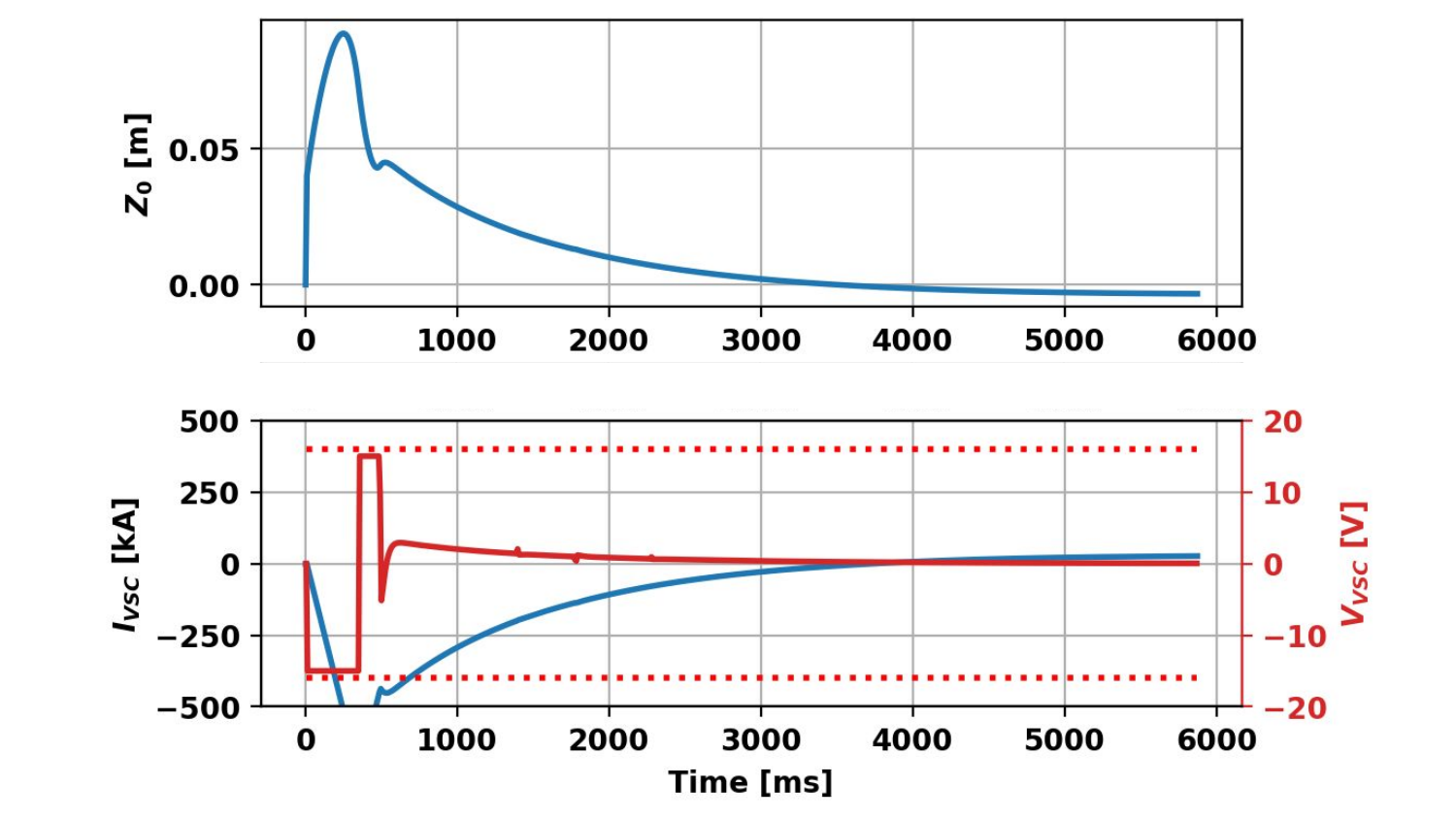}
    \footnotesize{
    \put(3, 45){\textbf{(b)}}
    \put(3, 18){\textbf{(c)}}}
    \end{overpic}
    \caption{\textbf{(a)} Vertical control coil voltage needed to stabilize plasma for a given initial perturbation scale. \textbf{(b)} Vertical position of the plasma over time in the case of a 14.5$\%$ perturbation. \textbf{(c)} Control coil voltage and current over time in the case of a 14.5$\%$ perturbation.}
    \label{fig:VDE}
\end{figure}

Vertical stability is assessed through TokaMaker time-dependent stability simulations, further explained in \cite{guizzoAssessmentVerticalStability2024a}, with the simulation results and vertical stabilization hardware shown in Figure \ref{fig:VDE}. Figure \ref{fig:VDE}b) shows that with the addition of targeted tungsten passive plates and assuming a control coil maximum voltage of 15~V, an estimated 14.5\% perturbation of the minor radius, or 10.4~cm, can be stabilized. This is a larger value than most existing tokamaks \cite{nelson_2024a}, suggesting CENTAUR is sufficiently vertically stable. The 15~V limit for AC superconducting coils is estimated from \cite{sanabria_development_2024}. 

A metric for characterizing the controllability of VDEs is the feedback capability parameter, or the product between VDE growth rate $\gamma$ and wall diffusion time $\tau_W$. CENTAUR has a VDE growth rate $\gamma \approx 12$~s$^{-1}$ and a wall diffusion time $\tau_W \approx 30$~ms, leading to $\gamma \tau_W \approx 0.36$, significantly more stable than a typical target controllable feedback capability parameter of $\gamma \tau_W \lessapprox 1.5$ \cite{guizzoAssessmentVerticalStability2024a}. Overall, the simulations suggest CENTAUR is sufficiently vertically stable despite the presence of the vacuum vessel and shielding between the control coils and the plasma. As a result, significant margin likely exists to further increase elongation for possible performance improvements. 

\section{Power Handling and Divertor Design}
\label{sec:power}
Any break-even magnetic confinement device must contend with managing substantial particle and heat loads from the plasma \cite{Krieger_2025_divertors, DOE2015PMI, Stangeby2000, loartePlasmaDetachmentJET1998, Leonard2018Detachment, Krasheninnikov2017Detachment, Asakura2023DivertorReview, Federici2001PMI}. Most tokamak designs concentrate the bulk of these loads onto a specifically designed part of the device, known as the divertor. Typically, the divertor uses wall materials that are designed to withstand the heat loads and radiative power exhaust (such as impurity radiation) to reduce the target loads. Negative triangularity devices are envisioned to operate in conditions more comparable to L-mode than H-mode. As a pedestal is not required, $P_{\mathrm{SOL}}$ does not need to be more than the L-to-H-mode transition power ($P_\mathrm{LH}$), so higher impurity fractions can be used \cite{Austin_2021}. With increased impurity radiation, advanced divertor techniques like strike line sweeping may no longer be needed to keep the maximum perpendicular heat flux below 10~MW$/$m$^2$ on the tungsten divertor plates \cite{Pitts_2007}.  
\begin{table}[tb]
    \centering
    \caption{UEDGE Simulation parameters and $\lambda_{q}$ comparison. Eich H-mode\cite{Eich_2013} scaling yields $\lambda_{q,\mathrm{OMP}}=0.37\ \mathrm{mm}$, Scarabosio L-mode\cite{Scarabosio_2013} scaling yields $\lambda_{q,\mathrm{OMP}}=3.28\ \mathrm{mm}$.}
    \label{tab:uedge}
    \begin{tabular}{|c|c|c|c|} \hline
    Parameter & Simulation value \\ \hline
    $P_{\rho=0.96}$ & $17.84$ MW \\ \hline
    $n_{i,\rho=0.96}$ & $2.15\times 10^{20}\ \mathrm{m^{-3}}$  \\ \hline
    $\lambda_{q,\mathrm{OMP}}$ & $1.56\ \mathrm{mm}$  \\ \hline
    \end{tabular}
\end{table}
\subsection{Scrape-off Layer Simulation and Core-Edge Integration}
\begin{figure}[htbp]
    \centering
    \includegraphics[width=0.65\linewidth]{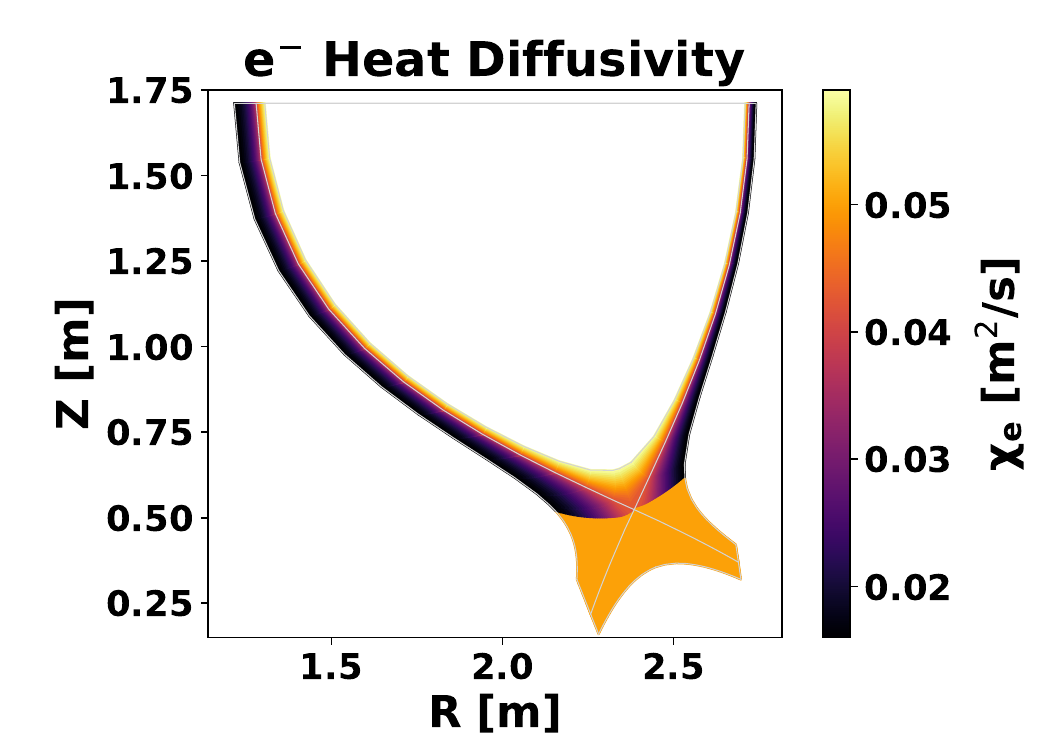}\\
    \includegraphics[width=0.65\linewidth]{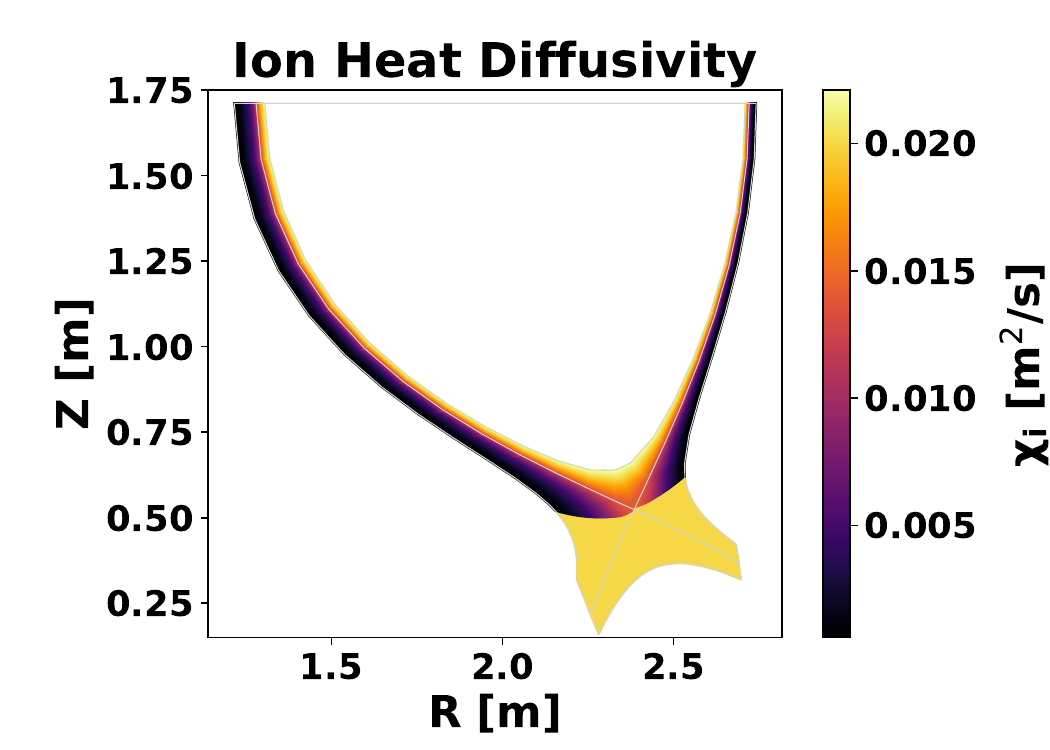}\\
    \includegraphics[width=0.65\linewidth]{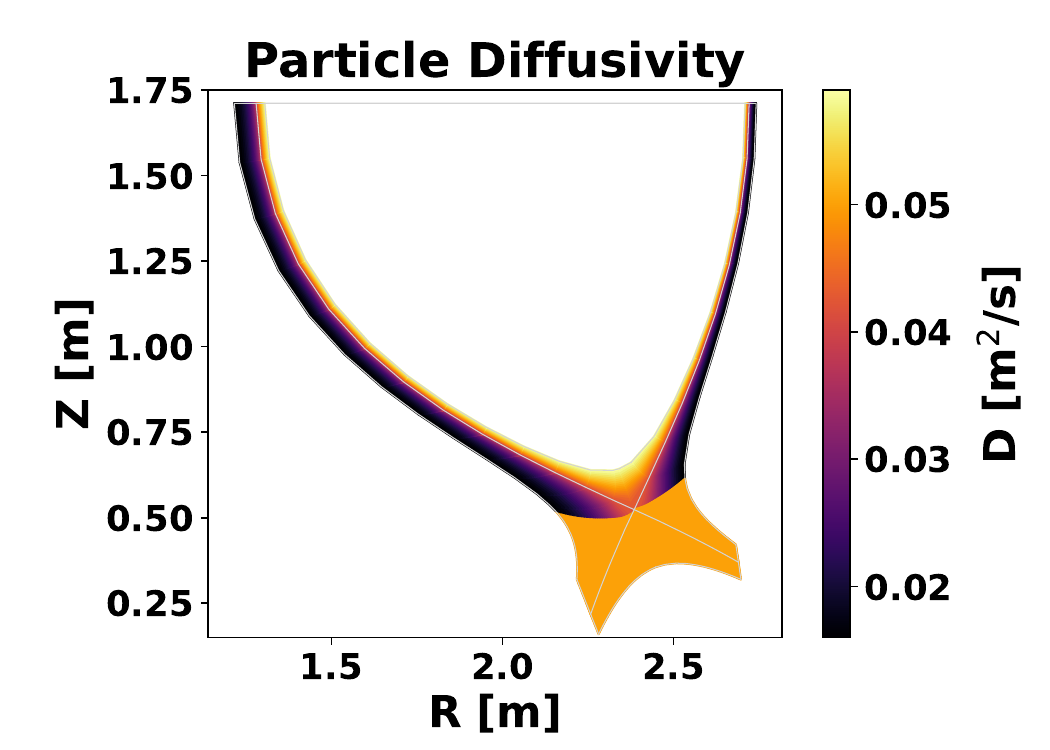}
    \caption{Imposed electron heat, ion heat, and particle cross-field diffusivity coefficient profiles, set cell-by-cell, incrementally decreasing over radial indices.}
    \label{fig:diffusivity}
\end{figure}

The UEDGE 2-D multi-fluid, edge-plasma code models the scrape off layer and self-consistently predicts the particle and heat loads on the divertor plates \cite{rognlienFullyImplicitTime1992,uedgeman}. UEDGE solves the Braginskii fluid equations, including ionization, recombination, and impurity radiation. The code also includes anomalous cross-field diffusion to approximate turbulent cross-field transport. Cross-field particle and heat transport coefficients are manually/iteratively matched to follow empirical scaling laws for the heat flux width \cite{Eich_2013,Scarabosio_2013}. Due to a lack of experimental characterization of diffusivity in the NT edge, diffusivity coefficient profiles in the SOL are set to be constant along magnetic field lines with stepwise changes perpendicular to the magnetic field, see Figure \ref{fig:diffusivity}. The amplitude of this profile over the mesh is decreased until a moderately H-mode biased $\lambda_q$ for the no impurity seeding base case is achieved, and is held constant in subsequent case iterations.

\begin{figure}[htbp]
    \centering
    \includegraphics[width=0.5\linewidth]{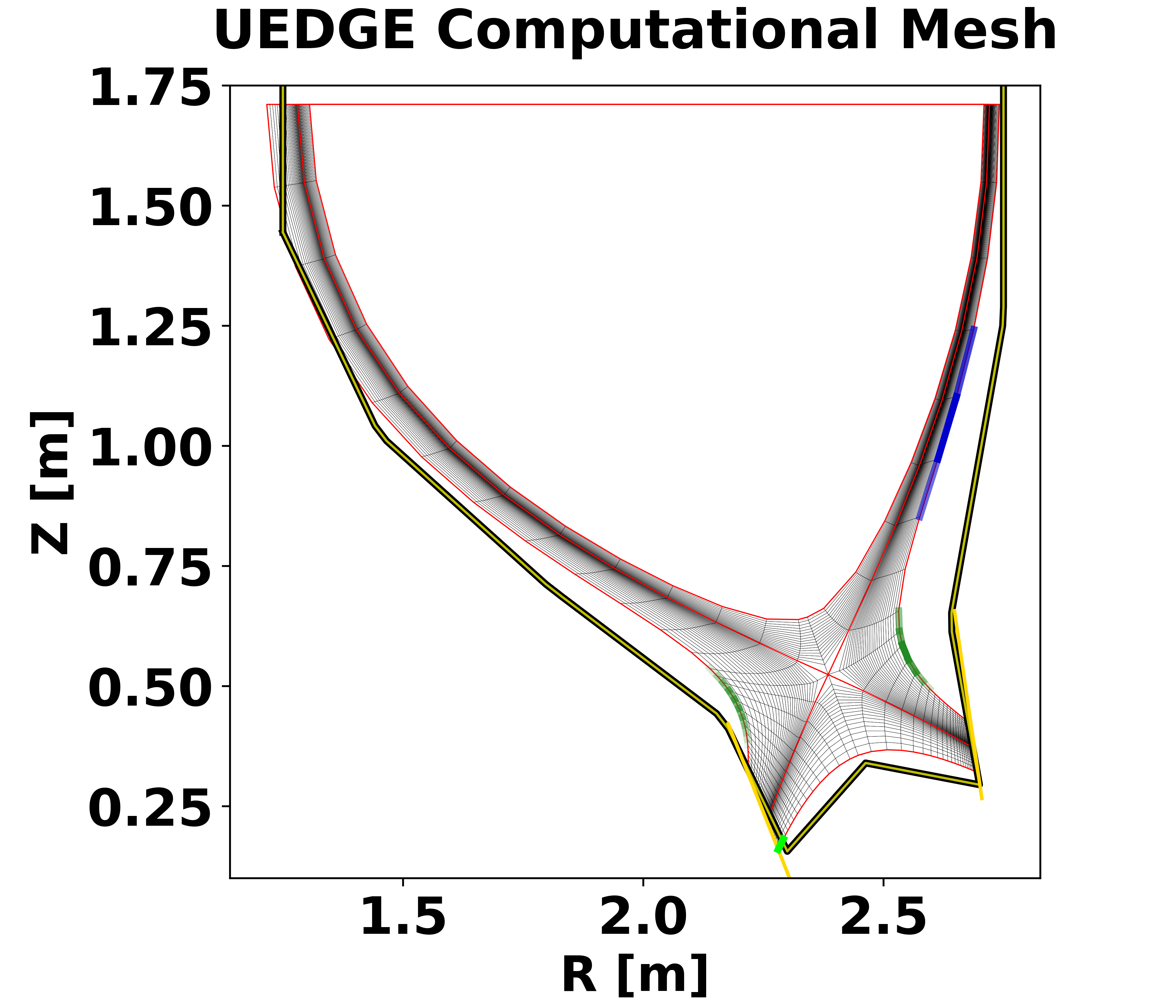}
    \caption{The UEDGE Computational mesh, configured to be from normalized poloidal flux $\psi_N\approx0.93$ to $\psi_N\approx1.078$ on the outboard side. Limiter (black \& yellow) is shown along with the computational divertor plates (yellow). The separatrix is in red. Deuterium gas fueling is in dark blue. Both neon gas puffs (next to the X-point) are shown in dark green. Species pumping (light green) is in the low field side (inboard side) private flux region.}
    \label{fig:uedge-mesh}
\end{figure}

The UEDGE mesh (Figure \ref{fig:uedge-mesh}) is generated with input magnetic equilibrium from TokaMaker~\cite{hansen_tokamaker_2024}. Each divertor plate is approximated as a line segment created at the mesh-vacuum interface in the divertor region. To save computational cost, only the lower half of the device has been simulated, and the input power is halved accordingly. Note that $\nabla B$ drift biases the power split away from 50/50, but in operation, the plasma can be moved vertically to recover a 50/50 split.

The inner mesh boundary is positioned at $\rho=\sqrt{\psi_N}=0.96$. The primary UEDGE inputs (power and density) at this core-edge boundary are produced by ASTRA and detailed in Table \ref{tab:centaur-params}. The NT-like $T_e$ pedestal is matched with the resulting UEDGE outer midplane electron temperature, $T_{e,omp}$ value at $\rho=0.99$. 
The BALOO infinite-n ballooning stability code also validates the pedestal. An exponential fit to the UEDGE outer midplane electron heat profile is used to characterize $\lambda_q$ (and is shown in Table \ref{tab:centaur-params} and Figure \ref{fig:mp}). NT $\lambda_q$ has been observed to lie between the upper bound of L-mode scaling and the lower bound of H-mode scaling, skewed toward H-mode. In NT scenarios on DIII-D, $\lambda_q$ has been experimentally measured between 1.5 mm and 3.5 mm. These values fall within the range of H-mode and L-mode scalings, with a bias toward H-mode width. \cite{Scotti_2024}.  Given the device parameters, the Eich et al. \cite{Eich_2013}\ H-mode scaling yields 
$\lambda_q = 0.37$ mm, while the Scarabosio et al. \cite{Scarabosio_2013}\ L-mode scaling yields $\lambda_q = 3.28$ mm. These two values are used as upper and lower bounds for the $\lambda_q$ space, respectively. 

A linear radial extrapolation boundary condition is used for n$_i$, T$_i$, and T$_e$ on the outer wall. The ion density is fixed to $n_{i,PFR}=10^{18}$~m$^{-3}$ at the private flux region wall. The private flux region is the region below the X-point and between the strike lines. The ion and electron temperature also use a linear radial extrapolation boundary condition on the private flux region wall. 

\subsubsection{Fixed Fraction Impurity Model}
Initial UEDGE simulations employ the fixed-fraction model for neon in the plasma edge. In this model, the impurity density is set to be a fraction of the ion density, $n_{imp} = n_i/f_{imp}=0.032$, and only accounts for neon radiation. The model does not consider any impact on particle balance and leaves $Z_{eff}$ unchanged. User-specified $f_{imp}$ can be spatially varying distributions. Collisional-radiative effective impurity rates are used to assess the enhanced radiation due to impurities. We follow the MANTA design approach by seeding different noble gases in the core and edge to optimize radiative cooling. Neon has been chosen as the impurity species in the edge region for all configurations due to its strong radiation properties and lower atomic number (Z) compared to other species, such as Argon. We assume that the impurities do not significantly mix: i.e. core-injected Argon is assumed to remain largely in the core, while the divertor puffed Ne is fully ionized before reaching the core. Under these conditions, the dual-species approach is an acceptable modeling approximation to achieve the desired radiation distribution. Detailed impurity transport modeling to refine these assumptions should be included in future work.

The model is set up to fuel the plasma with deuterium on the high field side lower wall, and user-placed pumps can be used for particle balance. For this design, the pump has been placed at the corner of the private flux region (PFR) and the outer divertor plate, as shown in Figure \ref{fig:uedge-mesh}. 

\subsubsection{Force Balance Model}
Impurities in UEDGE can also be treated as additional ion and gas species for each charge state and include physics like ionization and recombination between charge states, inter-species friction, and thermal forces. This model, coined force balance (for the parallel velocity equation when impurity inertia and viscosity are ignored \cite{igitkhanov}), can better model the separation between neon charge states and adds control over where impurities are added to the system as a gas species. Figure \ref{fig:uedge-mesh} shows the two neon gas puffs in green near the X-point. 

Similarly to deuterium, another pump is added in the private flux region near the outer divertor plate that only removes neon. The two neon puffs now give control over how much power is radiated in the outer and inner half of the SOL, and this ratio can be modified by changing the ratio of the gas puff amplitude. The deuterium fueling and pumping location was kept the same as the Fixed Fraction model. Scotti et al. \cite{Scotti_2025} investigated the out/in power asymmetry in DIII-D for different magnetic field directions and different triangularities. 

The neon charge states have the same core momentum and current boundary condition as deuterium. The outer and PFC wall boundary condition uses a gradient scale length approximation, with the minimum wall density being $10^{7}$ $m^{-3}$. This boundary condition gives more freedom than the extrapolation boundary condition. The fluid neutral model limits the convective transport to the free-streaming flux using flux limiters. The neutral gas flux limiters are adjusted to attain numerical stability of the convergence

\subsection{Power Deposition}
\begin{figure}[htbp]
    \centering
    \includegraphics[width=0.7\linewidth]{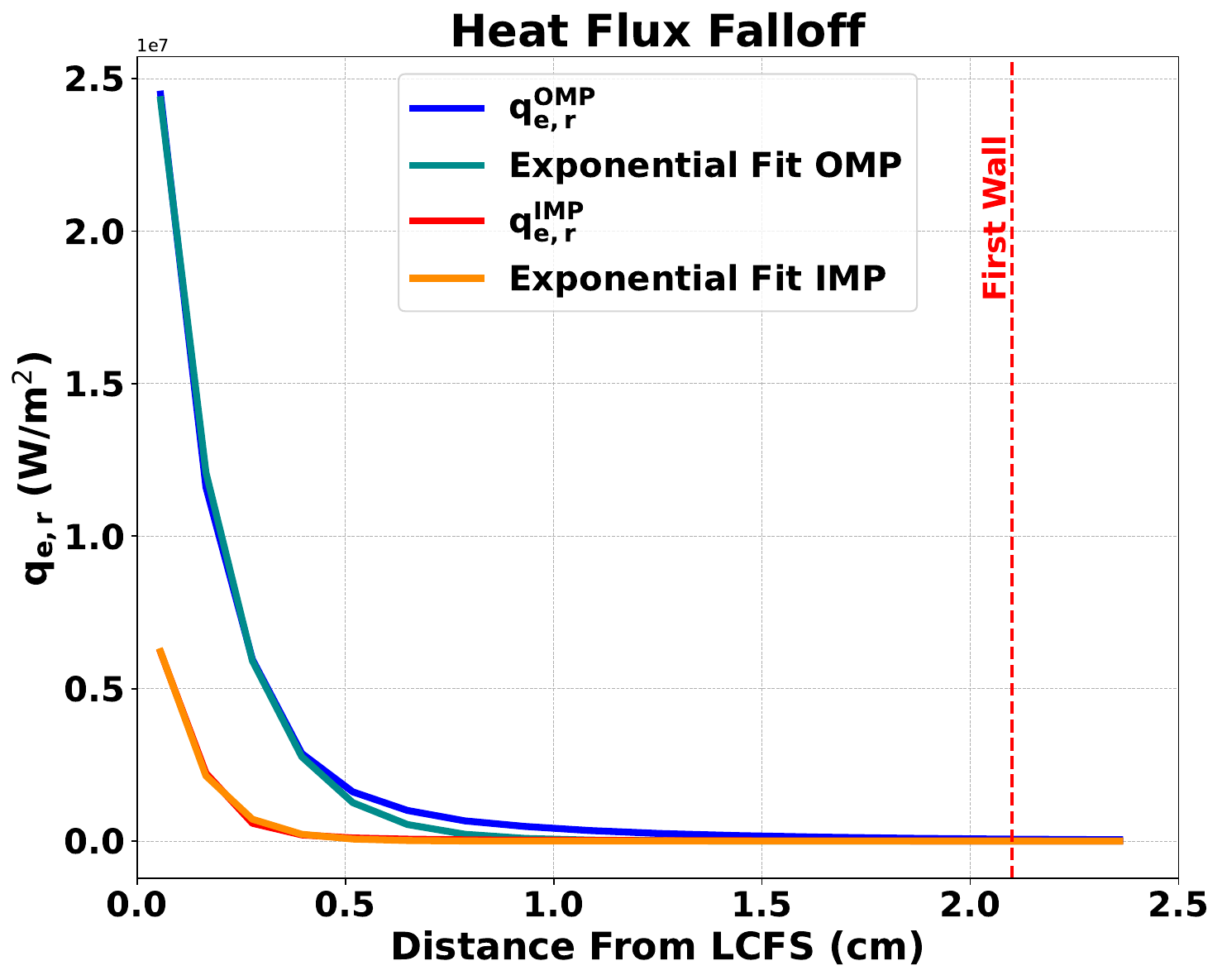} \\
    \includegraphics[width=0.7\linewidth]{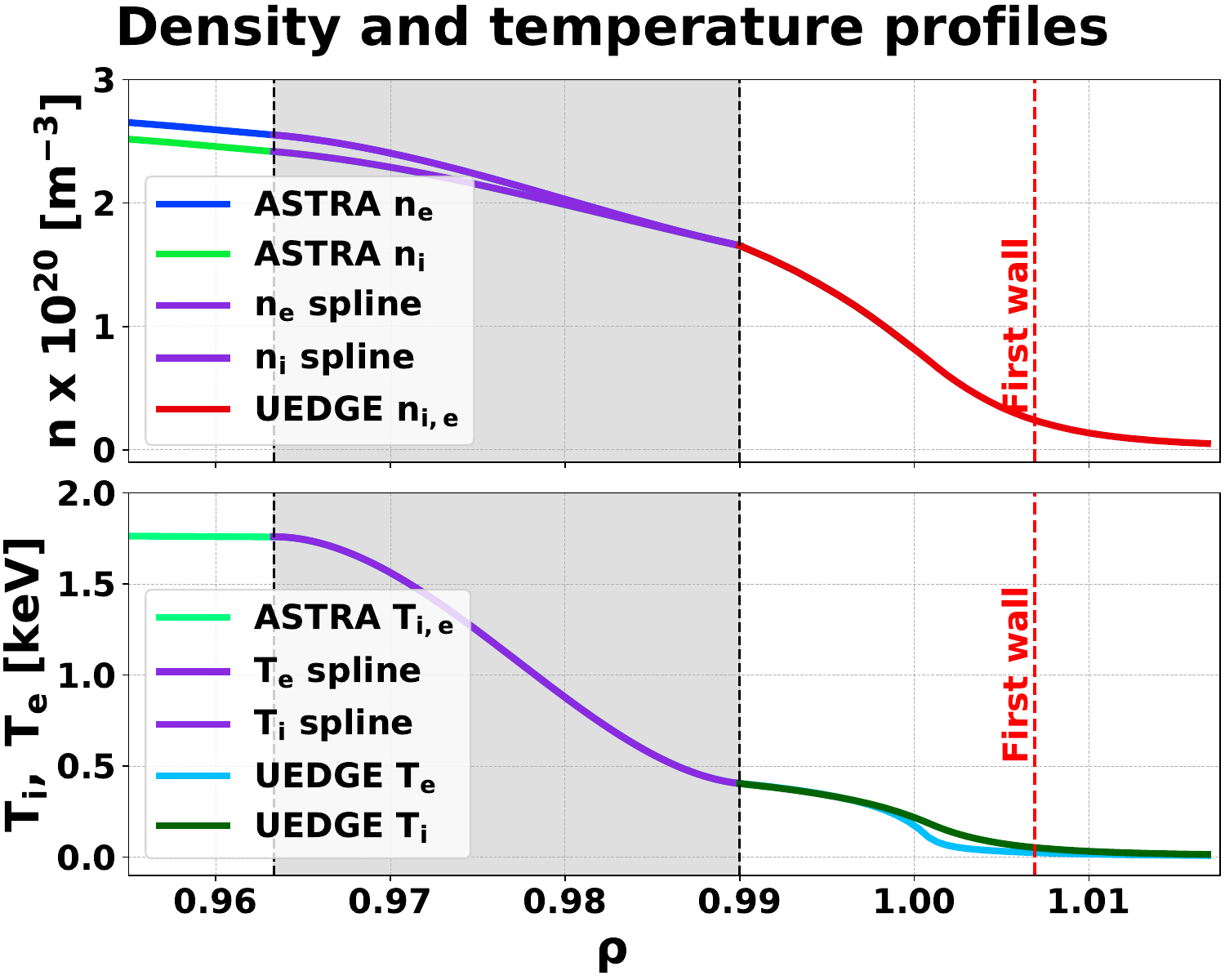}
    \caption{Electron heat radial falloff outside the LCFS at the midplane (left). $\lambda_q$ was found by applying an exponential fit (also shown) to the electron heat flux profile. Outer midplane wall kinetic profiles (right), including the cubic spline (grey shaded region bounded by dashed lines) that stitches ASTRA and UEDGE.}
    \label{fig:mp}
\end{figure}

In the final fixed fraction impurity scenario, an exponential fit to the OMP radial electron heat flux profile in Figure \ref{fig:mp} yields $\lambda_{q,\mathrm{OMP}} = 1.56$~mm, within the bounds of L-mode and H-mode scalings. The smallest plasma-wall separation occurs at the inner midplane (IMP): 2.1~cm. $\lambda_{q,\mathrm{IMP}} = 1.02$~mm, resulting in $T_{e,\mathrm{IMP\ wall}} = 7.48$~eV. This is below both the 10~eV tungsten sputtering limit and the 8~eV retention energy threshold \cite{sput, ROSZELL2013S1084}. $T_{e,\mathrm{IMP\ wall}}$ is a direct result of the radial temperature falloff length. Consideration of interchange, magnetic curvature, turbulence, and sheath physics in the edge is necessary for a more physics-informed wall temperature.

UEDGE is used to scan inner and outer divertor plate angles to identify the angles that minimize heat flux and neutral backflow into the plasma. The length of the divertor plates is also a factor. Closing the divertor more forces the plates to either intersect plasma near the X-point or intersect the existing vacuum vessel design. The optimal inner and outer plate angles given spatial constraints are $45^\circ$ and $58^\circ$, respectively. 

\begin{figure}[htbp]
    \centering
    \includegraphics[width=0.48\linewidth]{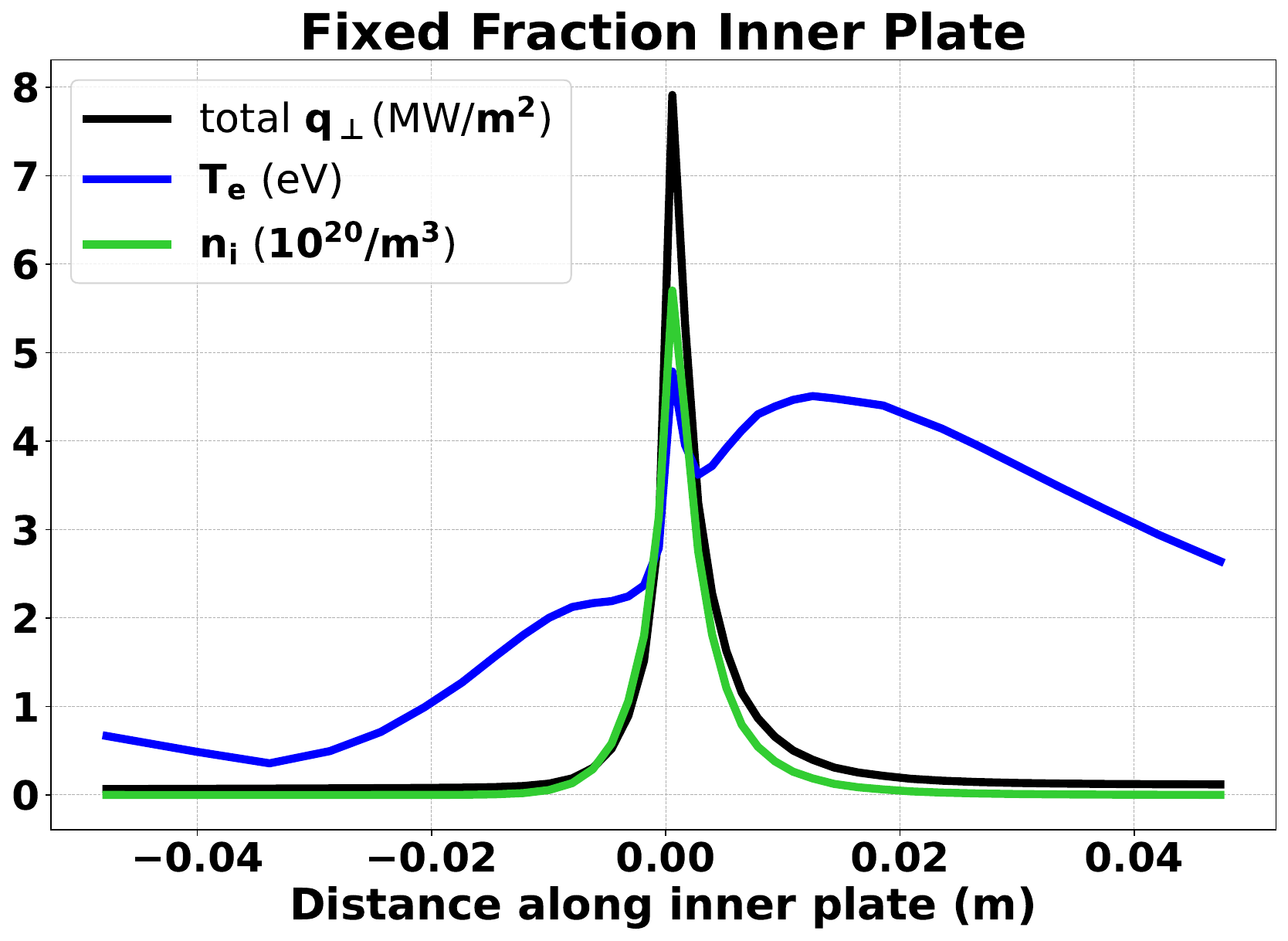} \quad
    \includegraphics[width=0.48\linewidth]{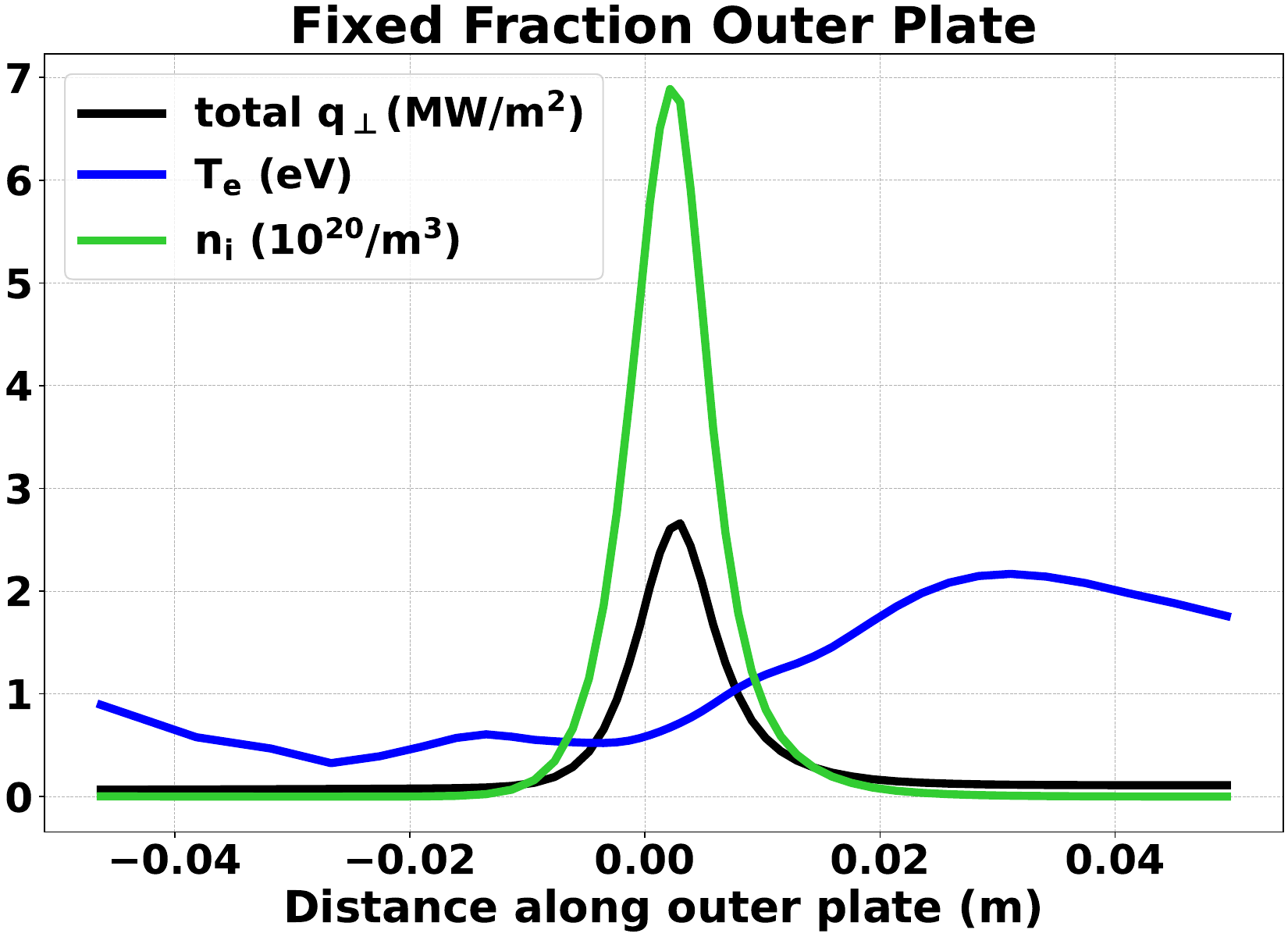} \quad
    \includegraphics[width=0.48\linewidth]{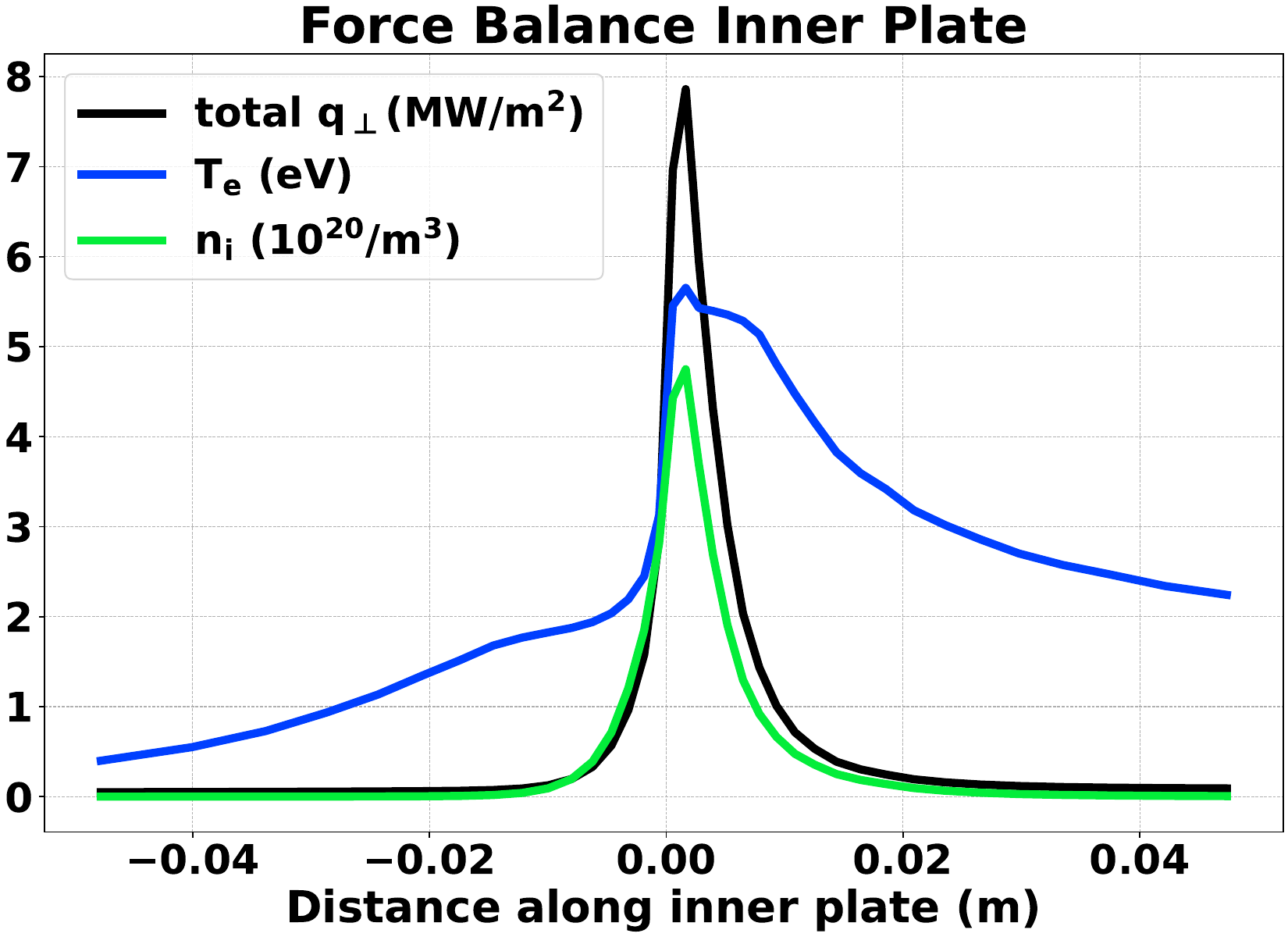}\quad
    \includegraphics[width=0.48\linewidth]{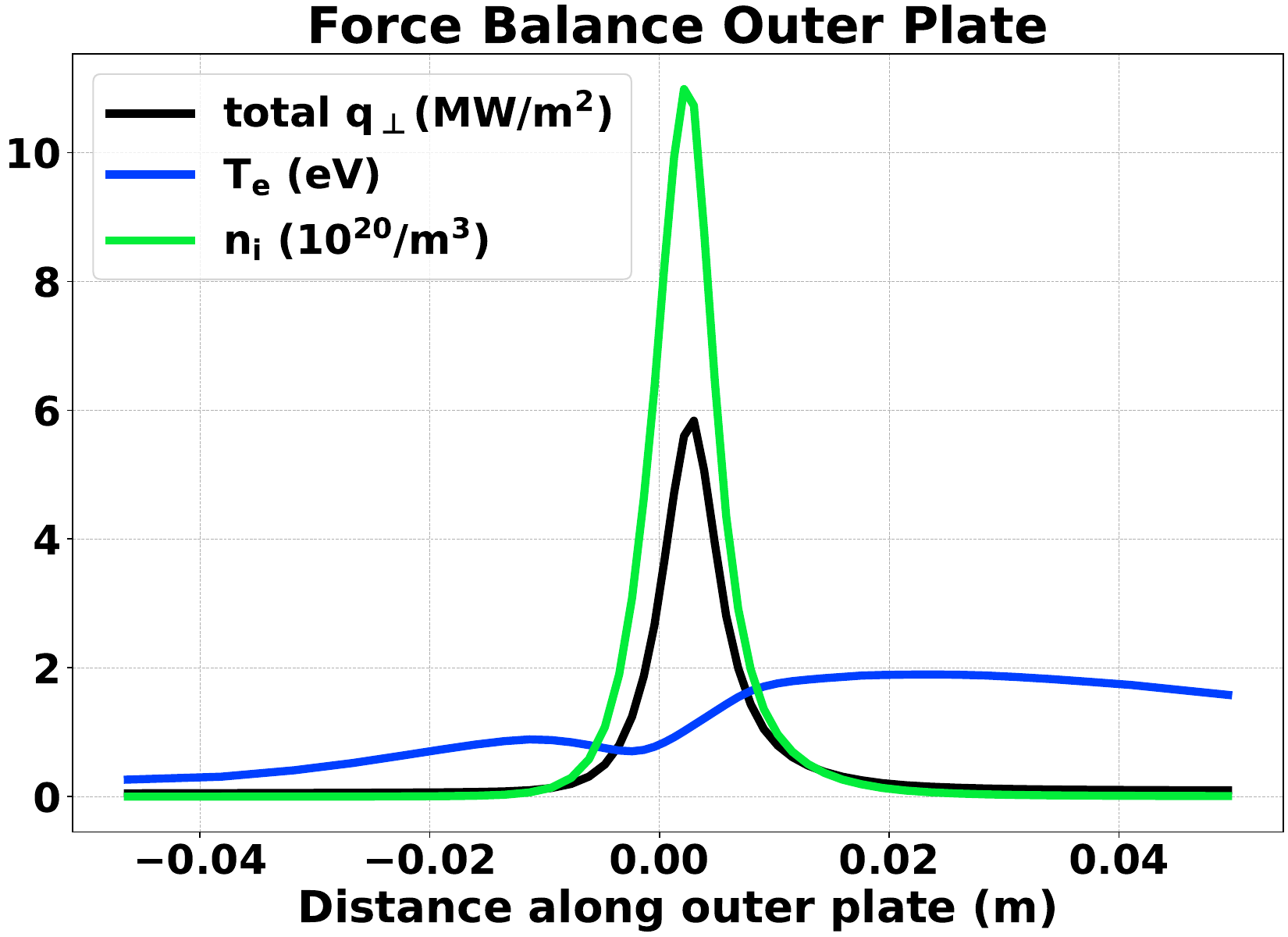} 
    \caption{Inner (left column) and outer (right column) divertor plate perpendicular heat flux, electron temperature and main ion density profiles along the respective plate. Top row is profiles from fixed fraction 3.2\% Ne seeding. Bottom row is profiles from the force balance model. Note the differences in imposing a power splitting ratio via localized impurity seeding in the force balance case.}
    \label{fig:both_div_profs}
\end{figure}

The final inner and outer plate heat fluxes are shown in Figure \ref{fig:both_div_profs}, with the results from the fixed fraction and the force balance model shown on the top and bottom rows, respectively. The peak heat flux of 7.9~MW/m$^2$ is below the steady-state tungsten limit of 10~MW/m$^2$. $P_{\mathrm{SOL}} = 9.424$~MW and $f_{\mathrm{rad,\;UEDGE}} = P_{\mathrm{rad,\;UEDGE}}/P_{\mathrm{fus}} = 20.6\%,$ for the fixed fraction model. In contrast, 4.83 MW of power is radiated by impurities and hydrogen in the force balance model (inside the whole computational domain). Using this value, the injected power and alpha power, the model finds $f_{\mathrm{rad,\;UEDGE}}=13.7\%.$ For both models, the ion density is higher in the outer divertor, while the electron temperature is lower. Even in the hotter inner divertor, the peak electron temperature is below the $\approx$13~eV sputtering and $\approx$8~eV retention energy thresholds \cite{sput} \cite{ROSZELL2013S1084}. 
Both plates are also lower than the tungsten deuteride molecule sputtering energy threshold \cite{ZHANG2022101265}. It's worth noting that the temperature is a Maxwellian, and the closer it gets to the sputtering threshold, the more of the distribution tail will exceed the threshold energy. 

\begin{figure}[htbp]
    \centering
    \includegraphics[width=0.75\linewidth]{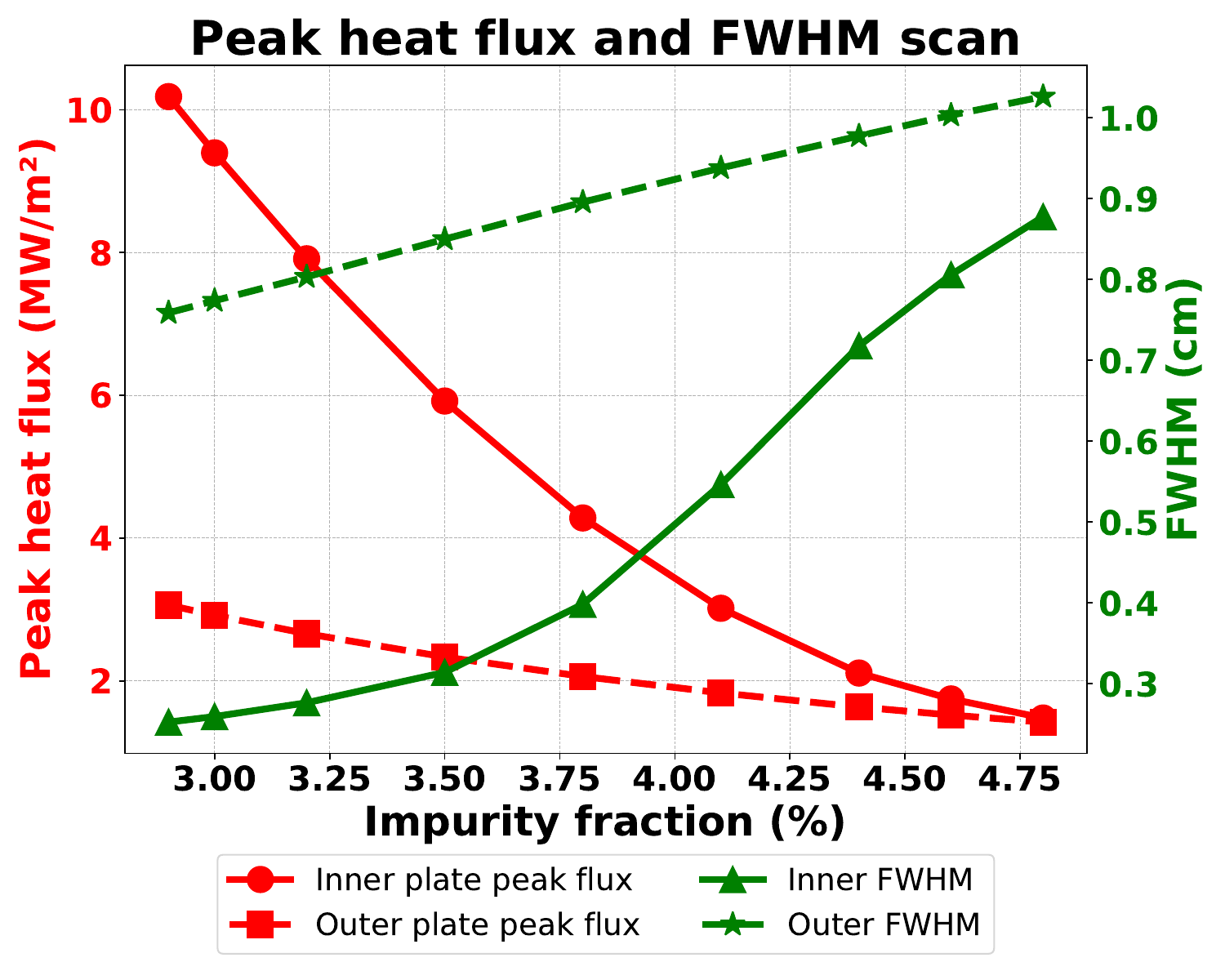}
    \caption{Peak perpendicular heat flux (red) and full-width-half-max (FWHM) of heat flux profile (green) as Ne seeding (in the fixed fraction model) is increased.}
    \label{fig:detachment}
\end{figure}

One solution to the heat and particle flux challenge is to reduce the energy and momentum of the plasma via radiation and collisions. In this regime, called the radiative plasma detachment regime, a large part of the plasma neutralizes and slows down in a layer above (detached from) the target plate \cite{soukhanovskiiReviewRadiativeDetachment2017, stangebyPlasmaBoundaryMagnetic2000}. While there are many signals for plasma detachment, it has been found that the target heat flux profile broadens and decreases in amplitude \cite{faitschBroadeningPowerFalloff2021, renExperimentalObservationHeat2021, soukhanovskiiDivertorHeatFlux, wangIntegrationFullDivertor2021}. Initial detachment studies used the fixed fraction model to quickly scan the global neon fraction. Figure \ref{fig:detachment} shows this scan of peak perpendicular heat flux and heat flux full width half max (FWHM) over neon fraction. As more neon is seeded, both peak heat fluxes drop. At the same time, the width of the heat flux profiles increases, point to the possibility of divertor detachment. More detailed detachment studies are left for future work and will likely require the force balance model to look at local neon and deuterium density build-up near the plate, as well as recombination.  

\begin{figure}[htbp]
    \centering
    \includegraphics[width=0.75\linewidth]{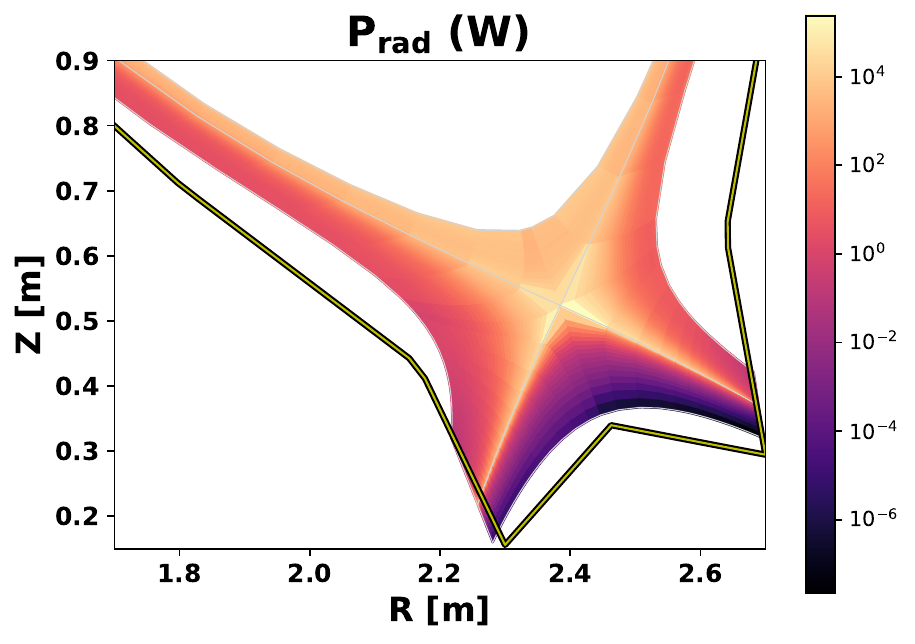}
    \caption{Power radiated in the computational domain calculated by the UEDGE force balance model.}
    \label{fig:prad}
\end{figure}

Figure \ref{fig:prad} shows the power radiated by neon in the whole domain using the force balance model. As expected, most of the power is radiated near the X-point and along the strike line. The effective charge $Z_{eff}\approx1.66$ along the separatrix, which is close to ASTRA's prescribed $Z_{eff}=1.6$. Further work should include changing the core boundary condition and neon seeding location to match the separatrix $Z_{eff}$ between UEDGE and ASTRA. Matching $Z_{eff}$ would also require more accurate neon upstream transport modeling.

\subsection{Heat Transfer Modeling}
The peak heat flux into the divertor plates is $<$10 MW/m$^2$, less than the ITER steady-state limit as expressed in the ITER divertor physics baseline \cite{pitts_2019}, so conditions in the divertor region are expected to be below critical material limits for tungsten plates. A Gaussian flux profile fit to the heat flux profile from UEDGE serves as input to a COMSOL heat transfer model of the divertor plates and substructure. The time dependence of the flux over the shot is represented via a piecewise linear function.

The heat transfer model predicts a maximum temperature on the divertor plate of $\sim$310~C over a thirty second pulse. The assumption of simple rectangular block components offers sufficient model accuracy while allowing for flexibility in the final design of the experiment. 
\begin{figure}[htbp]
    \centering
    \includegraphics[width=0.7\linewidth]{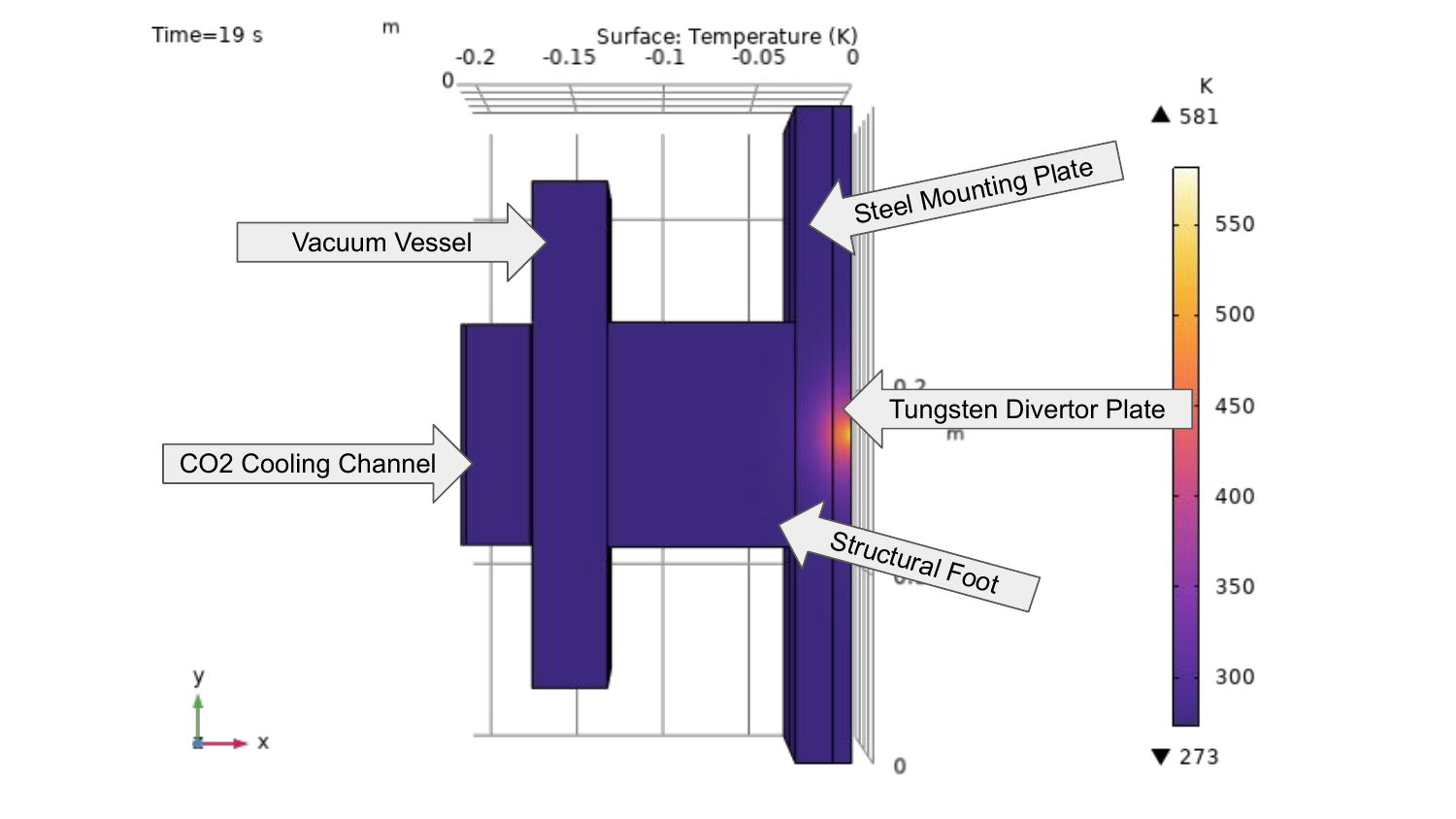}
    \caption{Simplified representation of the region between the divertor plate and the cooling channel outside the vaccuum vessel showing the temperature at its maximum value in a time-dependent simulation of heat transfer in a full-performance shot. The tungsten plate remains well below its recrystallization temperature.}
    \label{fig:heatmap}
\end{figure}
Figure \ref{fig:heatmap} displays the temperature map of the structure from divertor plate through to an external passive cooling channel at peak operational temperatures. The COMSOL geometry spans from the 1~cm tungsten divertor plate, through a steel mounting plate, a structural foot, the vaccum vessel, and ends at a passive carbon dioxide cooling channel to deliver coolant between shots.  The tungsten plasma facing components operate with a maximum temperature below 300~C, well below the recrystallization temperature of 1500~C \cite{suslova_2014}. This operating state ensures that the device will withstand the conditions of full-performance pulses. The strong divertor performance seen here is a feature of the highly radiative NT divertor region and implies that future iterations of this device could withstand higher performance operating points those explored here.

\section{Magnets and Device Structure}
\label{sec:magnet}

The CENTAUR magnets are designed to meet the high magnetic field requirements of the plasma core scenario while remaining within practical engineering and cost constraints. All major magnet components employ Rare-Earth Barium Copper Oxide (REBCO) high-temperature superconductors (HTS). REBCO has been utilized in the production of high-current, high-field magnets for fusion applications, demonstrating impressive performance at strong magnetic fields \cite{hartwig_sparc_2024, whyte_experimental_2024}. The final system consists of 18 toroidal field coils, a central solenoid, and a set of six poloidal field coils that together provide plasma confinement, shaping, and Ohmic current drive.

\subsection{Toroidal field coils} 
\label{subsec:tfcoils}
The toroidal field coils achieve an on-axis magnetic field of 10.9~T, with a maximum field of 23~T on the magnets themselves. The peak magnetic field ripple on the plasma boundary is 0.34\% at the outboard midplane (OMP).  Each high-temperature superconducting winding pack is composed of discretely wound pancakes stacked within a steel case, similar to the design of the SPARC Toroidal Field Model Coil \cite{hartwig_sparc_2024}. The dimensions of the CENTAUR toroidal field coils allow for 15 pancakes per winding pack, with 16 turns of superconducting tape stacks per pancake. Each tape stack is composed of 240 REBCO tapes. Each tape has dimensions of 4~mm by 14.4~mm.

To produce a 10.9~T field at the plasma magnetic axis, each toroidal field coil carries 6.06~MA of current, with an operating current of 25.25~kA per HTS tape stack. The resulting HTS current density of 440~A/mm$^2$ has a 56\% margin on the critical current density of 1000~A/mm$^2$ at 20~K and 25~T perpendicular magnetic field, determined from experimental data on commercially available superOx tapes \cite{molodyk_development_2021}. Using the approximate leading order dependence of the critical current density on temperature from \cite{wolf_critical_2018} and fitting the scaling law to the reported performance of superOX tapes at 4.2~K and 20~K from \cite{molodyk_development_2021}, the critical temperature at a 25~T field is estimated to be approximately 33~K. Over a 10~s plasma discharge, neutron heating is expected to raise the HTS temperature to a maximum of 27~K without active cooling, as discussed further in section~\ref{sec:neutron}. The HTS is therefore likely to remain below its critical current throughout the discharge. Future work will involve modeling the active cooling of the toroidal field coil magnets to obtain more robust predictions for the HTS operating margin. A summary of the major toroidal field coil parameters can be found in Table~\ref{tab:tfcoil}

\begin{table}[tb]
    \centering
    \caption{Toroidal field coil parameters.}
    \label{tab:tfcoil}
    \begin{tabular}{|c|c|} \hline
 Parameter & Value \\ \hline \hline
 Number of TFs & 18  \\ \hline
 Winding pack radial thickness & 0.4 m  \\ \hline
Winding pack azimuthal thickness & 0.24 m  \\ \hline
Pancakes per TF & 15  \\ \hline
Pancakes azimuthal thickness & 12 mm  \\ \hline
Turns per pancake & 16  \\ \hline
REBCO tapes per turn & 240  \\ \hline
Width of each REBCO stack & 14.4 mm  \\ \hline
Height of each REBCO stack & 4.0 mm  \\ \hline
Winding pack current & 6.06 MA  \\ \hline
Current per turn & 25.25 kA  \\ \hline
Operating current density & 440 A/$\mathrm{mm^2}$  \\ \hline
Peak magnetic field on winding pack & 23 T  \\ \hline
    \end{tabular}
\end{table}

A traditional \enquote{Princeton Dee} \cite{file_large_1971} toroidal field coil shape is not considered for CENTAUR due to the large magnet volumes it would require to accommodate the negative triangularity shape. Instead, both oval and conformal \enquote{negative triangularity} magnet shapes are analyzed for stress. COMSOL simulations suggest that the conformal magnet experiences higher peak stresses in the magnet lobes than the oval magnet, and an oval magnet is thus chosen for the final toroidal field coil design.

\subsection{Central solenoid}

In contrast to toroidal field coils, the central solenoid leverages PIT VIPER technology, where REBCO tape stacks are arranged in petals to form cables for use in pulsed HTS magnets.  PIT VIPER cables have been tested under several loading conditions and are predicted to sustain 50~kA per cable at 20~K and 25~T magnetic field \cite{sanabria_development_2024}.

The design of the central solenoid is primarily constrained by the flux swing requirements of the plasma core scenario. Using the analytic approximations from \cite{sugihara_plasma_1982}, an estimated 40.9~Wb of flux is required for startup in CENTAUR. An additional 7.7~Wb is needed to sustain the 10~s target flattop. This value is calculated by estimating the loop voltage from core plasma parameters using Spitzer resistivity, including corrections for $Z_{\mathrm{eff}}$ and neoclassical effects. Future work will involve higher fidelity start-up simulations in order to refine the plasma flux swing requirements.  The overall requirement of 48.6~Wb is challenging given the constrained inboard radial build. Furthermore, to produce a negative triangularity shape, the flux swing produced by the poloidal field coils detracts from the flux swing produced by the central solenoid, in contrast to positive triangularity where the effect is additive \cite{leuer_talk_2020}. To produce additional flux, the central solenoid is extended to include additional windings above and below the toroidal field coils at larger major radius, which is enabled by the elliptical TF shape. 

The outer radius of the solenoid at the midplane is limited to $R = 0.68$~m by the position of the toroidal field coils. To determine the optimal dimensions of the extended central solenoid, we perform a 2D scan of the solenoid inner radius and the outer radius of the solenoid extension and quantify the available flux swing, taking into account the field produced by the poloidal field coils. The results in figure~\ref{fig:solenoid-a} map out the flux swing produced by several geometries, and indicate which designs result in peak fields larger than the PIT Viper testing conditions \cite{sanabria_development_2024}. The chosen design point, shown in figure~\ref{fig:solenoid-b} has an inner radius of $R = 0.51$~m, an extended outer radius of $R = 0.75$~m, and produces a total flux swing of 52~Wb. When accounting for the flux detracted by the ramp-up of the poloidal field coils, the flux available for the plasma is just sufficient to meet the plasma startup and flattop requirement. While modest, the extension to the central solenoid is necessary for the CENTAUR design, with ~30\% of the flux being produced by the windings away from the midplane.  At full solenoid current of 89.8~MA, The maximum field on the solenoid is 25.8~T, which is marginally higher than the maximum field tested on PIT Viper cables. Further optimization of the central solenoid dimensions, including scans of the central solenoid height, may open opportunities to reduce the peak field while maintaining the available flux swing.

\begin{figure}[htbp]
    \centering
    \includegraphics[width=1.0\linewidth]{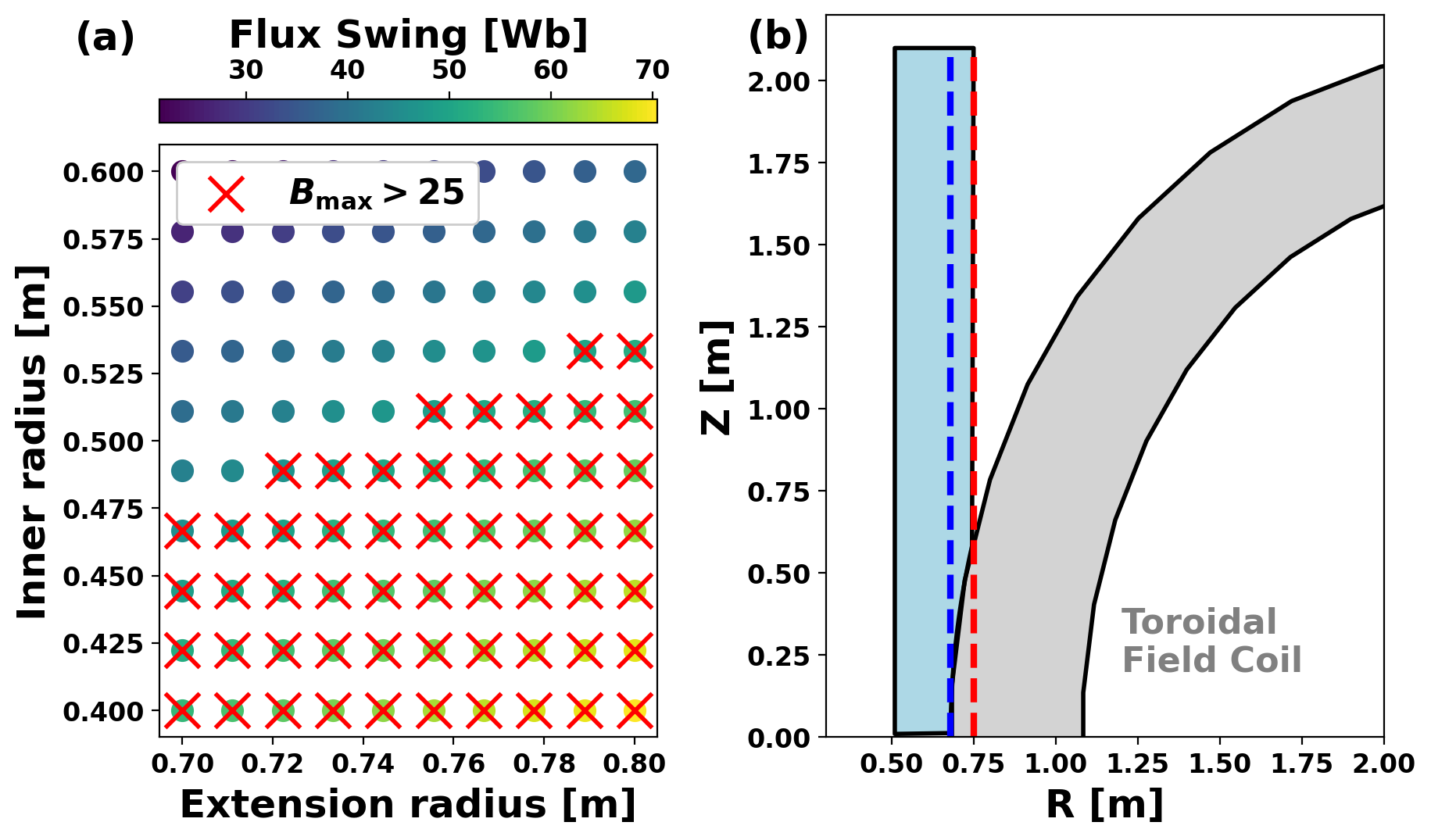}
    \caption{ (a) Solenoid flux swing as a function of solenoid inner radius and outer radius of the solenoid extension and (b) schematic of the CENTAUR central solenoid, which is extended with additional turns around the toroidal field coils between R = 0.68~m and R = 0.75 ~m in order to meet the flux requirement.}
    \label{fig:solenoid}%
    \labelphantom{fig:solenoid-a}
    \labelphantom{fig:solenoid-b}
\end{figure}

\subsection{Poloidal Field Coils}

The poloidal field (PF) coils are also designed as PIT VIPER HTS multi-turn coils. The PF coil set includes six coils with up-down symmetry. In order to use HTS PF coils, they must be positioned outside of the TF coils to allow machine assembly without requiring jointed HTS magnets, which are not a mature technology. This relieves any PF coil position restriction based on neutron heating because the center stack and TF coils are both shown to be able to sustain the higher neutron flux they will experience closer to the plasma. 

\begin{figure}[htbp]
    \centering
    \includegraphics[width=0.85\linewidth]{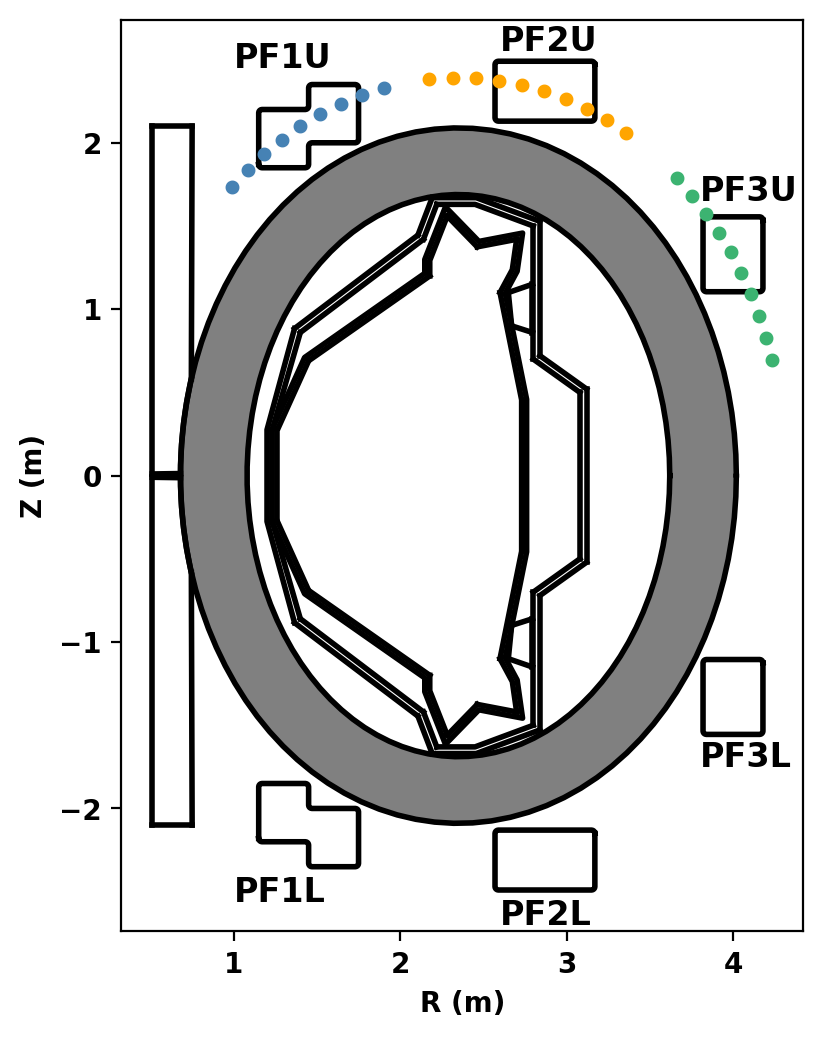}
    \caption{Cross section of CENTAUR with PF coils locations where blue, orange, and green circles show a set of locations sampled for PF1U/L, PF2U/L, and PF3U/L, respectively. Up-down symmetry was maintained, meaning the PF coil pairs were moved together. Each combination of coil positions was used to produce an equilibrium solution with TokaMaker, and the locations producing the lowest maximum coil current while producing the required core plasma parameters was selected.}
    \label{fig:PF coil locations}
\end{figure}

The PF coils are modeled using TokaMaker as regions of uniform current density in the 2D cross section, shown in Figure \ref{fig:PF coil locations}. The positions of the coils are optimized via a brute-force scan of possible coil positions while maintaining up-down symmetry. The goal of this scan is to minimize the largest PF coil current while producing the core scenario equilibrium targets. Figure \ref{fig:PF coil locations} shows an example of sampled PF coil locations. Every combination of PF coil positions (where the up-down pairs were moved together to maintain symmetry) is used to produce an equilibrium. The solution with the smallest maximum current in any PF coil where the target equilibrium parameters are achieved is chosen.

Using the same 50~kA per turn limit as the center solenoid and VIPER cable dimensions of 2.3~cm by 2.3~cm, a current density limit is calculated to be 95~MA/m$^2$ \cite{sanabria_development_2024}. The coil cross-sections are configured to minimize currents by bringing the current carrying regions closest to the plasma, without exceeding the current density limit or requiring angled cross sections which would produce conical magnet casings. The TokaMaker model is run with three fixed CS coil currents representing the extrema and the average states over the flux swing: $+$89.8, 0, and $-$89.8 MA. PF coil design details, currents, and current densities at the three CS states are shown in table \ref{tab:pfcoils}. All coils under all conditions modeled thus have significant margin before the current density limit discussed above. Future work will include systematic optimization of the position and design of the PF coils over an entire simulated pulse.

\begin{table}[htbp]
    \centering
    \caption{Poloidal field coil cross-sectional areas, currents, and current densities at three center stack coil currents. The PF coils are up-down symmetric, as is the plasma, so these data are presented for Upper (U) and Lower (L) pairs. Figure \ref{fig:PF coil locations} shows the locations of the coil.}
    \label{tab:pfcoils}
\begin{tabular}{|c|c|c|c|c|}
\hline
$I_{CS}$ [MA] & Quantity & PF1U/L & PF2U/L & PF3U/L \\ \hline \hline
 & Area [$m^2$] & 0.195 & 0.216 & 0.162 \\ \hline
$-89.8$& $I_{PF}$ [MA] & -16.4 & 13.9 & -10.1 \\ \hline
$0$ & $I_{PF}$ [MA] & -14.2 & 14.9 & -9.6 \\ \hline
$+89.8$& $I_{PF}$ [MA] & -12.0 & 15.9 & -9.1 \\ \hline
\end{tabular}
\end{table}

\subsection{Structure and mechanical stresses}

\begin{figure}[htbp]
    \centering
    \includegraphics[width=0.5\linewidth]{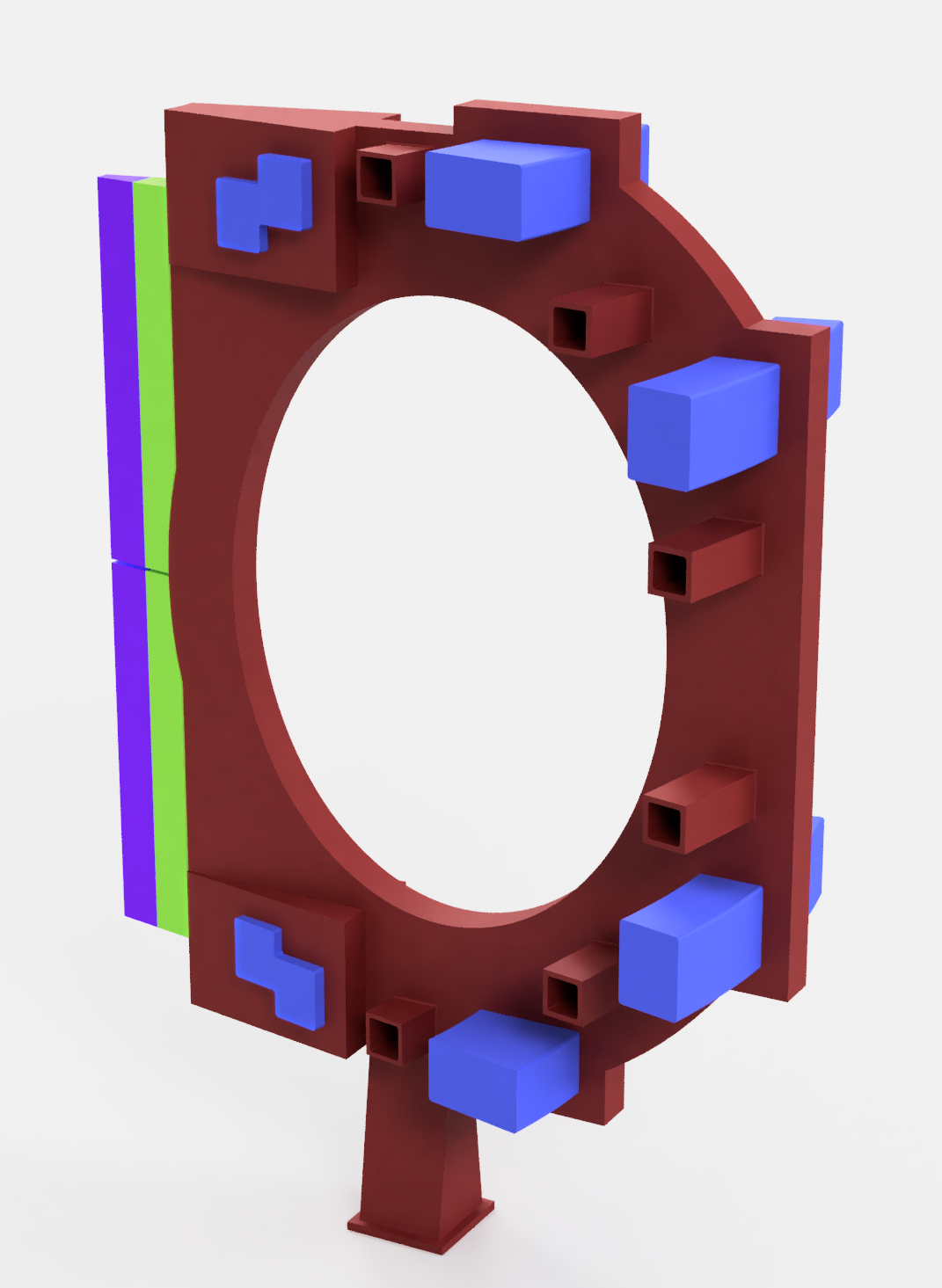}
    \caption{Magnet and structural build of one toroidal section. The CS resin plug is shown in purple, left of the green CS. TF and PF structural supports are shown in maroon, holding the blue PFs. The TF is embedded inside the ovular section of the maroon support structure. Eighteen equivalent toroidal sections encircle the device center-line to form the full CENTAUR build.}
    \label{fig:magnetsandstructure}
\end{figure}

Each toroidal field coil is surrounded by a case of Nitronic 40, a favorable structural material due to its high yield stress. The case is approximately 2.7~cm thick on all sides. Additional Nitronic 40 structure is included around the poloidal field coils and in between the toroidal field coils and central solenoid, as shown in Figure \ref{fig:magnetsandstructure}. The toroidal field coils are bucked against the solenoid to withstand the radially inward centering force. A plug made of non-magnetic resin is included in the central solenoid to combat the large radially inward force on the solenoid when it has no current. Additional Nitronic 40 brackets are included between toroidal field coils to withstand the overturning force.

COMSOL is used to assess the Von Mises stress on the magnet system components. For the purpose of this analysis, two loading conditions are considered. In the first, the central solenoid is at it's maximum current, and in the second, the central solenoid has zero current. For both conditions, the toroidal field coils, poloidal field coils, and plasma have maximum current. In the COMSOL model, the plasma is represented as a single filament at the magnetic axis. These two loading conditions are expected to create the maximum stresses on the magnets and structural components.

\begin{figure}[htbp]
    \centering
    \includegraphics[width=\linewidth]{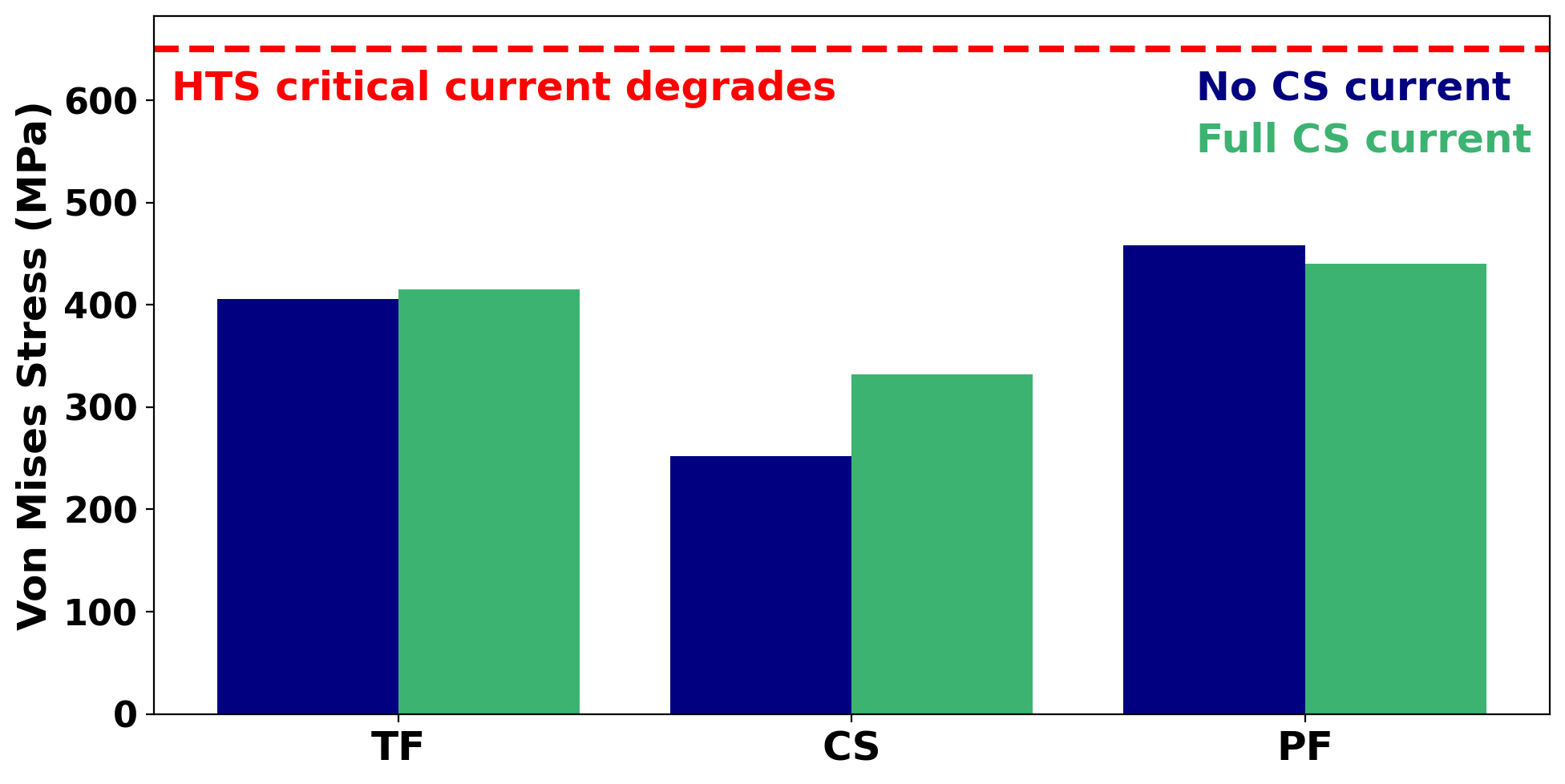}
    \caption{Von Mises stress on the toroidal field coils, central solenoid,and poloidal field coils for no central solenoid current and maximum central solenoid current.}
    \label{fig:COMSOLStructureResults}
\end{figure}

Results from these COMSOL simulations are depicted in Figure \ref{fig:COMSOLStructureResults}. Between both loading conditions, the maximum stresses on the toroidal field coils, central solenoid, and poloidal field coils are 415~MPa, 332~MPa, and 458~MPa, respectively. HTS critical current is not expected to degrade until 650~MPa \cite{Barth2015}, giving about a 40\% minimum margin of safety on all HTS components.

Preliminary COMSOL simulations also suggest that the peak loads on the toroidal field coil structure are below the 1400 MPa yield stress of Nitronic 40 for both loading conditions. Future solid mechanics simulations will be used to further optimize the mechanical structure and validate stress predictions.

\subsection{Vacuum vessel stresses from disruption-induced current quenches}

\begin{figure}[htbp]
    \centering
    \includegraphics[width=0.75\linewidth]{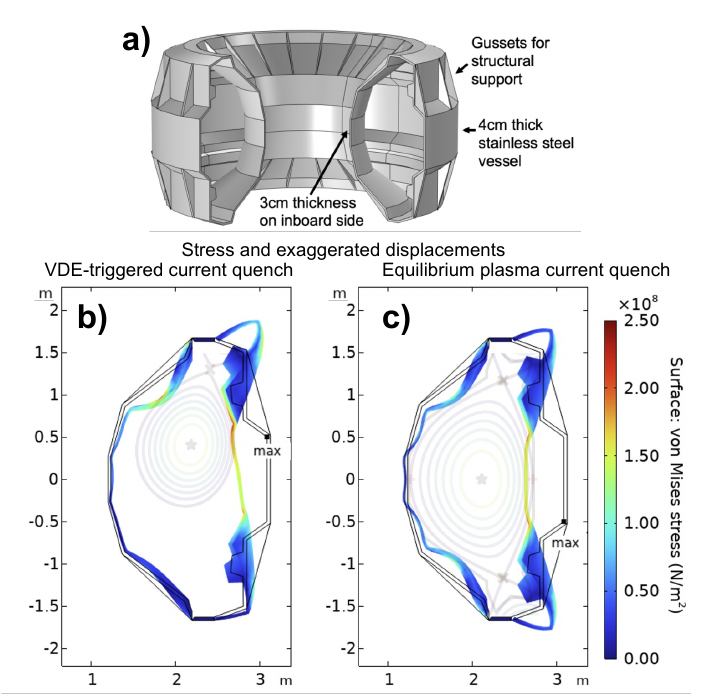}
    \caption{Current quench stresses on the vacuum vessel from VDE and vertically symmetric disruptions. a) shows the vacuum vessel design including gussets for structural support. b) shows the electromagnetic stress on the vacuum vessel from a VDE, and c) is the stress from a symmetric current quench. The Tokamaker equilibria before a disruption are shown in a lighter shade.}
    \label{fig:cq}
\end{figure}

TokaMaker is used to simulate current quenches (CQ) from VDE-induced disruptions and from symmetric disruptions. VDE simulations are described in section \ref{sec:core}. To simulate VDE disruptions, the coil currents are turned off when the plasma touches the limiter, and the consequent current-quench evolution is simulated in TokaMaker with fixed current distribution. For symmetric disruptions, the coil currents are turned off from the base equilibrium instead. In both cases, a fast CQ time of 3~ms is assumed using the ITPA disruptions database scaling \cite{eidietis_itpa_2015}. As the plasma evolves, the induced current density in the vacuum vessel (VV) and the resulting $\boldsymbol{J}\times \boldsymbol{B}$ forces are found as a function of time. The peak force on the VV is then used to calculate the maximum stress on the VV.

The $\boldsymbol{J}\times \boldsymbol{B}$ forces are applied to a realistic 3D VV geometry shown in \ref{fig:cq} a) to calculate electromagnetic stresses, shown in \ref{fig:cq} b) and \ref{fig:cq} c). The maximum stresses, located in the outboard side marked "max", are 249~MPa for VDE CQs and 230~MPa for symmetric CQs. According to experimental steel fatigue studies \cite{mohammadFatigueBehaviorAustenitic2012}, this results in 100,000 disruptions and 230,000 disruptions before VV structural failure, respectively. Therefore, the VV is unlikely to fail from disruption electromagnetic forces within the design lifetime. 

However, for added safety, this analysis led to the addition of gussets to the corners of the VV, which improve VV strength and are thin enough that they don't significantly affect the blanket's ability to shield neutrons. Further fidelity can be achieved by simulating eddy currents in a complete 3D VV geometry, including ports and other structures \cite{Hansen_2025}.

\section{Neutronics Analysis}
\label{sec:neutron}
Comprehensive lifetime analysis was completed using the open-source Monte Carlo neutron simulation code OpenMC \cite{Romano2015OpenMC}. As design points were iterated, a workflow was developed that allows transforming geometric parameters in TokaMaker to OpenMC objects. The neutron source was set as the plasma, based on the core and power team design points, with corresponding neutron outputs that were propagated throughout the whole device. As successive plasma designs were generated in TokaMaker, they were implemented via a framework directly into OpenMC modeling to create a forward feedback loop for the overall design.

\subsection{Lifetime design goals}
The constraining design goals for neutronics analysis are that the device must survive neutron damage from the fluence of 3000 full-power DT pulses and that the superconducting magnets remain below 33~K, so that they avoid quenching during each 10-second pulse. The pulse number is set from the outset, and the 33~K limit is set by considering the critical current and magnetic field constraints set by the magnet team and based on empirical studies of HTS (as noted in section \ref{subsec:tfcoils}) \cite{molodyk_development_2021,Fischer2018FastNeutronIrradiation}.

Our simulations validate the general assumptions on neutron activity in  fusion devices. Generally, fusion power plant designs such as MANTA \cite{the_manta_collaboration_manta_2024} and ARC \cite{sorbom2015arc} are thought to be limited by neutron damage, or displacement per atom (DPA), rather than nuclear heating. This is due to the bulk of the heat deposition being into the blanket layer, rather than in and around the HTS magnets. Also, power plants are designed to survive across much longer time scales, with much higher neutron fluences. However, in test facilities such as CENTAUR and SPARC \cite{Creely2020SPARC}, which operate shorter pulses with less cumulative neutron flux, the key limiting factor is neutron heating rather than DPA. This ultimately is a critical factor in the design process, since the inboard build required more significant heat shielding than originally planned for. Our modeling supports this outlook, and in the final CENTAUR design all HTS device components are predicted to survive over 10 times the 3000 pulse limit considering solely the DPA. This corresponds to $3\times10^{22}$~neutrons/m$^2$. 

\subsection{Radial build and shielding thickness}
Given that the primary concern is heat shielding, a variety of materials are tested in the OpenMC workflow to determine the best choice for CENTAUR. A set of fusion power and plasma shapes are taken and used to run identical high-fidelity simulations ($10^7$ particles) of different materials. The best materials tested are shown in table \ref{tab:material_comp} compared with the HTS heating resulting from the absence of shielding material.

\renewcommand{\arraystretch}{1.5}
\begin{table}[htbp]
    \centering
    \begin{tabular}{c|c}
    Shielding Material &
    Percent of Vacuum Heating \\
    \hline
    $\text{B}_4\text{C}$     & 42.2$\%$  \\
    WC     & 58.5$\%$ \\
    HDPE & 73.5$\%$ 
    \end{tabular}
    \caption{Comparison of shielding materials used to protect the HTS magnets from quenching. HDPE is high-density polyethylene.}
    \label{tab:material_comp}
\end{table}
\renewcommand{\arraystretch}{1.5}

\begin{figure}[htbp]
    \centering
    \includegraphics[width=1.1\linewidth]{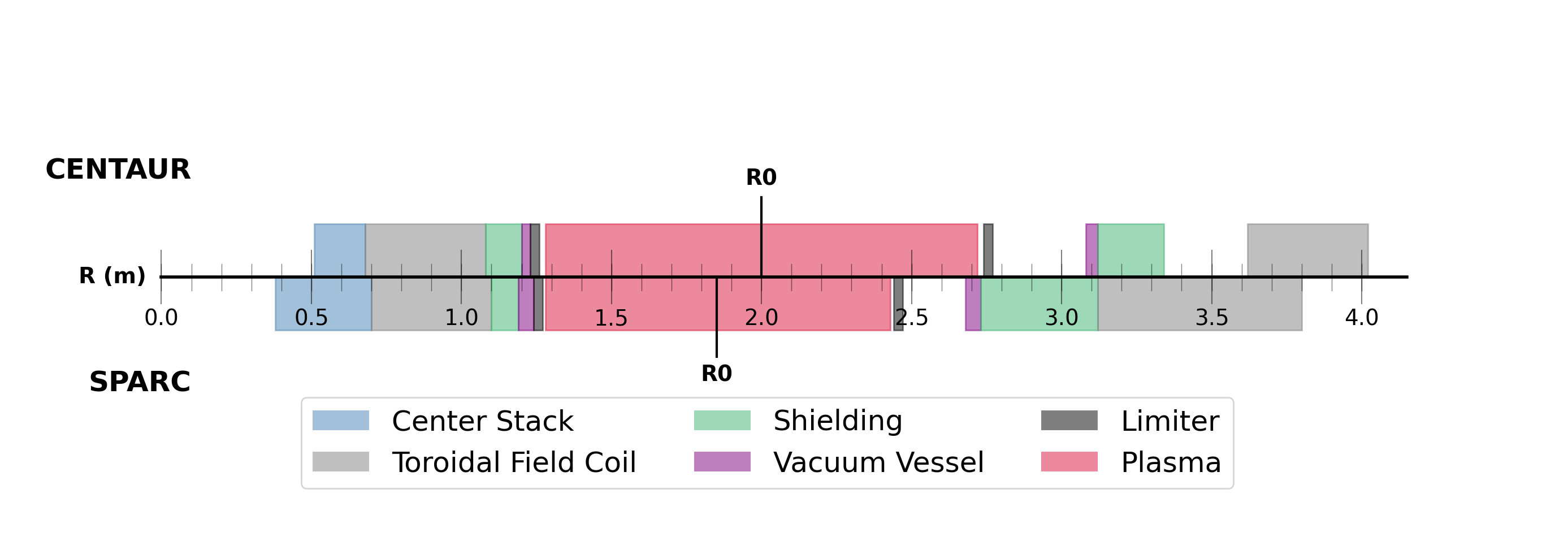}
    \caption{CENTAUR radial build compared to SPARC\cite{rodriguez-fernandez2022} at the midplane of both devices.}
    \label{fig:rad_build}
\end{figure}

Considering the maximum heating reduction provided by $\text{B}_4\text{C}$ and its planned use in other tokamak designs \cite{WANG_OPENMC}, the decisions on shielding thickness and placements are completed using only $\text{B}_4\text{C}$. Iterations between shielding width and fusion output produce a satisfactory operating point for core performance without quenching the magnets via neutron heating. This design point is one of the key constraining factors on the radial build of CENTAUR, seen in figure \ref{fig:rad_build}. 

Note that this design is not actively cooled during the shots, so the shielding and operating point is conservative and could be improved given modeling and inclusion of active cooling. While the plasma is considerably larger than SPARC radially, the maximum fusion power for CENTAUR is 40~MW while SPARC is planning on fusion power of around 140~MW.

\begin{figure}[htbp]
    \centering
    \includegraphics[width=1\linewidth]{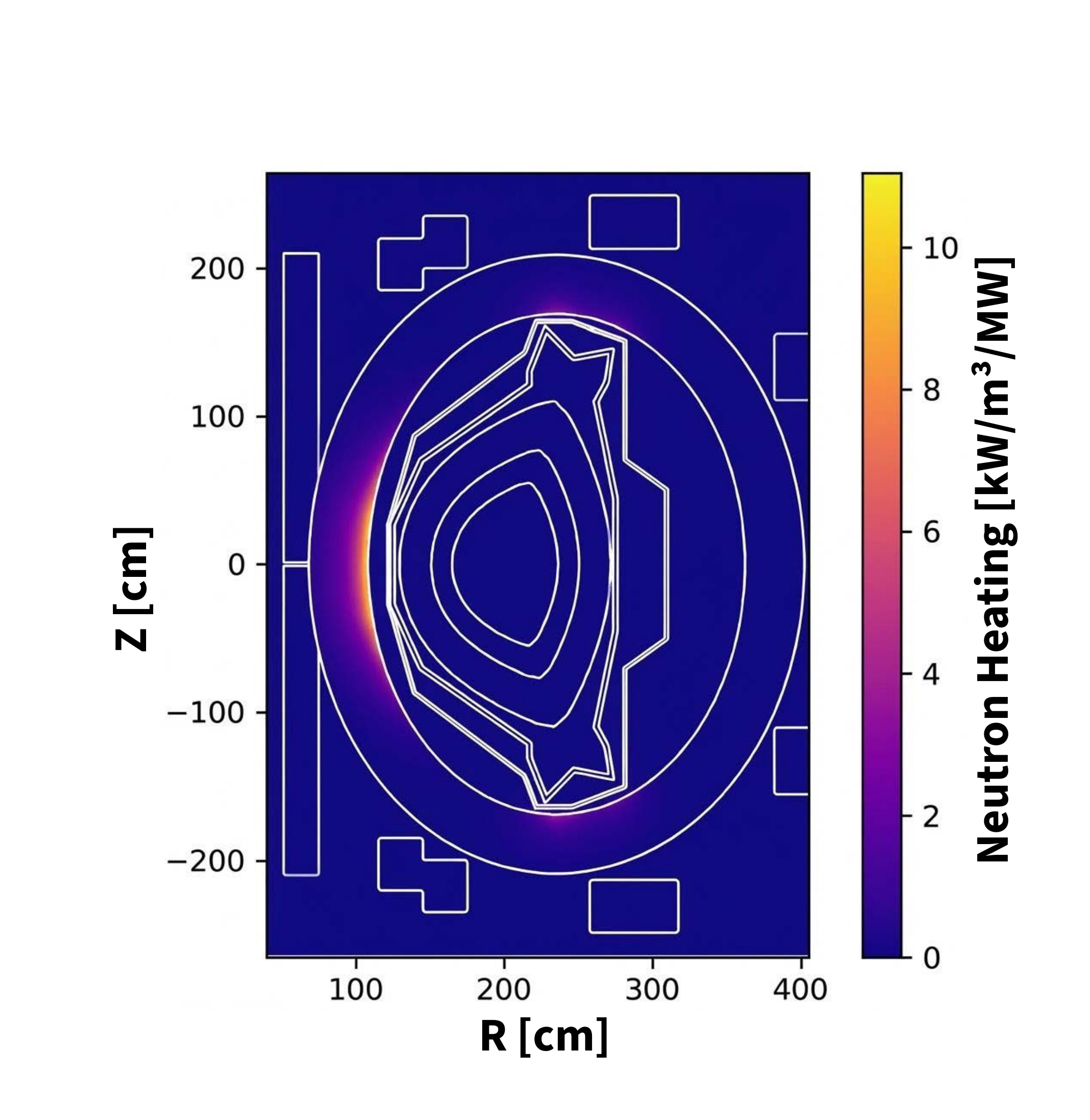}
    \caption{Representative OpenMC Monte Carlo neutron heating simulation using 12 cm of $\text{B}_4\text{C}$ shielding. The bulk of the heating is concentrated at the inboard midplane.}
    \label{fig:openmc}
\end{figure}

\begin{figure}[htbp]
    \centering
    \includegraphics[width=.8\linewidth]{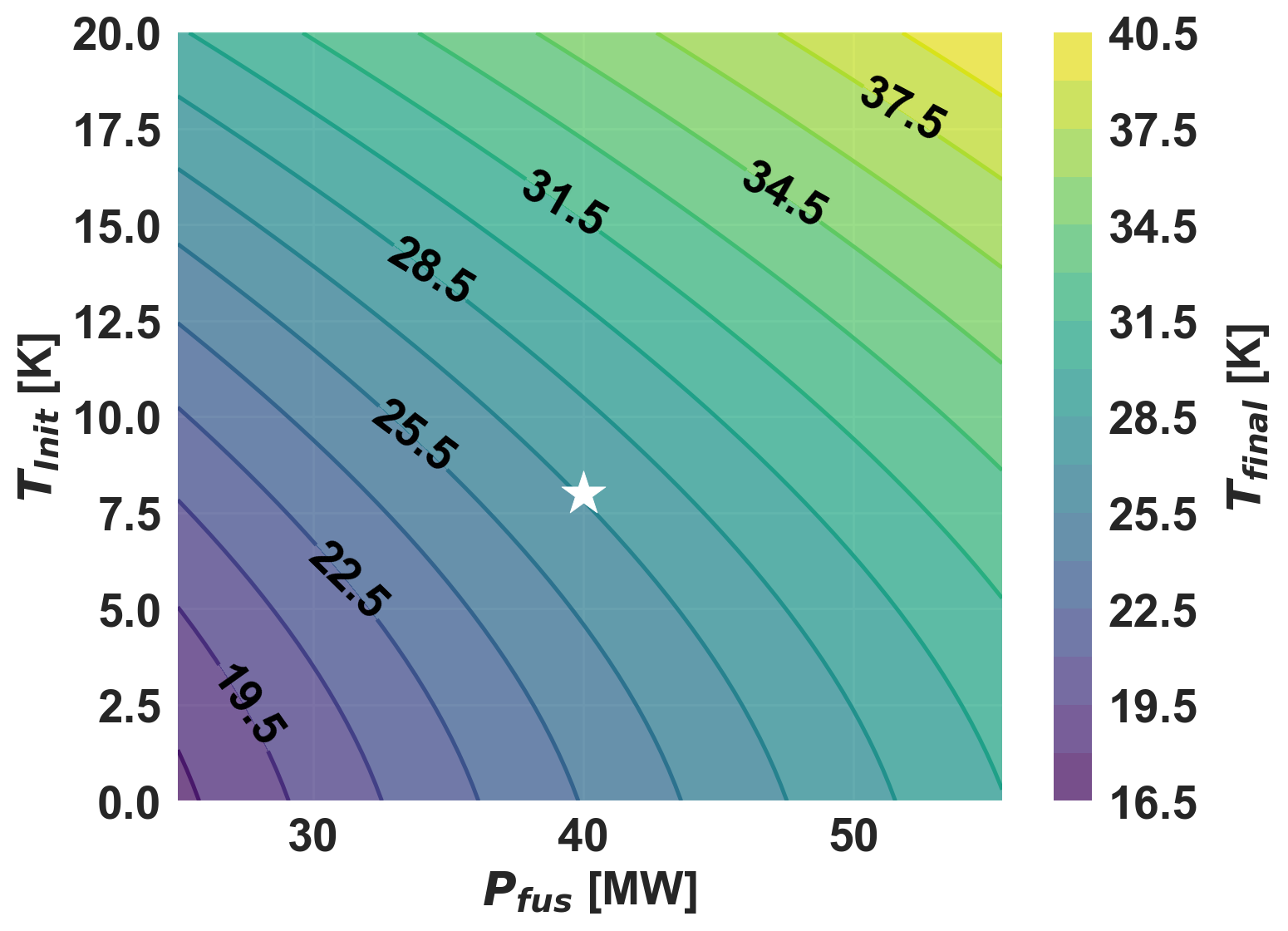}
    \caption{HTS magnet neutron heating over a 10-second shot. The star represents the CENTUAR operating point: initial HTS temperature of 8K, 12 cm of $B_4C$ shielding, and a 10 second flatop at 40 MW $P_{fus}$ with final HTS temperature below 33K.}
    \label{fig:neutron_heating}
\end{figure}

During the modeling process, a method was developed to scope out the operating regime in terms of initial temperature, fusion power, and final temperature. Since the neutron heating is concentrated at the inboard mid plane, numerous OpenMC simulations are run to relate fusion power to local volumetric heating (kW/m$^3$/MW). For example, an OpenMC simulation of the design point with 12~cm of $B_4C$ can be found in figure \ref{fig:openmc}. Then, the specific heat and density of HTS is used to convert the volumetric heating into a change in temperature. In figure \ref{fig:neutron_heating}, the operating point can be seen analyzed with this method, and is below the 33~K temperature limit at which the HTS magnets would quench from heating during operation. This is calculated based on an initial HTS temperature of 8~K with 12~cm of $B_4C$ and a plasma flat top of 10 seconds.

Considering the two main lifetime limitations from neutrons, DPA and heating, this modeling provides confidence that CENTAUR will meet its neutronics design goals. The DPA modeling shows the HTS will survive well beyond the 3000 DT pulse limit and the heating model shows the HTS will survive the neutron heating over the course of each shot. 

\section{Economic Analysis} \label{sec:econ}

Economic modeling is a critical piece of the design process for FPPs. CENTAUR is constrained to remain within a \$2B overnight cost target and maintain radioactive material compliance for on-site tritium inventory, in line with current restrictions on particle accelerators\cite{EnergyPolicyAct2005}. The costing model uses integrated system pricing based on materials and fabrication costs that adjust to the geometric and performance characteristics of the designed reactor similar to the MANTA costing model\cite{the_manta_collaboration_manta_2024}. CENTAUR achieves a total overnight cost of \$1.6B $\pm$0.2B based on the costing model developed in this work, maintaining the initial goal of a maximum budget of \$2B with a cost breakdown of 95$\%$ direct reactor costs and 5$\%$ indirect costs. Direct costs refer to systems that are directly involved in or necessary for the confinement and performance of the plasma while indirect costs are all systems that are needed to enable the direct cost systems or that manage and maintain the device.

\subsection{Device Costs}
\begin{figure}[htbp]
    \centering
    \includegraphics[width=0.8
    \linewidth]{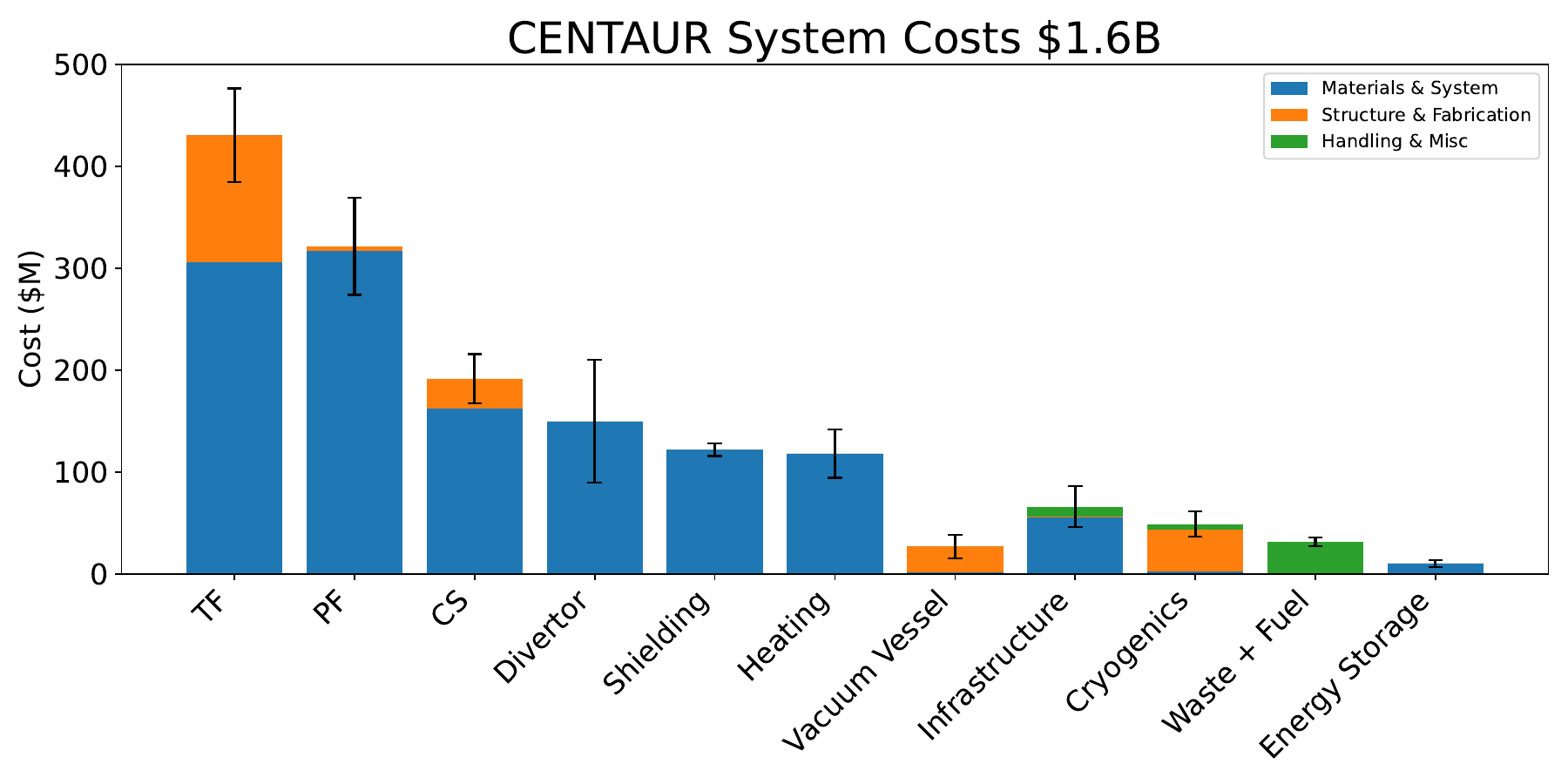}
    \caption{Total cost breakdown of CENTAUR by system.}
    \label{fig:system_costs_bar}
\end{figure}

Direct device costs constitute the majority of the total cost and are most significantly driven by the TF, PF, and CS magnet systems. Magnet pricing is calculated using the geometry of the TF and PF coils and the total length of HTS tape (606km$\pm$90km). The uncertainty calculation of the magnet systems is highly dependent on the price of HTS, which is difficult to determine given the quantity of HTS needed is not well documented at scale\cite{grant_sheahen_hts_costs}. Sensitivity analysis reveals that the PF coil system is the most affected by variation in the HTS price followed by the CS and TF magnets. The sensitivity to the price of HTS per kA-m is \$15.2M for the entire device split between  \$3.7M for the TF system, \$7.5M for the PF system, and \$4.1M for the CS system . Figure \ref{fig:system_costs_bar} illustrates how HTS costs dominate the cost of all magnet systems and are the most significant line items of the design.

Structural and fabrication costs are also significant factors in the TF, CS, vacuum vessel, and cryogenics systems. While structural components are not as variable, fabrication costs are the second most variable cost uncertainty after HTS tape. Fabrication costs affect the vacuum vessel costs most and are a significant source of uncertainty due to the lack of public quotes for other similar devices. Calculations for the overall cost of the vacuum vessel were linearly scaled based on the volume difference between CENTAUR and ITER \cite{ITERVacuumVesselCost}. Additionally, infrastructure costs are significantly higher compared to MANTA \cite{the_manta_collaboration_manta_2024} despite the omission of an electricity conversion plant in CENTAUR's design. This difference is due in large part to additional consideration of the location and construction costs of an operational site using high density concrete for neutron shielding of personnel and the public, though local legal restrictions on radiation sources could diminish construction costs significantly depending on the location site. Furthermore, land, electrical infrastructure, and construction costs were also estimated assuming a 0.25 km$^2$ campus within Claiborne, MS due to its proximity to the Grand Gulf Nuclear station and available electrical grid infrastructure. The location of the cost estimates was chosen only as an example case study and is not intended as an endorsement of the placement of a device like CENTAUR.

\subsubsection{Tritium Management}
The time evolution of tritium inventories within the fuel cycle was modeled using a system-level simulation implemented in Python, following the residence time method formalism established by Meschini \cite{meschini2023}. The computational framework tracks tritium mass across ten coupled components: the primary storage system, tritium extraction system, first wall, divertor, heat exchanger, detritiation system, vacuum pumps, fuel cleanup system, isotope separation system, and membrane separation unit. Each component is characterized by a residence time $\tau$ that governs the rate of tritium throughput. The tritium mass balance for each component obeys the differential equation
\begin{equation}
\frac{dI_i}{dt} = \sum_{j} \frac{I_j f_{j \rightarrow i}}{\tau_j} - \frac{(1 + \epsilon) I_i}{\tau_i} - \lambda I_i,
\end{equation}
where $I_i$ denotes the tritium inventory in component $i$, the coefficients $f_{j \rightarrow i}$ represent fractional flows between components, $\epsilon = 10^{-4}$ accounts for non-radioactive losses and $\lambda = 1.73 \times 10^{-9}~\mathrm{s}^{-1}$ is the radioactive decay constant corresponding to the 12.33 year tritium half-life. The coupled system of ordinary differential equations is integrated using the backward differentiation formula method, which provides numerical stability for the stiff dynamics arising from residence times that span several orders of magnitude, ranging from approximately 100 seconds in the membrane separation unit to 24 hours in the tritium extraction system. The baseline simulation assumes a tritium burn rate of $\dot{N} = 9.3 \times 10^{-7}~\mathrm{kg/s}$ and a tritium burn efficiency of 2$\%$, with component processing efficiencies uniformly set to 95$\%$\cite{meschini2023}. This implementation models only the inner fuel cycle dynamics and omits the breeding blanket, as the present analysis concerns fuel processing and recirculation pathways rather than tritium breeding.

\subsection{Costing Model Validation}
The CENTAUR costing model calculates costs for all the items listed in table \ref{tab:total_receipt} by system. Table \ref{tab:cost_model_comp} shows the comparison of the CENTAUR costing model overnight cost predictions against the ARIES \cite{aries_model}, and Sheffield \cite{sheffield_model} costing model for CENTAUR and SPARC designs. The CENTAUR costing model is more consistent with the ARIES model for CENTAUR's design, though it over-predicts the cost of SPARC in comparison. This difference in cost is due to both model's overemphasis on the cost of the poloidal field coils. Both models price the PF system differently than the toroidal field coils for the same field strength. illustrates this lopsided costing for SPARC. Despite the smaller discrepancy for SPARC, overall agreement between all costing models gives confidence to the overall \$1.6B estimate of the entire device.

\begin{table}[htbp]
    \centering
    \begin{tabular}{c|c|c}
         Model & CENTAUR & SPARC  \\
         \hline
         CENTAUR & $\$$1.6B & $\$$1.7B\\
         ARIES\cite{aries_model} & $\$$1.8B & $\$$1.6B\\
         Sheffield\cite{sheffield_model} & $\$$2.1B & $\$$1.8B \\
    \end{tabular}
    \caption{This table shows the overnight costs of CENTAUR and SPARC using the CENTAUR, ARIES, and Sheffield costing models.}
    \label{tab:cost_model_comp}
\end{table}

Economic analysis of CENTUAR shows the design to be cost effective and within initial goal of $\leq$\$2B overnight cost. The economic analysis using the CENTAUR costing model shows agreement with other historic costing models such as the ARIES and Sheffield models. Furthermore, the CENTAUR design costing model demonstrate that negative triangularity is not a cost prohibitive design choice for construction and operation. The success in economic projection is due in large part to the co-design process since the CENTAUR model allows for rapid iteration and individual system pricing that can inform system design.

\begin{table}[htbp]
    \centering
    \begin{tabular}{c|c|c}
         Material & Cost per unit [$\$$ (USD)] & System\\
         \hline
         REBCO (HTS) & 40 [Estimate]/kAm & TF,PF,CS \\
         nitronic 60 (steel)  &25/kg & TF,PF,CS\\
         Tungsten Carbide  &85/kg & Shielding\\
         Liquid He  &30/L & Cryo\\
         Tritium  & 30,000/g & Fueling\\
         Copper & 8/kg& Heating, Cryo, Infrastr. \\
         Deuterium   &11/g & Fueling\\
         Chromium-Vanadium  & 43/kg & Vacuum Vessel\\
         High density Concrete & 37.4/ft$^3$ & Infrastr.\\
    \end{tabular}
    \caption{Table of materials cost used for pricing of CENTAUR.}
    \label{mat cost table}
\end{table}

\begin{table}[htbp]
    \centering
    \begin{tabular}{c c c c}
         Item &Cost ($\$M$) &System $\&$ Category& Costing Method\\
         \hline
         TF HTS tape & 148 &TF  & Materials/System \\
         TF structural supports& 125 &TF & Structure $\&$ Fabrication\\
         TF resistive Leads& 158&TF & Materials/System \\
         PF HTS tape & 302&PF  & Materials/System \\
         PF structural supports& 4&PF & Structure $\&$ Fabrication\\
         PF resistive Leads& 16&PF & Materials/Systems \\
         CS HTS tape & 162&CS  & Materials/System \\
         CS structural supports& 30&CS & Structure $\&$ Fabrication\\
         Divertor materials & 150& Divertor& Materials/System\\
         Divertor coolants & $<$1&Divertor & Materials/System\\
         ICRH & 118&Heating&  Materials/System\\
         1st and 2nd wall& 6& VV &Materials/System \\
         VV fabrication&86 &VV & Structure $\&$ Fabrication \\
         Land& $<$1&Infrastructure & Handling $\&$ Misc. \\
         Electrical substation &4&Infrastructure & Handling $\&$ Misc. \\
         Electrical power lines &5&Infrastructure & Handling $\&$ Misc.\\
         Plant building materials&56 &Infrastructure & Materials/System\\
         Magnet Coolant costs & 3&Cryo & Materials/System\\
         Cryostat structure & 41&Cryo & Structure $\&$ Fabrication \\
         Coolant recycling & 5&Cryo & Handling $\&$ Misc.\\
         Neutron Shielding & 122&Shielding & Materials/System\\
         Tritium Inventory & $<$1& Fueling & Materials/System\\
         Tritium Handling & 31& Fueling & Handling $\&$ Misc.\\
         Energy Storage batteries & 10&Energy Storage & Materials/System\\
         
    \end{tabular}
    \caption{This table shows every cost calculated for CENTAUR using its costing model.}
    \label{tab:total_receipt}
\end{table}

\section{Conclusion}
CENTAUR offers a novel break even fusion device with a robust power handling scheme, largely due to advantages from strong negative triangularity plasma shaping. High-field from superconducting magnets ensures sufficient confinement for $Q>1$ operation, outside of H-mode and ELM-free. The large wetted surface and high radiative fraction of the NT divertor region allows for simple divertor geometry and a high margin of safety for plasma facing components. The HTS magnet systems produce a high toroidal field and sufficient flux swing in a compact design while remaining below HTS critical current limits.  The neutronics analysis confirms the viability of the device lifetime constraints and the necessary shielding to prevent unacceptable heat rise in the HTS magnets. Further, costing analysis shows that this device can be constructed, operated, and decommissioned within the budget limitations prescribed for the design study. This design study shows that CENTAUR  provides the necessary stepping stone between current-day tokamaks and a fusion pilot plant by demonstrating the negative triangularity operational scheme in Q$>$1 DT plasma conditions. The self-consistent analysis presented demonstrates that CENTAUR is designed to achieve scientific gain of 1.3 at an overnight cost under \$2B while employing available technologies and traditional heat exhaust divertor schemes 

Future work would most likely include a continued optimization of the operating scenario to extract maximal plasma performance. Further, the inclusion of self-consistent ion cyclotron resonant heating models would improve understanding of our expected core scenario. One of the principal areas of uncertainty in this design is magnet heating due to neutron fluence. A follow-up study to determine the effects of active cooling on superconducting TF coils would be of high value.

\label{sec:conc}

\section*{Acknowledgements}

The authors are grateful to all other course members of Columbia APPH 9142, including Princeton students who participated through the Inter-University Engineering Doctoral Consortium, and for the expert advice of Dr. Michael Bergmann. This work was supported in part by Columbia Univerity, US DOE Grants DE-SC0024898, DE-SC0022270 and DE-SC0022272, US DOE contract DE-AC52 07NA27344, and NSF award 2401039.

\printbibliography

@article{Hansen_2025,
title = {ThinCurr: An open-source 3D thin-wall eddy current modeling code for the analysis of large-scale systems of conducting structures},
journal = {Computer Physics Communications},
volume = {315},
pages = {109713},
year = {2025},
issn = {0010-4655},
doi = {https://doi.org/10.1016/j.cpc.2025.109713},
url = {https://www.sciencedirect.com/science/article/pii/S0010465525002152},
author = {Christopher Hansen and Alexander Battey and Anson Braun and Sander Miller and Michael Lagieski and Ian Stewart and Ryan Sweeney and Carlos Paz-Soldan}
}

@misc{leuer_talk_2020,
  author       = {Leuer, Jim},
  title        = {Ramp-Up Flux Consumption of +- Triangularity Plasmas},
  howpublished = {Private communication},
  year         = {2020},
  month        = 5
}

@article{ASTRA_Reference_Paper,
	title        = {{ASTRA} {Automated System for TRansport Analysis} in a Tokamak},
	author       = {Pereverzev, G. and Yushmanov, P. N.},
	year         = 2002,
	journal      = {Max Planck Institute for Plasma Physics, Garching, Germany},
	address      = {Germany},
	url          = {https://pure.mpg.de/rest/items/item_2138238/component/file_2138237/content}
}

@article{Staebler_2007,
    author = {Staebler, G. M. and Kinsey, J. E. and Waltz, R. E.},
    title = "{A theory-based transport model with comprehensive physicsa)}",
    journal = {Physics of Plasmas},
    volume = {14},
    number = {5},
    pages = {055909},
    year = {2007},
    issn = {1070-664X},
    doi = {10.1063/1.2436852},
    url = {https://doi.org/10.1063/1.2436852}                             
}

@article{Fable_2013,
doi = {10.1088/0741-3335/55/12/124028},
url = {https://dx.doi.org/10.1088/0741-3335/55/12/124028} ,
year = {2013},
publisher = {IOP Publishing},
volume = {55},
number = {12},
pages = {124028},
author = {E. Fable and others},
title = {Novel free-boundary equilibrium and transport solver with theory-based models and its validation against ASDEX Upgrade current ramp scenarios},
journal = {Plasma Phys. Control. Fusion}
}

@article{Houlberg_1982,
  title = {Contour Analysis of Fusion Reactor Plasma Performance},
  author = {Houlberg, W.A. and Attenberger, S.E. and Hively, L.M.},
  date = {1982-07},
  journaltitle = {Nuclear Fusion},
  volume = {22},
  number = {7},
  pages = {935},
  doi = {10.1088/0029-5515/22/7/006},
  url = {https://doi.org/10.1088/0029-5515/22/7/006}
}

@article{10.1063/1.872193,
  title = {Stable Equilibria for Bootstrap-Current-Driven Low Aspect Ratio Tokamaks},
  author = {Miller, R. L. and Lin-Liu, Y. R. and Turnbull, A. D. and Chan, V. S. and Pearlstein, L. D. and Sauter, O. and Villard, L.},
  date = {1997-04},
  journaltitle = {Physics of Plasmas},
  volume = {4},
  number = {4},
  eprint = {https://pubs.aip.org/aip/pop/article-pdf/4/4/1062/19079719/1062_1_online.pdf},
  pages = {1062--1068},
  issn = {1070-664X},
  doi = {10.1063/1.872193},
  url = {https://doi.org/10.1063/1.872193}
}

@article{10.1063/1.2044587,
  title = {Gyro-{{Landau}} Fluid Equations for Trapped and Passing Particles},
  author = {Staebler, G. M. and Kinsey, J. E. and Waltz, R. E.},
  date = {2005-10},
  journaltitle = {Physics of Plasmas},
  volume = {12},
  number = {10},
  eprint = {https://pubs.aip.org/aip/pop/article-pdf/doi/10.1063/1.2044587/14688354/102508_1_online.pdf},
  pages = {102508},
  issn = {1070-664X},
  doi = {10.1063/1.2044587},
  url = {https://doi.org/10.1063/1.2044587}
}

@techreport{DOE2015PMI,
  title        = {Report on Science Challenges and Research Opportunities in Plasma Materials Interactions},
  author       = {{U.S. Department of Energy}},
  institution  = {U.S. Department of Energy},
  year         = {2015},
  type         = {Technical Report},
  address      = {Washington, DC, USA}
}

@article{loartePlasmaDetachmentJET1998,
  title = {Plasma Detachment in {{JET Mark I}} Divertor Experiments},
  author = {Loarte, A. and Monk, R. D. and {Mart{\'i}n-Sol{\'i}s}, J. R. and Campbell, D. J. and Chankin, A. V. and Clement, S. and Davies, S. J. and Ehrenberg, J. and Erents, S. K. and Guo, H. Y. and Harbour, P. J. and Horton, L. D. and Ingesson, L. C. and J{\"a}ckel, H. and Lingertat, J. and Lowry, C. G. and Maggi, C. F. and Matthews, G. F. and McCormick, K. and O'Brien, D. P. and Reichle, R. and Saibene, G. and Smith, R. J. and Stamp, M. F. and Stork, D. and Vlases, G. C.},
  year = 1998,
  month = {3},
  journal = {Nuclear Fusion},
  volume = {38},
  number = {3},
  pages = {331},
  issn = {0029-5515},
  doi = {10.1088/0029-5515/38/3/303},
  urldate = {2026-04-03},
  langid = {english}
}

@article{Federici2001PMI,
  title        = {Plasma–material interactions in current tokamaks and their implications for next step devices},
  author       = {Federici, G. and Skinner, C.},
  journal      = {Journal of Nuclear Materials},
  year         = {2001},
  note         = {Review of underlying physical processes and experimental database},
}

@article{Leonard2018Detachment,
  title        = {Plasma detachment in divertor tokamaks},
  author       = {Leonard, A. W.},
  journal      = {Plasma Physics and Controlled Fusion},
  year         = {2018},
  doi          = {10.1088/1361-6587/aaa7a9},
}

@article{Krasheninnikov2017Detachment,
  title        = {Physics of ultimate detachment of a tokamak divertor plasma},
  author       = {Krasheninnikov, S. I.},
  journal      = {Journal of Plasma Physics},
  year         = {2017},
}

@article{Asakura2023DivertorReview,
  title        = {Recent progress of plasma exhaust concepts and divertor designs for tokamak DEMO reactors},
  author       = {Asakura, N. and Hoshino, K. and others},
  journal      = {Nuclear Materials and Energy},
  year         = {2023},
  doi          = {10.1016/j.nme.2023.101446},
}

@book{Stangeby2000,
  title     = {The Plasma Boundary of Magnetic Fusion Devices},
  author    = {Stangeby, P. C.},
  year      = {2000},
  publisher = {Institute of Physics Publishing}
}

@article{Krieger_2025_divertors,
doi = {10.1088/1741-4326/adaf42},
url = {https://doi.org/10.1088/1741-4326/adaf42},
year = {2025},
month = {3},
publisher = {IOP Publishing},
volume = {65},
number = {4},
pages = {043001},
author = {Krieger, K. and Brezinsek, S. and Coenen, J.W. and Frerichs, H. and Kallenbach, A. and Leonard, A.W. and Loarer, T. and Ratynskaia, S. and Vianello, N. and Asakura, N. and Bernert, M. and Carralero, D. and Ding, R. and Douai, D. and Eich, T. and Gasparyan, Y. and Hakola, A. and Hatano, Y. and Jakubowski, M. and Kobayashi, M. and Krasheninnikov, S. and Masuzaki, S. and Nakano, T. and Neu, R. and Pitts, R.A. and Rapp, J. and Schmid, K. and Schmitz, O. and Tskhakaya, D. and Wang, L. and Wauters, T. and Wiesen, S.},
title = {Scrape-off layer and divertor physics: Chapter 5 of the special issue: on the path to tokamak burning plasma operation},
journal = {Nuclear Fusion}
}

@article{Thome_2024,
  title = {Overview of Results from the 2023 {{DIII-D}} Negative Triangularity Campaign},
  author = {Thome, K E and Austin, M E and Hyatt, A and Marinoni, A and Nelson, A O and Paz-Soldan, C and Scotti, F and Boyes, W and Casali, L and Chrystal, C and Ding, S and Du, X D and Eldon, D and Ernst, D and Hong, R and McKee, G R and Mordijck, S and Sauter, O and Schmitz, L and Barr, J L and Burke, M G and Coda, S and Cote, T B and Fenstermacher, M E and Garofalo, A and Khabanov, F O and Kramer, G J and Lasnier, C J and Logan, N C and Lunia, P and McLean, A G and Okabayashi, M and Shiraki, D and Stewart, S and Takemura, Y and Truong, D D and Osborne, T and Van Zeeland, M A and Victor, B S and Wang, H Q and Watkins, J G and Wehner, W P and Welander, A S and Wilks, T M and Yang, J and Yu, G and Zeng, L and Team, the DIII-D},
  date = {2024-09},
  journaltitle = {Plasma Physics and Controlled Fusion},
  volume = {66},
  number = {10},
  pages = {105018},
  publisher = {IOP Publishing},
  doi = {10.1088/1361-6587/ad6f40},
  url = {https://doi.org/10.1088/1361-6587/ad6f40}
}

@misc{pereverzev2002astra,
  title={ASTRA. Automated System for TRansport Analysis in a tokamak},
  author={Pereverzev, Gregorij V and Yushmanov, e PN},
  year={2002},
  publisher={Max-Planck-Institut f{\"u}r Plasmaphysik}
}

@article{HANSEN2024109111,
  title = {{{TokaMaker}}: {{An}} Open-Source Time-Dependent {{Grad-Shafranov}} Tool for the Design and Modeling of Axisymmetric Fusion Devices},
  author = {Hansen, C. and Stewart, I.G. and Burgess, D. and Pharr, M. and Guizzo, S. and Logak, F. and Nelson, A.O. and Paz-Soldan, C.},
  date = {2024},
  journaltitle = {Computer Physics Communications},
  volume = {298},
  pages = {109111},
  issn = {0010-4655},
  doi = {10.1016/j.cpc.2024.109111},
  url = {https://www.sciencedirect.com/science/article/pii/S0010465524000341}
}

@article{rognlienFullyImplicitTime1992,
  title = {A Fully Implicit, Time Dependent 2-{{D}} Fluid Code for Modeling Tokamak Edge Plasmas},
  author = {Rognlien, T. D. and Milovich, J. L. and Rensink, M. E. and Porter, G. D.},
  year = 1992,
  month = 12,
  journal = {Journal of Nuclear Materials},
  series = {Plasma-{{Surface Interactions}} in {{Controlled Fusion Devices}}},
  volume = {196--198},
  pages = {347--351},
  issn = {0022-3115},
  doi = {10.1016/S0022-3115(06)80058-9},
  urldate = {2026-04-06},
  url = {https://www.sciencedirect.com/science/article/pii/S0022311506800589}
}

@manual{uedgeman,
  title={Users manual for the UEDGE edge-plasma transport code},
  author={T. D. Rognlien and M. E. Rensink},
  year={2023},
  publisher={LLNL Report},
  url={https://github.com/llnl/UEDGE/releases/tag/v8.1.1-patch.0}
}

@book{stangebyPlasmaBoundaryMagnetic2000,
  title = {The Plasma Boundary of Magnetic Fusion Devices},
  author = {Stangeby, Peter C.},
  year = 2000,
  series = {Plasma Physics Series},
  publisher = {Institute of Physics Publ},
  address = {Bristol [ u.a. ]},
  isbn = {978-0-7503-0559-4},
  langid = {english}
}

@article{wangIntegrationFullDivertor2021,
  title = {Integration of Full Divertor Detachment with Improved Core Confinement for Tokamak Fusion Plasmas},
  author = {Wang, L. and Wang, H. Q. and Ding, S. and Garofalo, A. M. and Gong, X. Z. and Eldon, D. and Guo, H. Y. and Leonard, A. W. and Hyatt, A. W. and Qian, J. P. and Weisberg, D. B. and McClenaghan, J. and Fenstermacher, M. E. and Lasnier, C. J. and Watkins, J. G. and Shafer, M. W. and Xu, G. S. and Huang, J. and Ren, Q. L. and Buttery, R. J. and Humphreys, D. A. and Thomas, D. M. and Zhang, B. and Liu, J. B.},
  year = 2021,
  month = 3,
  journal = {Nature Communications},
  volume = {12},
  number = {1},
  pages = {1365},
  publisher = {Nature Publishing Group},
  issn = {2041-1723},
  doi = {10.1038/s41467-021-21645-y},
  urldate = {2026-04-08},
  copyright = {2021 The Author(s)},
  langid = {english}
}

@article{soukhanovskiiDivertorHeatFlux,
  title = {Divertor {{Heat Flux Reduction}} and {{Detachment}} in {{NSTX}}},
  author = {Soukhanovskii, V A and Maingi, R and Raman, R and Bell, R E and Bush, C and Kaita, R and Kugel, H W and Lasnier, C J and LeBlanc, B P and Menard, J E and Paul, S F and Roquemore, A L},
  langid = {english}
}

@article{renExperimentalObservationHeat2021,
  title = {Experimental Observation of Heat Flux Mitigation during Divertor Detachment in the {{DIII-D}} Small Angle Slot Divertor},
  author = {Ren, J. and Donovan, D. C. and Watkins, J. G. and Wang, H. Q. and Thomas, D. M. and Boivin, R.},
  year = 2021,
  month = 3,
  journal = {Nuclear Materials and Energy},
  volume = {26},
  pages = {100887},
  issn = {2352-1791},
  doi = {10.1016/j.nme.2020.100887},
  urldate = {2026-04-08}
}

@article{faitschBroadeningPowerFalloff2021,
  title = {Broadening of the Power Fall-off Length in a High Density, High Confinement {{H-mode}} Regime in {{ASDEX Upgrade}}},
  author = {Faitsch, M. and Eich, T. and Harrer, G. F. and Wolfrum, E. and Brida, D. and David, P. and Griener, M. and Stroth, U.},
  year = 2021,
  month = 3,
  journal = {Nuclear Materials and Energy},
  volume = {26},
  pages = {100890},
  issn = {2352-1791},
  doi = {10.1016/j.nme.2020.100890},
  urldate = {2026-04-08}
}

@article{soukhanovskiiReviewRadiativeDetachment2017,
  title = {A Review of Radiative Detachment Studies in Tokamak Advanced Magnetic Divertor Configurations},
  author = {Soukhanovskii, V A},
  year = 2017,
  month = 4,
  journal = {Plasma Physics and Controlled Fusion},
  volume = {59},
  number = {6},
  pages = {064005},
  publisher = {IOP Publishing},
  issn = {0741-3335},
  doi = {10.1088/1361-6587/aa6959},
  urldate = {2026-04-08},
  langid = {english}
}

@article{mcclenaghan2024examining,
  title={Examining transport and integrated modeling predictive capabilities for negative-triangularity scenarios},
  author={McClenaghan, J and Marinoni, A and Nelson, AO and Neiser, T and Lao, LL and Staebler, GM and Smith, SP and Meneghini, OM and Lyons, BC and Snyder, PB and others},
  journal={Plasma Physics and Controlled Fusion},
  volume={66},
  number={11},
  pages={115008},
  year={2024},
  publisher={IOP Publishing}
}

@article{Romano2015OpenMC,
  author       = {Romano, Paul K. and Horelik, Nicholas E. and Herman, Bryan R. and Nelson, Adam G. and Forget, Benoit and Smith, Kord},
  title        = {OpenMC: A state-of-the-art Monte Carlo code for research and development},
  journal      = {Annals of Nuclear Energy},
  volume       = {82},
  pages        = {90--97},
  year         = {2015},
  publisher    = {Elsevier},
  doi          = {10.1016/j.anucene.2014.07.048},
  url          = {https://doi.org/10.1016/j.anucene.2014.07.048}
}

@article{sorbom2015arc,
  title        = {ARC: A compact, high‐field, fusion nuclear science facility and demonstration power plant with demountable magnets},
  author       = {Sorbom, B.~N. and Ball, J. and Palmer, T.~R. and Mangiarotti, F.~J. and Sierchio, J.~M. and Bonoli, P. and Kasten, C. and Sutherland, D.~A. and Barnard, H.~S. and Haakonsen, C.~B. and Goh, J. and Sung, C. and Whyte, D.~G.},
  journal      = {Fusion Engineering and Design},
  volume       = {100},
  pages        = {378--405},
  year         = {2015},
  doi          = {10.1016/j.fusengdes.2015.07.008}
}

@article{Creely2020SPARC,
  author    = {A.~J. Creely and M.~J. Greenwald and S.~B. Ballinger and D.~Brunner and
               J.~Canik and J.~Doody and T.~F{\"u}l{\"o}p and D.~T. Garnier and
               R.~Granetz and T.~K. Gray and C.~Holland and N.~T. Howard and
               J.~W. Hughes and J.~H. Irby and V.~A. Izzo and G.~J. Kramer and
               A.~Q. Kuang and B.~LaBombard and Y.~Lin and B.~Lipschultz and
               N.~C. Logan and J.~D. Lore and E.~S. Marmar and K.~Montes and
               R.~T. Mumgaard and C.~Paz-Soldan and C.~Rea and M.~L. Reinke and
               P.~Rodriguez-Fernandez and K.~S{\"a}rkim{\"a}ki and F.~Sciortino and
               S.~D. Scott and A.~Snicker and P.~B. Snyder and B.~N. Sorbom and
               R.~Sweeney and R.~A. Tinguely and E.~A. Tolman and M.~Umansky and
               O.~Vallhagen and J.~Varje and D.~G. Whyte and J.~C. Wright and
               S.~J. Wukitch and J.~Zhu and SPARC Team},
  title     = {Overview of the SPARC tokamak},
  journal   = {Journal of Plasma Physics},
  volume    = {86},
  number    = {5},
  pages     = {865860502},
  year      = {2020},
  doi       = {10.1017/S0022377820001257},
  publisher = {Cambridge University Press}
}

@article{Fischer2018FastNeutronIrradiation,
  author       = {D. X. Fischer and R. Prokopec and J. Emhofer and M. Eisterer},
  title        = {The effect of fast neutron irradiation on the superconducting properties of REBCO coated conductors with and without artificial pinning centers},
  journal      = {Superconductor Science and Technology},
  volume       = {31},
  number       = {4},
  pages        = {044006},
  year         = {2018},
  doi          = {10.1088/1361-6668/aaadf2},
}

@book{EnergyPolicyAct2005,
  title        = {Energy Policy Act of 2005},
  year         = {2005},
  note         = {H.R. 6, 109th Cong. (Aug. 8, 2005), became Public Law 109-58},
  publisher    = {U.S. Government Printing Office},
  url          = {https://www.congress.gov/bill/109th-congress/house-bill/6},
  author       = {United States Congress} 
}

@article{meschini2023,
    author = {Samuele Meschini, Sara E. Ferry, Rémi Delaporte-Mathurin and Dennis G. Whyte},
    title = {Modeling and analysis of the tritium fuel cycle for ARC- and STEP-class D-T fusion power plants},
    journal = {Nuclear Fusion},
    year = {2023}
}

@article{sheffield_model,
author = {J. Sheffield and R. A. Dory and S. M. Cohn and J. G. Delene and L. Parsly and D. E. T. F. Ashby and W. T. Reiersen},
title = {Cost Assessment of a Generic Magnetic Fusion Reactor},
journal = {Fusion Technology},
volume = {9},
number = {2},
pages = {199--249},
year = {1986},
publisher = {Taylor \& Francis},
doi = {10.13182/FST9-2-199},
URL = {https://doi.org/10.13182/FST9-2-199},
}

@techreport{aries_model,
  author      = {Waganer, L. M.},
  title       = {ARIES Cost Account Documentation},
  institution = {University of California, San Diego},
  year        = {2013}
}

@article{Scotti_2025,
doi = {10.1088/1361-6587/adf881},
url = {https://doi.org/10.1088/1361-6587/adf881},
year = {2025},
publisher = {IOP Publishing},
volume = {67},
number = {9},
pages = {095030},
author = {Scotti, F and Marinoni, A and McLean, A G and Nelson, A O and Paz-Soldan, C and Thome, K E and Zhao, M and Allen, S and Austin, M and Burke, M G and Bykov, I and Chrystal, C and Eldon, D and Fenstermacher, M and Hyatt, A and Glass, F and Lasnier, C J and Lore, J and Leonard, A and Murphy, C and Osborne, T and Sauter, O and Truong, D and Wang, H Q and Wilcox, R},
title = {Divertor characterization and access to dissipative divertor conditions in negative triangularity discharges in DIII-D},
journal = {Plasma Physics and Controlled Fusion},
}

@article{Slendebroek_2026,
  title   = {Exploring the fusion power plant design space: comparative analysis of positive and negative triangularity tokamaks through optimization},
  author  = {Slendebroek, T. and Nelson, A. O. and Meneghini, O. M. and Dose, G. and Ghiozzi, A. G. and Harvey, J. and Lyons, B. C. and McClenaghan, J. and Neiser, T. F. and Weisberg, D. B. and Yoo, M. G. and Bursch, E. and Holland, C.},
  journal = {Nuclear Fusion},
  volume  = {66},
  number  = {2},
  year    = {2026},
  doi     = {10.1088/1741-4326/ae27e6},
  url     = {https://doi.org/10.1088/1741-4326/ae27e6}
}

@article{igitkhanov,
author = {Igitkhanov, Yu. L.},
title = {Impurity Transport at Arbitrary Densities in the Divertor Plasma},
journal = {Contributions to Plasma Physics},
volume = {28},
number = {4-5},
pages = {477-482},
doi = {https://doi.org/10.1002/ctpp.2150280435},
url = {https://onlinelibrary.wiley.com/doi/abs/10.1002/ctpp.2150280435},
eprint = {https://onlinelibrary.wiley.com/doi/pdf/10.1002/ctpp.2150280435},
year = {1988}
}

@article{ROSZELL2013S1084,
title = {Deuterium retention in single-crystal tungsten irradiated with 10–500 eV/D+},
journal = {Journal of Nuclear Materials},
volume = {438},
pages = {S1084-S1087},
year = {2013},
note = {Proceedings of the 20th International Conference on Plasma-Surface Interactions in Controlled Fusion Devices},
issn = {0022-3115},
doi = {https://doi.org/10.1016/j.jnucmat.2013.01.238},
url = {https://www.sciencedirect.com/science/article/pii/S0022311513002468},
author = {J.P. Roszell and J.W. Davis and V.Kh. Alimov and K. Sugiyama and A.A. Haasz}
}

@article{ZHANG2022101265,
title = {Spectroscopic investigation of the tungsten deuteride sputtering in the EAST divertor},
journal = {Nuclear Materials and Energy},
volume = {33},
pages = {101265},
year = {2022},
issn = {2352-1791},
doi = {https://doi.org/10.1016/j.nme.2022.101265},
url = {https://www.sciencedirect.com/science/article/pii/S2352179122001466},
author = {Q. Zhang and F. Ding and S. Brezinsek and L. Yu and L.Y. Meng and P.A. Zhao and D.W. Ye and Z.H. Hu and Y. Zhang and R. Ding and L. Wang and G.-N. Luo}
}

@article{sput,
title = {Sputtering of beryllium, tungsten, tungsten oxide and mixed W–C layers by deuterium ions in the near-threshold energy range},
journal = {Journal of Nuclear Materials},
volume = {266-269},
pages = {222-227},
year = {1999},
issn = {0022-3115},
doi = {https://doi.org/10.1016/S0022-3115(98)00819-8},
url = {https://www.sciencedirect.com/science/article/pii/S0022311598008198},
author = {M.I Guseva and A.L Suvorov and S.N Korshunov and N.E Lazarev}
}

@article{ITERVacuumVesselCost,
  author       = {David Dalton},
  title        = {Westinghouse Wins \$180 Million Contract For Assembly Of ITER Vacuum Vessel},
  journal      = {NucNet},
  year         = {2025},
  month        = {7},
  day          = {2},
  url          = {https://www.nucnet.org/news/westinghouse-wins-usd180-million-contract-for-assembly-of-iter-vacuum-vessel-7-3-2025},
  note         = {Accessed: 2026-02-10},
}

@article{sorbom2015,
  title = {{{ARC}}: {{A}} Compact, High-Field, Fusion Nuclear Science Facility and Demonstration Power Plant with Demountable Magnets},
  shorttitle = {{{ARC}}},
  author = {Sorbom, B.N. and Ball, J. and Palmer, T.R. and Mangiarotti, F.J. and Sierchio, J.M. and Bonoli, P. and Kasten, C. and Sutherland, D.A. and Barnard, H.S. and Haakonsen, C.B. and Goh, J. and Sung, C. and Whyte, D.G.},
  year = 2015,
  month = 11,
  journal = {Fusion Engineering and Design},
  volume = {100},
  pages = {378--405},
  issn = {09203796},
  doi = {10.1016/j.fusengdes.2015.07.008},
  urldate = {2024-02-18},
  langid = {english}
}

@article{rodriguez-fernandez2022,
  title = {Overview of the {{SPARC}} Physics Basis towards the Exploration of Burning-Plasma Regimes in High-Field, Compact Tokamaks},
  author = {{Rodriguez-Fernandez}, P. and Creely, A.J. and Greenwald, M.J. and Brunner, D. and Ballinger, S.B. and Chrobak, C.P. and Garnier, D.T. and Granetz, R. and Hartwig, Z.S. and Howard, N.T. and Hughes, J.W. and Irby, J.H. and Izzo, V.A. and Kuang, A.Q. and Lin, Y. and Marmar, E.S. and Mumgaard, R.T. and Rea, C. and Reinke, M.L. and Riccardo, V. and Rice, J.E. and Scott, S.D. and Sorbom, B.N. and Stillerman, J.A. and Sweeney, R. and Tinguely, R.A. and Whyte, D.G. and Wright, J.C. and Yuryev, D.V.},
  year = 2022,
  month = 9,
  journal = {Nuclear Fusion},
  volume = {62},
  number = {4},
  pages = {042003},
  issn = {0029-5515, 1741-4326},
  doi = {10.1088/1741-4326/ac1654},
  urldate = {2024-10-21},
  langid = {english}
}

@article{Austin_2021,
  title = {Diverted Negative Triangularity Plasmas on DIII-D: The Benefit of High Confinement without the Liability of an Edge Pedestal},
  author = {Austin, M. E. and Hyatt, A. W. and Saarelma, S. and Scotti, F. and Yan, Z. and others},
  date = {2021},
  journaltitle = {Nuclear Fusion},
  volume = {61},
  number = {11},
  pages = {116016},
  doi = {10.1088/1741-4326/ac1f90},
  url = {https://doi.org/10.1088/1741-4326/ac1f90}
}

@article{nelsonCharacterizationELMfreeNegative2024b,
  title = {Characterization of the {{ELM-free}} Negative Triangularity Edge on {{DIII-D}}},
  author = {Nelson, A O and Schmitz, L and Cote, T and Parisi, J F and Stewart, S and Paz-Soldan, C and Thome, K E and Austin, M E and Scotti, F and Barr, J L and Hyatt, A and Leuthold, N and Marinoni, A and Neiser, T and Osborne, T and Richner, N and Welander, A S and Wehner, W P and Wilcox, R and Wilks, T M and Yang, J and Team, the DIII-D.},
  date = {2024-09},
  journaltitle = {Plasma Physics and Controlled Fusion},
  shortjournal = {Plasma Phys. Control. Fusion},
  volume = {66},
  number = {10},
  pages = {105014},
  publisher = {IOP Publishing},
  issn = {0741-3335},
  doi = {10.1088/1361-6587/ad6a83},
  url = {https://doi.org/10.1088/1361-6587/ad6a83},
  urldate = {2026-02-13},
  langid = {english}
}

@article{Barth2015,
  author  = {Barth, Christian and Mondonico, Giorgio and Senatore, Carmine},
  title   = {Electro-mechanical properties of {REBCO} coated conductors from
             various industrial manufacturers at 77\,K, self-field and 4.2\,K, 19\,T},
  journal = {Superconductor Science and Technology},
  year    = {2015},
  volume  = {28},
  number  = {4},
  pages   = {045011},
  doi     = {10.1088/0953-2048/28/4/045011},
}

@article{Hansen_2024_tokamaker,
   title={TokaMaker: An open-source time-dependent Grad-Shafranov tool for the design and modeling of axisymmetric fusion devices},
   volume={298},
   ISSN={0010-4655},
   url={http://dx.doi.org/10.1016/j.cpc.2024.109111},
   DOI={10.1016/j.cpc.2024.109111},
   journal={Computer Physics Communications},
   publisher={Elsevier BV},
   author={Hansen, C. and Stewart, I.G. and Burgess, D. and Pharr, M. and Guizzo, S. and Logak, F. and Nelson, A.O. and Paz-Soldan, C.},
   year={2024},
   month=5, pages={109111} }

@article{Pitts_2007,
  title = {Chapter 5: Physics of the ITER divertor},
  author = {Pitts, R. A. and Carpentier, S. and Escourbiac, F. and Hirai, T. and Komarov, V. and Kukushkin, A. S. and Lisgo, S. and Loarte, A. and Merola, M. and Mitteau, R. and Raffray, A. R. and Shimada, M. and Stangeby, P. C.},
  date = {2007},
  journaltitle = {Nuclear Fusion},
  volume = {47},
  number = {6},
  pages = {S143--S194},
  doi = {10.1088/0029-5515/47/6/S05},
  url = {https://doi.org/10.1088/0029-5515/47/6/S05}
}

@article{Eich_2013,
  title = {Scaling of the tokamak near the scrape-off layer H-mode power width and implications for ITER},
  author = {Eich, T. and Leonard, A. W. and Pitts, R. A. and Fundamenski, W. and Gray, T. K. and Herrmann, A. and Kirk, A. and Kallenbach, A. and Kukushkin, A. S. and LaBombard, B. and Maingi, R. and Makowski, M. A. and Scarabosio, A. and Sieglin, B. and Terry, J. L. and Thornton, A. J.},
  date = {2013},
  journaltitle = {Nuclear Fusion},
  volume = {53},
  number = {9},
  pages = {093031},
  doi = {10.1088/0029-5515/53/9/093031},
  url = {https://doi.org/10.1088/0029-5515/53/9/093031}
}

@article{Scarabosio_2013,
  title = {Scaling of the tokamak scrape-off layer power fall-off length in L-mode plasmas},
  author = {Scarabosio, A. and Eich, T. and Fundamenski, W. and Leonard, A. W. and Pitts, R. A. and Sieglin, B. and Thornton, A. J. and others},
  date = {2013},
  journaltitle = {Nuclear Fusion},
  volume = {53},
  number = {11},
  pages = {113002},
  doi = {10.1088/0029-5515/53/11/113002},
  url = {https://doi.org/10.1088/0029-5515/53/11/113002}
}

@misc{lunia2025energyconfinementtimescaling,
      title={Energy Confinement Time Scaling Law Derived from Paz-Soldan NF 2024}, 
      author={P. Lunia and A. O. Nelson and C. Paz-Soldan},
      year={2025},
      eprint={2509.04279},
      archivePrefix={arXiv},
      primaryClass={physics.plasm-ph},
      url={https://arxiv.org/abs/2509.04279}, 
}

@article{10.1063/1.5091802,
  title = {H-Mode Grade Confinement in {{L-mode}} Edge Plasmas at Negative Triangularity on {{DIII-D}}},
  author = {Marinoni, A. and Austin, M. E. and Hyatt, A. W. and Walker, M. L. and Candy, J. and Chrystal, C. and Lasnier, C. J. and McKee, G. R. and Odstrčil, T. and Petty, C. C. and Porkolab, M. and Rost, J. C. and Sauter, O. and Smith, S. P. and Staebler, G. M. and Sung, C. and Thome, K. E. and Turnbull, A. D. and Zeng, L. and Team, DIII-D},
  date = {2019-04},
  journaltitle = {Physics of Plasmas},
  volume = {26},
  number = {4},
  eprint = {https://pubs.aip.org/aip/pop/article-pdf/doi/10.1063/1.5091802/19778787/042515_1_online.pdf},
  pages = {042515},
  issn = {1070-664X},
  doi = {10.1063/1.5091802},
  url = {https://doi.org/10.1063/1.5091802}
}

@article{PhysRevLett.122.115001,
  title = {Achievement of Reactor-Relevant Performance in Negative Triangularity Shape in the DIII-D Tokamak},
  author = {Austin, M. E. and Marinoni, A. and Walker, M. L. and Brookman, M. W. and deGrassie, J. S. and Hyatt, A. W. and McKee, G. R. and Petty, C. C. and Rhodes, T. L. and Smith, S. P. and Sung, C. and Thome, K. E. and Turnbull, A. D.},
  journal = {Phys. Rev. Lett.},
  volume = {122},
  issue = {11},
  pages = {115001},
  numpages = {5},
  year = {2019},
  month = {3},
  publisher = {American Physical Society},
  doi = {10.1103/PhysRevLett.122.115001},
  url = {https://link.aps.org/doi/10.1103/PhysRevLett.122.115001}
}

@article{Aucone_2024,
  title = {Experiments and Modelling of Negative Triangularity {{ASDEX Upgrade}} Plasmas in View of {{DTT}} Scenarios},
  author = {Aucone, L and Mantica, P and Happel, T and Hobirk, J and Pütterich, T and Vanovac, B and Zimmermann, C F B and Bernert, M and Bolzonella, T and Cavedon, M and Dunne, M and Fischer, R and Innocente, P and Kappatou, A and McDermott, R M and Mariani, A and Muscente, P and Plank, U and Sciortino, F and Tardini, G and WPTE Team, the EUROfusion and Upgrade Team, the ASDEX},
  date = {2024-05},
  journaltitle = {Plasma Physics and Controlled Fusion},
  volume = {66},
  number = {7},
  pages = {075013},
  publisher = {IOP Publishing},
  doi = {10.1088/1361-6587/ad4d1c},
  url = {https://doi.org/10.1088/1361-6587/ad4d1c}
}

@article{Mariani_2024,
  title = {First-Principle Based Predictions of the Effects of Negative Triangularity on {{DTT}} Scenarios},
  author = {Mariani, A. and Balestri, A. and Mantica, P. and Merlo, G. and Ambrosino, R. and Balbinot, L. and Brioschi, D. and Casiraghi, I. and Castaldo, A. and Frassinetti, L. and Fusco, V. and Innocente, P. and Sauter, O. and Vlad, G.},
  date = {2024-02},
  journaltitle = {Nuclear Fusion},
  volume = {64},
  number = {4},
  pages = {046018},
  publisher = {IOP Publishing},
  doi = {10.1088/1741-4326/ad2abc},
  url = {https://doi.org/10.1088/1741-4326/ad2abc}
}

@article{WANG_OPENMC,
title = {An OpenMC model of the SPARC tokamak for the diagnostic scoping studies},
journal = {Fusion Engineering and Design},
volume = {221},
pages = {115390},
year = {2025},
issn = {0920-3796},
doi = {https://doi.org/10.1016/j.fusengdes.2025.115390},
url = {https://www.sciencedirect.com/science/article/pii/S0920379625005861},
author = {X. Wang and R. Gocht and J. Ball and S. Mackie and E. Panontin and E. Peterson and P. Raj and I. Holmes and A.A. Saltos and A. Johnson and A. Grieve and R.A. Tinguely}
}

@article{GUIZZO2025115257,
title = {Electromagnetic system conceptual design for a negative triangularity tokamak},
journal = {Fusion Engineering and Design},
volume = {219},
pages = {115257},
year = {2025},
issn = {0920-3796},
doi = {https://doi.org/10.1016/j.fusengdes.2025.115257},
url = {https://www.sciencedirect.com/science/article/pii/S0920379625004533},
author = {S. Guizzo and M.A. Drabinskiy and C. Hansen and A.G. Kachkin and E.N. Khairutdinov and A.O. Nelson and M.R. Nurgaliev and M. Pharr and G.F. Subbotin and C. Paz-Soldan}
}

@article{wilsonCharacterizingNegativeTriangularity2025a,
  title = {Characterizing the Negative Triangularity Reactor Core Operating Space with Integrated Modeling},
  author = {Wilson, H S and Nelson, A O and McClenaghan, J and Rodriguez-Fernandez, P and Parisi, J and Paz-Soldan, C},
  date = {2025-01-31},
  journaltitle = {Plasma Physics and Controlled Fusion},
  shortjournal = {Plasma Phys. Control. Fusion},
  volume = {67},
  number = {1},
  pages = {015026},
  issn = {0741-3335, 1361-6587},
  doi = {10.1088/1361-6587/ad9be5},
  url = {https://iopscience.iop.org/article/10.1088/1361-6587/ad9be5},
  urldate = {2026-03-06},
  langid = {english}
}

@techreport{grant_sheahen_hts_costs,
  title        = {Cost Projections for High Temperature Superconductors},
  author       = {Grant, Paul M. and Sheahen, Thomas P.},
  institution  = {Electric Power Research Institute (EPRI) and Science Applications International Corporation (SAIC)},
  address      = {Palo Alto, CA and Gaithersburg, MD},
  year         = {2002},
  note         = {EPRI, Palo Alto, CA 94304; SAIC, Gaithersburg, MD 20878}
}

@article{mohammadFatigueBehaviorAustenitic2012,
    title = {Fatigue behavior of {Austenitic} {Type} {316L} {Stainless} {Steel}},
    volume = {36},
    issn = {1757-899X},
    url = {https://iopscience.iop.org/article/10.1088/1757-899X/36/1/012012},
    doi = {10.1088/1757-899X/36/1/012012},
    language = {en},
    urldate = {2026-03-22},
    journal = {IOP Conference Series: Materials Science and Engineering},
    author = {Mohammad, K A and Ali, Aidy and Sahari, B B and Abdullah, S},
    month = 9,
    year = {2012},
    pages = {012012},
}

@misc{guizzoAssessmentVerticalStability2024a,
    title = {Assessment of vertical stability for negative triangularity pilot plants},
    url = {http://arxiv.org/abs/2401.15217},
    doi = {10.48550/arXiv.2401.15217},
    language = {en},
    urldate = {2026-03-23},
    publisher = {arXiv},
    author = {Guizzo, S. and Nelson, A. O. and Hansen, C. and Logak, F. and Paz-Soldan, C.},
    month = 1,
    year = {2024},
    note = {arXiv:2401.15217 [physics]},
}

@MISC{Body2024-bn,
  title     = "cfs-energy/cfspopcon: v7.0.2",
  author    = "Body, Tom and Hasse, Christoph and Saltzman, Audrey and Wang,
               Allen and {IsaacSavona} and Nelson, Oak and Looby, Tom",
  publisher = "Zenodo",
  year      =  2024
}

@article{Scotti_2024,
    title = {High performance power handling in the absence of an H-mode edge in negative triangularity DIII-D plasmas},
    volume = {64},
    number = {9},
    issn = {0029-5515},
    url = {https://doi.org/10.1088/1741-4326/ad5f41},
    doi = {10.1088/1741-4326/ad5f41},
    language = {en},
    urldate = {2026-03-31},
    journal = {Nuclear Fusion},
    author = {Scotti, F. and Sabbagh, S. A. and Snyder, P. and Ferron, J. R. and DIII-D Team},
    month = 9,
    year = {2024},
    pages = {094001},
}

@article{fajardoAnalyticalModelCombined2023,
  title = {Analytical Model for the Combined Effects of Rotation and Collisionality on Neoclassical Impurity Transport},
  author = {Fajardo, D and Angioni, C and Casson, F J and Field, A R and Maget, P and Manas, P and Team, the ASDEX Upgrade and Contributors, J. E. T.},
  date = {2023-02},
  journaltitle = {Plasma Physics and Controlled Fusion},
  shortjournal = {Plasma Phys. Control. Fusion},
  volume = {65},
  number = {3},
  pages = {035021},
  publisher = {IOP Publishing},
  issn = {0741-3335},
  doi = {10.1088/1361-6587/acb0fc},
  url = {https://doi.org/10.1088/1361-6587/acb0fc},
  urldate = {2026-04-27},
  langid = {english}
}

@article{whyte_experimental_2024,
	title = {Experimental {Assessment} and {Model} {Validation} of the {SPARC} {Toroidal} {Field} {Model} {Coil}},
	volume = {34},
	language = {en},
	number = {2},
	journal = {IEEE TRANSACTIONS ON APPLIED SUPERCONDUCTIVITY},
	author = {Whyte, D G and LaBombard, B and Doody, J and Golﬁnopolous, T and Granetz, R and Lammi, C and Lane-Walsh, S and Michael, P and Mouratidis, T and Mumgaard, R and Muncks, J P and Nash, D and Riva, N and Santoro, F and Sattarov, A and Stillerman, J and Uppalapati, K and Vieira, R and Watterson, A and Wilcox, S and Hartwig, Z S},
	year = {2024},
}

@article{wolf_critical_2018,
	title = {Critical {Current} {Densities} of 482 {A}/mm$^{\textrm{2}}$ in {HTS} {CrossConductors} at 4.2 {K} and 12 {T}},
	volume = {28},
	copyright = {https://ieeexplore.ieee.org/Xplorehelp/downloads/license-information/IEEE.html},
	issn = {1051-8223, 1558-2515},
	url = {https://ieeexplore.ieee.org/document/8316945/},
	doi = {10.1109/TASC.2018.2815767},
	language = {en},
	number = {4},
	urldate = {2026-02-12},
	journal = {IEEE Transactions on Applied Superconductivity},
	author = {Wolf, Michael J. and Bagrets, Nadezda and Fietz, Walter H. and Lange, Christian and Weiss, Klaus-Peter},
	month = jun,
	year = {2018},
	pages = {1--4},
}

@article{eidietis_itpa_2015,
	title = {The {ITPA} disruption database},
	volume = {55},
	issn = {0029-5515},
	url = {https://doi.org/10.1088/0029-5515/55/6/063030},
	doi = {10.1088/0029-5515/55/6/063030},
	language = {en},
	number = {6},
	urldate = {2026-01-16},
	journal = {Nuclear Fusion},
	publisher = {IOP Publishing},
	author = {Eidietis, N.W. and Gerhardt, S.P. and Granetz, R.S. and Kawano, Y. and Lehnen, M. and Lister, J.B. and Pautasso, G. and Riccardo, V. and Tanna, R.L. and Thornton, A.J. and Participants, The ITPA Disruption Database},
	month = may,
	year = {2015},
	pages = {063030},
}

@article{hansen_tokamaker_2024,
	title = {{TokaMaker}: {An} open-source time-dependent {Grad}-{Shafranov} tool for the design and modeling of axisymmetric fusion devices},
	volume = {298},
	issn = {00104655},
	shorttitle = {{TokaMaker}},
	url = {https://linkinghub.elsevier.com/retrieve/pii/S0010465524000341},
	doi = {10.1016/j.cpc.2024.109111},
	language = {en},
	urldate = {2026-01-16},
	journal = {Computer Physics Communications},
	author = {Hansen, C. and Stewart, I.G. and Burgess, D. and Pharr, M. and Guizzo, S. and Logak, F. and Nelson, A.O. and Paz-Soldan, C.},
	month = may,
	year = {2024},
	pages = {109111},
}

@article{the_manta_collaboration_manta_2024,
	title = {{MANTA}: a negative-triangularity {NASEM}-compliant fusion pilot plant},
	volume = {66},
	issn = {0741-3335, 1361-6587},
	shorttitle = {{MANTA}},
	url = {https://iopscience.iop.org/article/10.1088/1361-6587/ad6708},
	doi = {10.1088/1361-6587/ad6708},
	language = {en},
	number = {10},
	urldate = {2026-01-14},
	journal = {Plasma Physics and Controlled Fusion},
	author = {{The MANTA Collaboration} and Rutherford, G and Wilson, H S and Saltzman, A and Arnold, D and Ball, J L and Benjamin, S and Bielajew, R and De Boucaud, N and Calvo-Carrera, M and Chandra, R and Choudhury, H and Cummings, C and Corsaro, L and DaSilva, N and Diab, R and Devitre, A R and Ferry, S and Frank, S J and Hansen, C J and Jerkins, J and Johnson, J D and Lunia, P and Van De Lindt, J and Mackie, S and Maris, A D and Mandell, N R and Miller, M A and Mouratidis, T and Nelson, A O and Pharr, M and Peterson, E E and Rodriguez-Fernandez, P and Segantin, S and Tobin, M and Velberg, A and Wang, A M and Wigram, M and Witham, J and Paz-Soldan, C and Whyte, D G},
	month = oct,
	year = {2024},
	pages = {105006},
}

@article{sanabria_development_2024,
	title = {Development of a high current density, high temperature superconducting cable for pulsed magnets},
	volume = {37},
	issn = {0953-2048, 1361-6668},
	url = {https://iopscience.iop.org/article/10.1088/1361-6668/ad7efc},
	doi = {10.1088/1361-6668/ad7efc},
	language = {en},
	number = {11},
	urldate = {2026-01-14},
	journal = {Superconductor Science and Technology},
	author = {Sanabria, Charlie and Radovinsky, Alexey and Craighill, Christopher and Uppalapati, Kiran and Warner, Alex and Colque, Julio and Allen, Elle and Evcimen, Sera and Heller, Sam and Chavarria, David and Metcalfe, Kristen and Lenzen, Saehan and Hubbard, Amanda and Watterson, Amy and Chamberlain, Sarah and Diaz-Pacheco, Rui and Weinreb, Benjamin and Brownell, Elizabeth and Nealey, Justin and Hughes, Annie and Laamanen, Eric and Vasudeva, Keshav and Nash, Daniel and McCormack, Colin and Salazar, Erica and Duke, Owen and Hicks, Matt and Adams, Jeremy and Kolb-Bond, Dylan and Liu, Timothy and Malhotra, Kara and Meichle, David P and Francis, Ashleigh and Cheng, Jl and Shepard, Maise and Greenberg, Aliya and Fry, Vinny and Kostifakis, Nick and Avola, Carl and Ljubicic, Paul and Palmer, Lex and Sundar Rajan, Gayatri and Padukone, Ronak and Kuznetsov, Sergey and Donez, Kai and Golfinopoulos, Theodore and Michael, Philip C and Vieira, Rui and Martovetsky, Nicolai and Badcock, Rodney and Davies, Mike and Hunze, Arvid and Ludbrook, Bart and Gupta, Ramesh and Joshi, Piyush and Joshi, Shresht and Yahia, Anis Ben and Bajas, Hugo and Jenni, Markus and Mueller, Christoph and Holenstein, Manuel and Sedlak, Kamil and Sorbom, Brandon and Brunner, Daniel},
	month = nov,
	year = {2024},
	pages = {115010},
}

@article{sugihara_plasma_1982,
	title = {Plasma {Design} {Considerations} of {Near} {Term} {Tokamak} {Fusion} {Experimental} {Reactor}},
	volume = {19},
	issn = {0022-3131},
	url = {https://doi.org/10.1080/18811248.1982.9734193},
	doi = {10.1080/18811248.1982.9734193},
	number = {8},
	urldate = {2026-01-14},
	journal = {Journal of Nuclear Science and Technology},
	publisher = {Taylor \& Francis},
	author = {SUGIHARA, Masayoshi and FUJISAWA, Nobol and UEDA, Kojyu and SAITO, Seiji and HATAYAMA, Akiyoshi and SHIMADA, Ryuichi},
	month = aug,
	year = {1982},
	note = {\_eprint: https://doi.org/10.1080/18811248.1982.9734193},
	pages = {628--637},
}

@article{file_large_1971,
	title = {Large {Superconducting} {Magnet} {Designs} for {Fusion} {Reactors}},
	volume = {18},
	issn = {1558-1578},
	url = {https://ieeexplore.ieee.org/document/4326354},
	doi = {10.1109/TNS.1971.4326354},
	number = {4},
	urldate = {2026-01-14},
	journal = {IEEE Transactions on Nuclear Science},
	author = {File, J. and Mills, R. G. and Sheffield, G. V.},
	month = aug,
	year = {1971},
	pages = {277--282},
}

@article{molodyk_development_2021,
	title = {Development and large volume production of extremely high current density {YBa2Cu3O7} superconducting wires for fusion},
	volume = {11},
	copyright = {2021 The Author(s)},
	issn = {2045-2322},
	url = {https://www.nature.com/articles/s41598-021-81559-z},
	doi = {10.1038/s41598-021-81559-z},
	language = {en},
	number = {1},
	urldate = {2026-01-14},
	journal = {Scientific Reports},
	publisher = {Nature Publishing Group},
	author = {Molodyk, A. and Samoilenkov, S. and Markelov, A. and Degtyarenko, P. and Lee, S. and Petrykin, V. and Gaifullin, M. and Mankevich, A. and Vavilov, A. and Sorbom, B. and Cheng, J. and Garberg, S. and Kesler, L. and Hartwig, Z. and Gavrilkin, S. and Tsvetkov, A. and Okada, T. and Awaji, S. and Abraimov, D. and Francis, A. and Bradford, G. and Larbalestier, D. and Senatore, C. and Bonura, M. and Pantoja, A. E. and Wimbush, S. C. and Strickland, N. M. and Vasiliev, A.},
	month = jan,
	year = {2021},
	pages = {2084},
}

@article{hartwig_sparc_2024,
	title = {The {SPARC} {Toroidal} {Field} {Model} {Coil} {Program}},
	volume = {34},
	issn = {1558-2515},
	url = {https://ieeexplore.ieee.org/document/10316582/},
	doi = {10.1109/TASC.2023.3332613},
	number = {2},
	urldate = {2026-01-14},
	journal = {IEEE Transactions on Applied Superconductivity},
	author = {Hartwig, Zachary S. and Vieira, Rui F. and Dunn, Darby and Golfinopoulos, Theodore and LaBombard, Brian and Lammi, Christopher J. and Michael, Philip C. and Agabian, Susan and Arsenault, David and Barnett, Raheem and Barry, Mike and Bartoszek, Larry and Beck, William K. and Bellofatto, David and Brunner, Daniel and Burke, William and Burrows, Jason and Byford, William and Cauley, Charles and Chamberlain, Sarah and Chavarria, David and Cheng, JL and Chicarello, James and Diep, Van and Dombrowski, Eric and Doody, Jeffrey and Doos, Raouf and Eberlin, Brian and Estrada, Jose and Fry, Vincent and Fulton, Matthew and Garberg, Sarah and Granetz, Robert and Greenberg, Aliya and Greenwald, Martin and Heller, Samuel and Hubbard, Amanda E. and Ihloff, Ernest and Irby, James H. and Iverson, Mark and Jardin, Peter and Korsun, Daniel and Kuznetsov, Sergey and Lane-Walsh, Stephen and Landry, Richard and Lations, Richard and Leccacorvi, Rick and Levine, Matthew and Mackay, George and Metcalfe, Kristen and Moazeni, Kevin and Mota, John and Mouratidis, Theodore and Mumgaard, Robert and Muncks, JP and Murray, Richard A. and Nash, Daniel and Nottingham, Ben and O'Shea, Colin and Pfeiffer, Andrew T. and Pierson, Samuel Z. and Purdy, Clayton and Radovinsky, Alexi and Ravikumar, Dhananjay K. and Reyes, Veronica and Riva, Nicolo and Rosati, Ron and Rowell, Michael and Salazar, Erica E. and Santoro, Fernando and Sattarov, Akhdiyor and Saunders, Wayne and Schweiger, Patrick and Schweiger, Shane and Shepard, Maise and Shiraiwa, Syun'ichi and Silveira, Maria and Snowman, FT and Sorbom, Brandon N. and Stahle, Peter and Stevens, Ken and Stillerman, Joshua and Tammana, Deepthi and Toland, Thomas L. and Tracey, David and Turcotte, Ronnie and Uppalapati, Kiran and Vernacchia, Matthew and Vidal, Christopher and Voirin, Erik and Warner, Alex and Watterson, Amy and Whyte, Dennis G. and Wilcox, Sidney and Wolf, Michael and Wood, Bruce and Zhou, Lihua and Zhukovsky, Alex},
	month = mar,
	year = {2024},
	pages = {1--16},
}

@article{nelson_2023,
	title = {Robust {Avoidance} of {Edge}-{Localized} {Modes} alongside {Gradient} {Formation} in the {Negative} {Triangularity} {Tokamak} {Edge}},
	volume = {131},
	issn = {0031-9007, 1079-7114},
	url = {https://link.aps.org/doi/10.1103/PhysRevLett.131.195101},
	doi = {10.1103/PhysRevLett.131.195101},
	language = {en},
	number = {19},
	urldate = {2026-05-02},
	journal = {Physical Review Letters},
	author = {Nelson, A. O. and Schmitz, L. and Paz-Soldan, C. and Thome, K. E. and Cote, T. B. and Leuthold, N. and Scotti, F. and Austin, M. E. and Hyatt, A. and Osborne, T.},
	month = nov,
	year = {2023},
	pages = {195101},
}

@article{paz-soldan_2024,
	title = {Simultaneous access to high normalized density, current, pressure, and confinement in strongly-shaped diverted negative triangularity plasmas},
	volume = {64},
	issn = {0029-5515},
	url = {https://doi.org/10.1088/1741-4326/ad69a4},
	doi = {10.1088/1741-4326/ad69a4},
	language = {en},
	number = {9},
	urldate = {2026-04-03},
	journal = {Nuclear Fusion},
	publisher = {IOP Publishing},
	author = {Paz-Soldan, C. and Chrystal, C. and Lunia, P. and Nelson, A.O. and Thome, K.E. and Austin, M.E. and Cote, T.B. and Hyatt, A.W. and Leuthold, N. and Marinoni, A. and Osborne, T.H. and Pharr, M. and Sauter, O. and Scotti, F. and Wilks, T.M. and Wilson, H.S.},
	month = aug,
	year = {2024},
	pages = {094002},
}

@article{nelson_2024a,
	title = {Implications of vertical stability control on the {SPARC} tokamak},
	volume = {64},
	issn = {0029-5515},
	url = {https://doi.org/10.1088/1741-4326/ad58f6},
	doi = {10.1088/1741-4326/ad58f6},
	language = {en},
	number = {8},
	urldate = {2026-04-01},
	journal = {Nuclear Fusion},
	publisher = {IOP Publishing},
	author = {Nelson, A.O. and Garnier, D.T. and Battaglia, D.J. and Paz-Soldan, C. and Stewart, I. and Reinke, M. and Creely, A.J. and Wai, J.},
	month = jun,
	year = {2024},
	pages = {086040},
}

@article{kuang_2020,
	title = {Divertor heat flux challenge and mitigation in {SPARC}},
	volume = {86},
	issn = {0022-3778, 1469-7807},
	url = {https://www.cambridge.org/core/journals/journal-of-plasma-physics/article/divertor-heat-flux-challenge-and-mitigation-in-sparc/A25A8CFADBBA33AD9AAC18F24E40A18E},
	doi = {10.1017/S0022377820001117},
	language = {en},
	number = {5},
	urldate = {2026-03-27},
	journal = {Journal of Plasma Physics},
	author = {Kuang, A. Q. and Ballinger, S. and Brunner, D. and Canik, J. and Creely, A. J. and Gray, T. and Greenwald, M. and Hughes, J. W. and Irby, J. and LaBombard, B. and Lipschultz, B. and Lore, J. D. and Reinke, M. L. and Terry, J. L. and Umansky, M. and Whyte, D. G. and Wukitch, S. and Team, the SPARC},
	month = oct,
	year = {2020},
	pages = {865860505},
}

@article{medvedev_2015,
	title = {The negative triangularity tokamak: stability limits and prospects as a fusion energy system},
	volume = {55},
	issn = {0029-5515},
	shorttitle = {The negative triangularity tokamak},
	url = {https://doi.org/10.1088/0029-5515/55/6/063013},
	doi = {10.1088/0029-5515/55/6/063013},
	language = {en},
	number = {6},
	urldate = {2026-03-05},
	journal = {Nuclear Fusion},
	publisher = {IOP Publishing},
	author = {Medvedev, S.Yu. and Kikuchi, M. and Villard, L. and Takizuka, T. and Diamond, P. and Zushi, H. and Nagasaki, K. and Duan, X. and Wu, Y. and Ivanov, A.A. and Martynov, A.A. and Poshekhonov, Yu.Yu. and Fasoli, A. and Sauter, O.},
	month = may,
	year = {2015},
	pages = {063013},
}

@article{pitts_2019,
	title = {Physics basis for the first {ITER} tungsten divertor},
	volume = {20},
	issn = {23521791},
	url = {https://linkinghub.elsevier.com/retrieve/pii/S2352179119300237},
	doi = {10.1016/j.nme.2019.100696},
	language = {en},
	urldate = {2026-03-05},
	journal = {Nuclear Materials and Energy},
	author = {Pitts, R.A. and Bonnin, X. and Escourbiac, F. and Frerichs, H. and Gunn, J.P. and Hirai, T. and Kukushkin, A.S. and Kaveeva, E. and Miller, M.A. and Moulton, D. and Rozhansky, V. and Senichenkov, I. and Sytova, E. and Schmitz, O. and Stangeby, P.C. and De Temmerman, G. and Veselova, I. and Wiesen, S.},
	month = aug,
	year = {2019},
	pages = {100696},
}

@article{suslova_2014,
	title = {Recrystallization and grain growth induced by {ELMs}-like transient heat loads in deformed tungsten samples},
	volume = {4},
	copyright = {2014 The Author(s)},
	issn = {2045-2322},
	url = {https://www.nature.com/articles/srep06845},
	doi = {10.1038/srep06845},
	language = {en},
	number = {1},
	urldate = {2026-02-16},
	journal = {Scientific Reports},
	publisher = {Nature Publishing Group},
	author = {Suslova, A. and El-Atwani, O. and Sagapuram, D. and Harilal, S. S. and Hassanein, A.},
	month = nov,
	year = {2014},
	pages = {6845},
}

@article{nelson_2024,
	title = {First access to {ELM}-free negative triangularity at low aspect ratio},
	volume = {64},
	issn = {0029-5515},
	url = {https://doi.org/10.1088/1741-4326/ad89db},
	doi = {10.1088/1741-4326/ad89db},
	language = {en},
	number = {12},
	urldate = {2026-02-10},
	journal = {Nuclear Fusion},
	publisher = {IOP Publishing},
	author = {Nelson, A.O. and Vincent, C. and Anand, H. and Lovell, J. and Parisi, J.F. and Wilson, H.S. and Imada, K. and Wehner, W.P. and Kochan, M. and Blackmore, S. and McArdle, G. and Guizzo, S. and Rondini, L. and Freiberger, S. and Paz-Soldan, C. and Team, the MAST-U.},
	month = oct,
	year = {2024},
	pages = {124004},
}
\end{document}